\documentclass[reprint,superscriptaddress,preprintnumbers,longbibliography,
amsmath,amssymb,aps,superscriptaddress,floatfix,tightenlines,prl
]{revtex4-2}
\usepackage{subfigure}
\usepackage{mathrsfs}
\usepackage[T1]{fontenc}
\usepackage{graphicx}
\usepackage{dcolumn}
\usepackage{bm}
\usepackage{color}
\usepackage{times}
\usepackage[colorlinks=true,
citecolor=magenta,
linkcolor=red,
anchorcolor=black,
urlcolor=purple]{hyperref}
\usepackage{overpic}
\usepackage{braket}
\usepackage{changes}
\definechangesauthor[name={syh},color=red]{syh}

\begin{document}
\title{Quantum simulation of topological zero modes on a 41-qubit superconducting processor}

\author{Yun-Hao Shi}
\thanks{These authors contributed equally to this work.}
\affiliation{Institute of Physics, Chinese Academy of Sciences, Beijing 100190, China}
\affiliation{School of Physical Sciences, University of Chinese Academy of Sciences, Beijing 100049, China}
\affiliation{Beijing Academy of Quantum Information Sciences, Beijing 100193, China}

\author{Yu Liu}
\thanks{These authors contributed equally to this work.}
\affiliation{Institute of Physics, Chinese Academy of Sciences, Beijing 100190, China}
\affiliation{School of Physical Sciences, University of Chinese Academy of Sciences, Beijing 100049, China}

\author{Yu-Ran Zhang}
\thanks{These authors contributed equally to this work.}
\affiliation{School of Physics and Optoelectronics, South China University of Technology, Guangzhou 510640, China}
\affiliation{Theoretical Quantum Physics Laboratory, Cluster for Pioneering Research, RIKEN, Wako-shi, Saitama 351-0198, Japan}
\affiliation{Center for Quantum Computing, RIKEN, Wako-shi, Saitama 351-0198, Japan}

\author{Zhongcheng Xiang}
\thanks{These authors contributed equally to this work.}
\affiliation{Institute of Physics, Chinese Academy of Sciences, Beijing 100190, China}
\affiliation{School of Physical Sciences, University of Chinese Academy of Sciences, Beijing 100049, China}

\author{Kaixuan Huang}
\affiliation{Beijing Academy of Quantum Information Sciences, Beijing 100193, China}

\author{Tao Liu}
\affiliation{School of Physics and Optoelectronics, South China University of Technology, Guangzhou 510640, China}

\author{Yong-Yi Wang}
\affiliation{Institute of Physics, Chinese Academy of Sciences, Beijing 100190, China}
\affiliation{School of Physical Sciences, University of Chinese Academy of Sciences, Beijing 100049, China}

\author{Jia-Chi Zhang}
\affiliation{Institute of Physics, Chinese Academy of Sciences, Beijing 100190, China}
\affiliation{School of Physical Sciences, University of Chinese Academy of Sciences, Beijing 100049, China}

\author{Cheng-Lin Deng}
\affiliation{Institute of Physics, Chinese Academy of Sciences, Beijing 100190, China}
\affiliation{School of Physical Sciences, University of Chinese Academy of Sciences, Beijing 100049, China}

\author{Gui-Han Liang}
\affiliation{Institute of Physics, Chinese Academy of Sciences, Beijing 100190, China}
\affiliation{School of Physical Sciences, University of Chinese Academy of Sciences, Beijing 100049, China}

\author{Zheng-Yang Mei}
\affiliation{Institute of Physics, Chinese Academy of Sciences, Beijing 100190, China}
\affiliation{School of Physical Sciences, University of Chinese Academy of Sciences, Beijing 100049, China}

\author{Hao Li}
\affiliation{Institute of Physics, Chinese Academy of Sciences, Beijing 100190, China}

\author{Tian-Ming Li}
\affiliation{Institute of Physics, Chinese Academy of Sciences, Beijing 100190, China}
\affiliation{School of Physical Sciences, University of Chinese Academy of Sciences, Beijing 100049, China}

\author{Wei-Guo Ma}
\affiliation{Institute of Physics, Chinese Academy of Sciences, Beijing 100190, China}
\affiliation{School of Physical Sciences, University of Chinese Academy of Sciences, Beijing 100049, China}

\author{Hao-Tian Liu}
\affiliation{Institute of Physics, Chinese Academy of Sciences, Beijing 100190, China}
\affiliation{School of Physical Sciences, University of Chinese Academy of Sciences, Beijing 100049, China}

\author{Chi-Tong Chen}
\affiliation{Institute of Physics, Chinese Academy of Sciences, Beijing 100190, China}
\affiliation{School of Physical Sciences, University of Chinese Academy of Sciences, Beijing 100049, China}

\author{Tong Liu}
\affiliation{Institute of Physics, Chinese Academy of Sciences, Beijing 100190, China}
\affiliation{School of Physical Sciences, University of Chinese Academy of Sciences, Beijing 100049, China}

\author{Ye Tian}
\affiliation{Institute of Physics, Chinese Academy of Sciences, Beijing 100190, China}

\author{Xiaohui Song}
\affiliation{Institute of Physics, Chinese Academy of Sciences, Beijing 100190, China}

\author{S. P. Zhao}
\affiliation{Institute of Physics, Chinese Academy of Sciences, Beijing 100190, China}
\affiliation{School of Physical Sciences, University of Chinese Academy of Sciences, Beijing 100049, China}
\affiliation{Songshan Lake  Materials Laboratory, Dongguan, Guangdong 523808, China}

\author{Kai Xu}
\email{kaixu@iphy.ac.cn}
\affiliation{Institute of Physics, Chinese Academy of Sciences, Beijing 100190, China}
\affiliation{School of Physical Sciences, University of Chinese Academy of Sciences, Beijing 100049, China}
\affiliation{Beijing Academy of Quantum Information Sciences, Beijing 100193, China}
\affiliation{Songshan Lake Materials Laboratory, Dongguan, Guangdong 523808, China}
\affiliation{CAS Center for Excellence in Topological Quantum Computation, UCAS, Beijing 100049, China}

\author{Dongning Zheng}
\email{dzheng@iphy.ac.cn}
\affiliation{Institute of Physics, Chinese Academy of Sciences, Beijing 100190, China}
\affiliation{School of Physical Sciences, University of Chinese Academy of Sciences, Beijing 100049, China}
\affiliation{Songshan Lake Materials Laboratory, Dongguan, Guangdong 523808, China}
\affiliation{CAS Center for Excellence in Topological Quantum Computation, UCAS, Beijing 100049, China}

\author{Franco Nori}
\email{fnori@riken.jp}
\affiliation{Theoretical Quantum Physics Laboratory, Cluster for Pioneering Research, RIKEN, Wako-shi, Saitama 351-0198, Japan}
\affiliation{Center for Quantum Computing, RIKEN, Wako-shi, Saitama 351-0198, Japan}
\affiliation{Physics Department, University of Michigan, Ann Arbor, Michigan 48109-1040, USA}

\author{Heng Fan}
\email{hfan@iphy.ac.cn}
\affiliation{Institute of Physics, Chinese Academy of Sciences, Beijing 100190, China}
\affiliation{School of Physical Sciences, University of Chinese Academy of Sciences, Beijing 100049, China}
\affiliation{Beijing Academy of Quantum Information Sciences, Beijing 100193, China}
\affiliation{Songshan Lake Materials Laboratory, Dongguan, Guangdong 523808, China}
\affiliation{CAS Center for Excellence in Topological Quantum Computation, UCAS, Beijing 100049, China}
\affiliation{Hefei National Laboratory, Hefei 230088, China}

\begin{abstract}
Quantum simulation of different exotic topological phases of quantum matter on a noisy intermediate-scale quantum (NISQ) processor is attracting growing interest. Here, we develop a one-dimensional 43-qubit superconducting quantum processor, named as \textit{Chuang-tzu}, to simulate and characterize emergent topological states. By engineering diagonal Aubry-Andr$\acute{\mathbf{e}}$-Harper (AAH) models, we experimentally demonstrate the Hofstadter butterfly energy spectrum. Using Floquet engineering, we verify the existence of the topological zero modes in the commensurate off-diagonal AAH models, which have never been experimentally realized before. Remarkably, the qubit number over 40 in our quantum processor is large enough to capture the substantial topological features of a quantum system from its complex band structure, including Dirac points, the energy gap's closing, the difference between even and odd number of sites, and the distinction between edge and bulk states. Our results establish a versatile hybrid quantum simulation approach to exploring quantum topological systems in the NISQ era.
\end{abstract}
\maketitle

The Aubry-Andr$\acute{\mathrm{e}}$-Harper (AAH) model~\cite{Harper1955,aubry1980} has been attracting considerable attention in various topics of  condensed matter physics, including Hofstadter butterfly~\cite{Hofstadter1976,Hatsugai1990}, Anderson localization~\cite{Anderson1958}, quasicrystals~\cite{Kraus2012}, and topological phases of matter~\cite{Hasan2010,Qi2011}. The incommensurate diagonal AAH model describes a one-dimensional (1D) tight bonding lattice with quasi-periodic potential. In this model
a localization transition is predicted \cite{aubry1980}, which has been observed experimentally \cite{Lahini2009,Li2022}.
Moreover, the diagonal AAH model can be exactly mapped to the two-dimensional (2D) Hofstadter model \cite{Hofstadter1976},
showing a 2D quantum Hall effect (QHE)
with topologically protected edge states, which have been observed in experiments \cite{Kraus2012,Xiang2022}.
The energy spectra of Bloch electrons in perpendicular magnetic fields versus
the dimensional perpendicular magnetic field $b$ form the Hofstadter butterfly \cite{Hofstadter1976,Hatsugai1990},
showing the splitting of energy bands for a specific value of $b$.
The Hofstadter butterfly energy spectrum has been measured in
quasi-periodic lattices \cite{Roushan2017,Ni2019,Hangleiter2021}, superlattices \cite{Dean2013,Lu2021,Rozhkov2016},
and Floquet dissipative quasicrystal \cite{Weidemann2022}.
A further generalization to commensurate off-diagonal AAH models,
with the hopping amplitude being cosine-modulated commensurate with the lattice,
indicates the existence of topological zero-energy edge states in the gapless regime \cite{Ganeshan2013}.
The topological zero modes differ from the edge states in the 1D diagonal AAH models (similar to the quantum Hall edge) and have never been observed in experiments before.

\begin{figure*}[t]
	\begin{center}
		\includegraphics[width=0.87\textwidth]{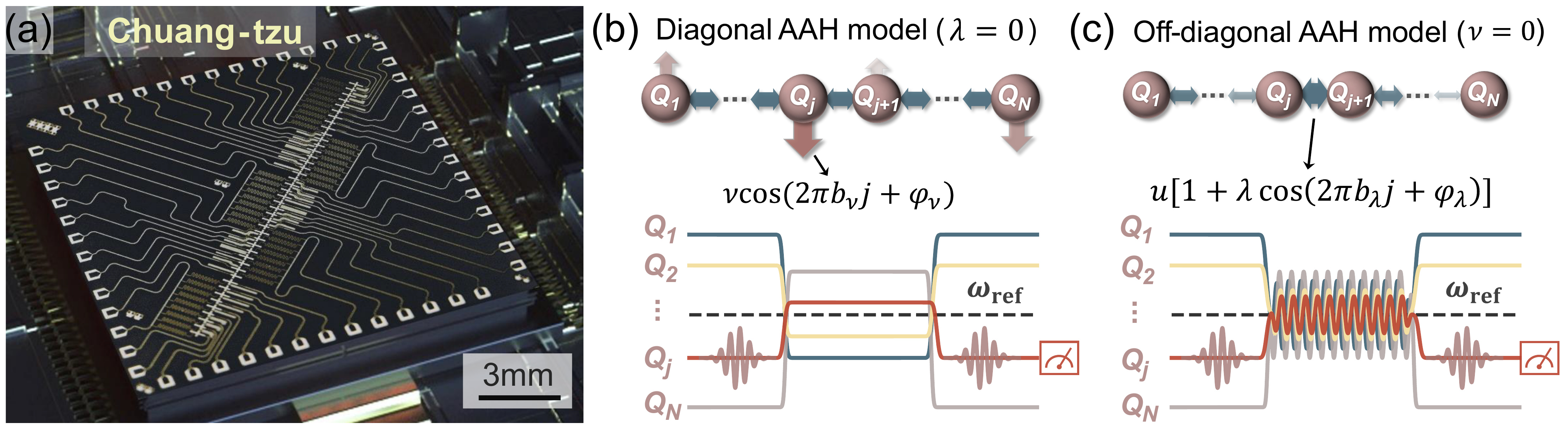}
		\caption{Device and pulse sequences. (a) Optical micrograph of the 43-qubit quantum chip.
			(b) Diagonal AAH model simulated by periodically tuning the qubit's frequency
			and the pulse sequence for its band structure spectroscopy.
			(c) Off-diagonal AAH model engineered by Floquet engineering qubit's frequency and
			the pulse sequence for its band structure spectroscopy. }
		\label{fig:1}
	\end{center}
\end{figure*}

Rapid developments in quantum techniques allow for programming non-trivial topological models and observing their topological states on quantum simulating platforms with a fast-growing number of qubits
\cite{Semeghini2021,Satzinger2021,Xiang2022,Mi2022}.
Even without fault tolerance, the programmability of a noisy intermediate-scale quantum (NISQ) processor helps to explore various topological phases that are still challenging in real materials \cite{Preskill2018, Nori2014,Daley2022,NP_Leefmans2022,Cheng2023}.
Here, we develop a 43-qubit superconducting quantum processor arranged in a 1D array,
named as \emph{Chuang-tzu} [Fig.~\ref{fig:1}(a)], to simulate the generalized 1D AAH model. The mean energy relaxation time and pure dephasing time of 41 qubits  in our experiments are 21.0 and 1.2$\mu$s, respectively. Since our processor is designed to fulfill the hard-core limit \cite{Yan2019,Xiang2022},
the effective Hamiltonian reads
\begin{equation}
	\label{eq:sample}
	\hat{H}_{0}=\sum_{j=1}^{N-1}g_{j,j+1}(\hat{a}^{\dagger}_{j}\hat{a}_{j+1}+\textrm{H.c.})+\sum_{j=1}^{N}\omega_j\hat{a}^{\dagger}_{j}\hat{a}_{j},
\end{equation}
where  $\hat{a}^{\dag}$  ($\hat{a}$) denotes the hard-core bosonic creation (annihilation) operator.
In our sample, the frequency $\omega_j$ of each qubit $Q_j$ is tunable, but the hopping strength $g_{j,j+1}$
between nearest-neighbor (NN) $Q_j$ and $Q_{j+1}$ cannot be tuned directly. Here
we use the Floquet engineering technique as demonstrated in \cite{Cai2019,Zhao2022,Denisov2014,Wu2018,Reagor2018,Lignier2007,Eckardt2017} to simulate
the generalized 1D AAH model with a form
\begin{eqnarray}
	\label{eq:AAH}
	\hat{H}_{\mathrm{gAAH}}&=&\sum_{j=1}^{N-1}u[1+\lambda\cos(2\pi b_\lambda j+\varphi_{\lambda})](\hat{a}^{\dagger}_{j}\hat{a}_{j+1}+\textrm{H.c.})\nonumber\\
	&&+\sum_{j=1}^{N}v\cos(2\pi b_vj+\varphi_v)\hat{a}^{\dagger}_{j}\hat{a}_{j},
\end{eqnarray}
with $\lambda\!=\!0$ and $v\!=\!0$ corresponding to the diagonal and off-diagonal AAH models, respectively. In our system, we can independently vary the effective on-site potential $\omega^{\textrm{eff}}_{j}$ and the effective hopping
strength $g^{\textrm{eff}}_{j,j+1}$ by the rectangle flux bias and time-periodic driving on
the Z control lines of qubits, respectively.
The effective  $g^{\textrm{eff}}_{j,j+1}$ can be adjusted from
about $-3.0$ to $7.6$MHz. Thus, the dynamics of the generalized AAH models
are simulated with an approximately effective Hamiltonian using Floquet engineering,
and our simulator behaves as a programmable hybrid analogue-digital quantum simulator
from the viewpoint in \cite{Daley2022}.
Details of tuning hopping strength via Floquet engineering are discussed in \cite{SM}.

\begin{figure*}[t]
	\begin{center}
		\includegraphics[width=0.86\textwidth]{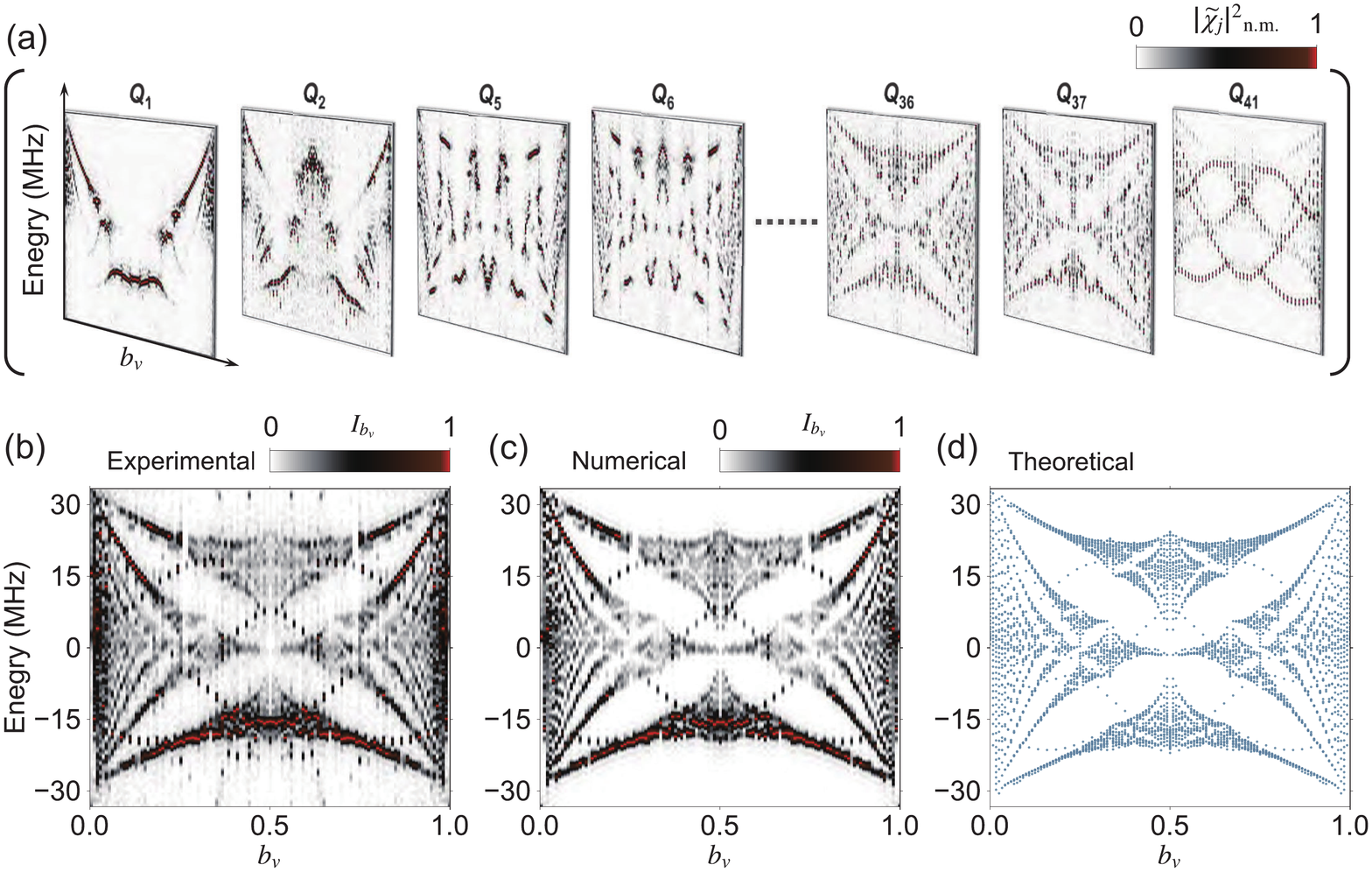}
		\caption{Hofstadter butterfly energy spectrum.
			By engineering various instances of AAH models, the energy spectrum of the Bloch electrons in perpendicular magnetic fields can be measured using band structure spectroscopy \cite{Roushan2017,Xiang2022}. Here we use $N=41$ qubits to simulate the quasi-periodic lattice. (a) Experimentally measured squared FT magnitudes $|\widetilde{\chi}_j|^2$ when choosing a target qubit $Q_j$.
			(b)-(d) Experimental data of $I_{b_v}\equiv\sum_{j}|\widetilde{\chi}_j|^2$ (b), the summation of the squared FT magnitudes, which is compared with the numerical data by simulating the dynamics of the system (c), and the theoretical prediction (d).
		}
		\label{fig:2}
	\end{center}
\end{figure*}

First, we engineer the diagonal AAH model~\cite{Satija2013,Degottardi2013} with $N=41$ qubits and measure the Hofstadter butterfly spectrum in the quasi-periodic lattices by setting $\lambda=0$ and tuning
the on-site potential as $v\cos(2\pi b_vj)$ with $v/(2\pi)\simeq 15.2$MHz and $\varphi_v=0$ [Fig.~\ref{fig:1}(b)].
We simulate 121 instances of diagonal AAH chains when varying $b_v$ from 0 to 1.
Using the band structure spectroscopic technique \cite{Roushan2017,Xiang2022},
we obtain the squared Fourier transformation (FT) magnitude $|\widetilde{\chi}_j|^2$ of the response function
$\chi_j(t)\equiv\langle\hat{\sigma}_j^x(t)\rangle+i\langle\hat{\sigma}_j^y(t)\rangle$,
after preparing a selected qubit $Q_j$ at $|+_j\rangle=(|0_j\rangle+|1_j\rangle)/\sqrt{2}$.
Figure~\ref{fig:2}(a) plots $|\widetilde{\chi}_j|^2$ for several selected qubits $Q_j$, and each
of them only contains partial information about the energy spectrum.
The summation of the squared FT magnitudes [Fig.~\ref{fig:2}(b)] of all chosen qubits $I_{b_v}\equiv\sum_{j}|\widetilde{\chi}_j|^2$
clearly shows the Hofstadter butterfly energy spectrum,
which agrees well with the numerical calculation by simulating the system's dynamics [Fig.~\ref{fig:2}(c)] and the theoretical prediction [Fig.~\ref{fig:2}(d)].
Note that the fractal structure of ``Hofstadter's butterfly'',
splitting of energy bands for several $b_v$, are clearly shown,
which is attributed to the sufficiently large qubit number of our quantum processor \cite{Ni2019}.
In addition, the wing-like gaps emerge because of the topological feature
of the diagonal AAH models, and the 2D integer QHE is characterized by
the Chern number \cite{Thouless1982}, which
has been experimentally investigated in \cite{Xiang2022} for $b_v={1}/{3}$.

\begin{figure*}[t]
	\begin{center}
		\includegraphics[width=0.86\textwidth]{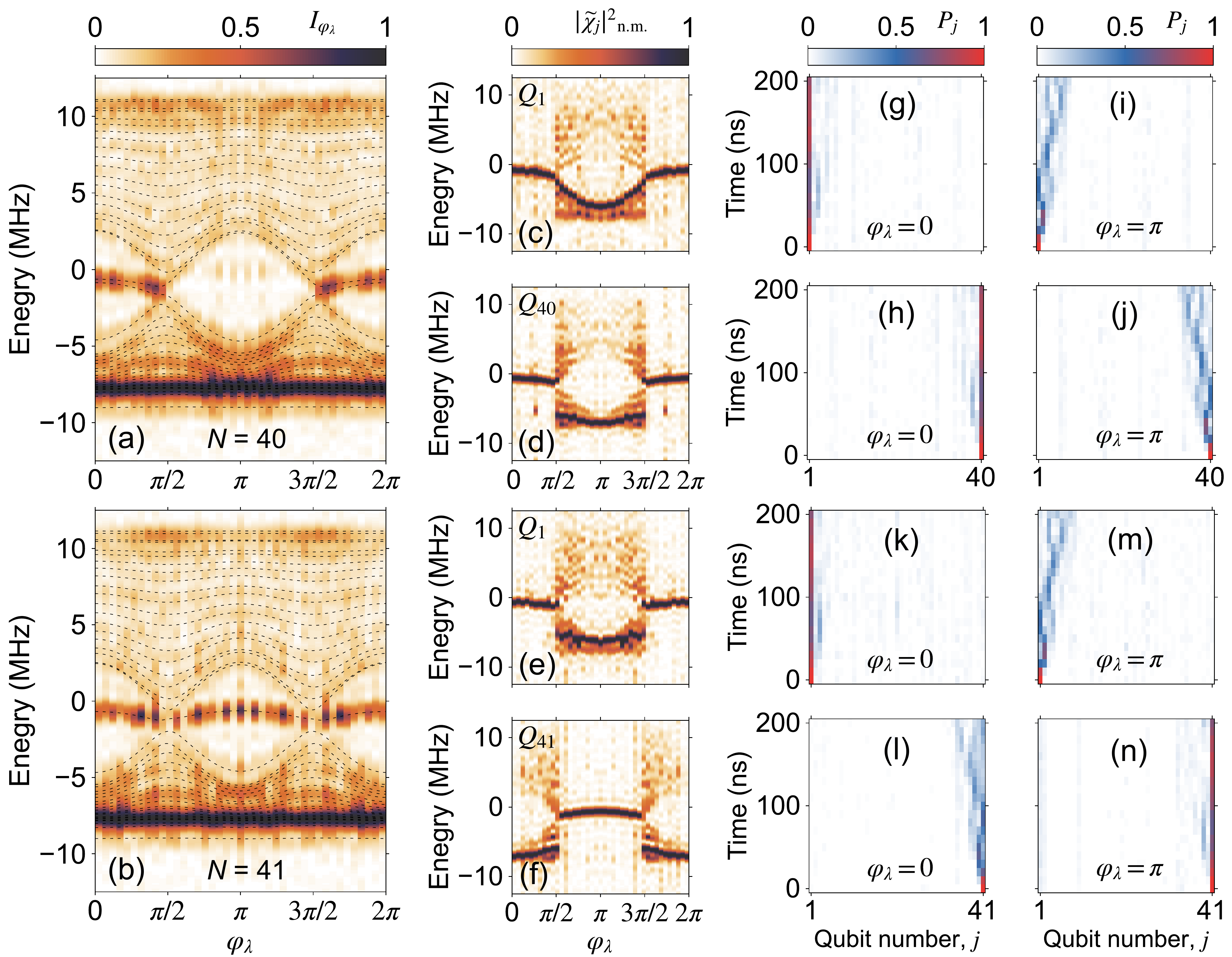}
		\caption{Experimental characterization of the topological zero-energy edge modes in commensurate off-diagonal AAH models for $\pi$-flux ($b_\lambda={1}/{2}$). Band structure spectroscopy of off-diagonal AAH models with even number $N=40$ (a) and odd number $N=41$ (b) of sites, which are compared with the theoretical projected band structures (dashed curves). Here $u/(2\pi)=4.78$MHz and $\lambda=0.4$. Normalized squared FT magnitudes $|\widetilde{\chi}_j|_{\textrm{n.m.}}^2$
			when choosing the leftmost qubit Q$_1$ (c) and the rightmost qubit Q$_{40}$ (d) as target qubits with
			$N=40$. $|\widetilde{\chi}_j|_{\textrm{n.m.}}^2$ for boundary qubits Q$_1$ (e) and
			Q$_{41}$ (f) as target qubits with $N=41$.
			(g--j) Time evolution of the excitation probability $P_j$ during the
			QWs of a single excitation initially prepared at the boundary qubits ($Q_1$ or $Q_{40}$)
			for $\varphi_\lambda=0$ (g,h) and $\varphi_\lambda=\pi$ (i,j) with $N=40$.
			(k--n) Time evolution of $P_j$ during the QWs of a single excitation initially placed at the boundary qubits
			($Q_1$ or $Q_{41}$) for $\varphi_\lambda=0$ (k,l) and $\varphi_\lambda=\pi$ (m,n) with $N=41$.
		}
		\label{fig:3}
	\end{center}
\end{figure*}

Next, we perform a hybrid analogue-digital quantum simulation of the off-diagonal
AAH models with $v=0$ and $\lambda\neq0$ using Floquet engineering \cite{Zhao2022} [Fig.~\ref{fig:1}(c)],
which show no QHE \cite{Ganeshan2013}.
With the bulk-edge correspondence \cite{Bansil2016,Xiang2022}, we characterize their topological zero-energy modes,
of which the experimental observation is still absent.
We first engineer the commensurate off-diagonal AAHs for $b_\lambda={1}/{2}$ that can
be mapped to a 2D Hofstadter model with $\pi$-flux per plaquette.
We experimentally extract the band structures of the lattices with $N=40$ (even) and $41$ (odd) sites
by measuring $I_{\varphi_\lambda}$ for $\varphi_\lambda\in[0,2\pi]$ as shown in
Fig.~\ref{fig:3}(a) and \ref{fig:3}(b), respectively,
which agrees well with the theoretical prediction (dashed curves).
The measured gapless band structure clearly shows two Dirac points with a linear dispersion, which is similar to those observed in graphene \cite{Ganeshan2013}.
On the lattice with $N=40$ (even) sites, two topological zero modes appear for $\varphi_\lambda\in(-{\pi}/{2},{\pi}/{2})$ [Fig.~\ref{fig:3}(a)], while the topological zero edge mode exists for the whole parameter regime [Fig.~\ref{fig:3}(b)] with $N=41$ (odd) sites. These exotic topological edge states are also verified from the experimentally measured squared FT magnitudes $|\widetilde{\chi}_j|^2$ for boundary qubits,
as shown in Fig.~\ref{fig:3}(c--f).
For even sites, the $|\widetilde{\chi}_1|^2$ [Fig.~\ref{fig:3}(c)] and $|\widetilde{\chi}_{40}|^2$ [Fig.~\ref{fig:3}(d)] for $Q_1$ and $Q_{40}$, respectively, both contain information of topological zero edge states in the regime $(-{\pi}/{2},{\pi}/{2})$.
However, for odd sites, the $|\widetilde{\chi}_1|^2$ [Fig.~\ref{fig:3}(e)]
for $Q_1$ shows left edge state for $(-{\pi}/{2},{\pi}/{2})$ and the $Q_{41}$'s $|\widetilde{\chi}_{41}|^2$
[Fig.~\ref{fig:3}(f)] implies the existence of the right edge mode for
$({\pi}/{2},{3\pi}/{2})$. The small shift of the zero energy of the edge state is attributed
to the existence of weak next-nearest-neighboring (NNN) hopping (with an average of about $0.7$MHz) of our sample that slightly breaks the particle-hole symmetry, 
see details in \cite{SM}.
Our experiments therefore verify the robustness of the topological zero-energy edge states in the commensurate off-diagonal AAH models.

\begin{figure*}[t]
	\begin{center}
		\includegraphics[width=0.86\textwidth]{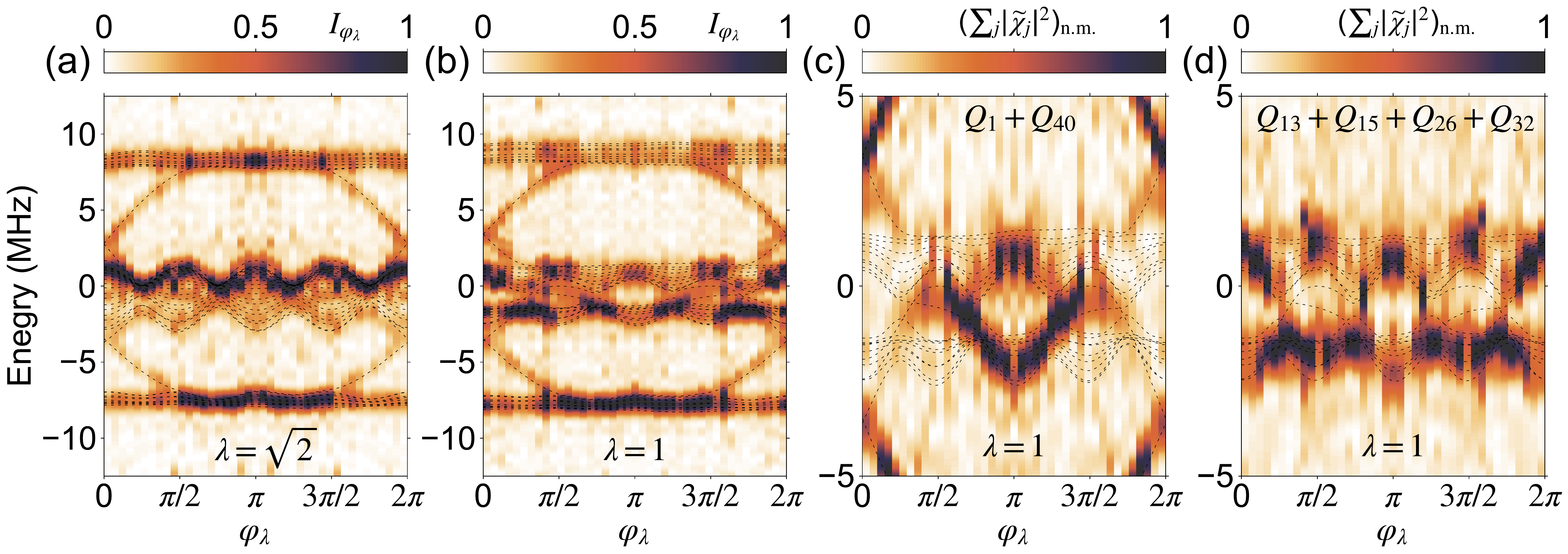}
		\caption{Band structure spectroscopy of generic commensurate off-diagonal AAH models
				with $N=40$ for $b_\lambda={1}/{4}$.
			(a) Experimental $I_{\varphi_\lambda}$ for $\lambda=\sqrt{2}$ and $u/(2\pi)=2.77$MHz.
			The gap between two central bands closes, and no topological edge states between these two bands are observed.
			(b) Experimental $I_{\varphi_\lambda}$ for $\lambda=1$ and $u/(2\pi)=3.35$MHz. The two central bands are clearly observed gapped without edge modes in the regimes $\varphi_\lambda\in(-{\pi}/{4},{\pi}/{4})$ and $({3\pi}/{4},{5\pi}/{4})$.
			(c) Normalized FT magnitudes of two boundary qubits $(|\widetilde{\chi}_1|^2+|\widetilde{\chi}_{40}|^2)_{\textrm{n.m.}}$,
			compared with the theoretical projected band structures (dashed curves).
			The topologically non-trivial zero-energy modes are observed between two central bands.
			(d) Four bulk qubits $(\sum_{j=13,15,26,32}|\widetilde{\chi}_j|^2)_{\textrm{n.m.}}$,
			compared to the theoretical projected band structures (dashed curves).
			Four band crossing points are observed near $\varphi_\lambda={\pi}/{4}$, ${3\pi}/{4}$,
			${5\pi}/{4}$, and ${7\pi}/{4}$.}
		\label{fig:4}
	\end{center}
\end{figure*}

Furthermore, the topological edge state can also be identified in real space by witnessing the localization of an edge excitation during its quantum walks (QWs) on the 1D qubit chain \cite{Cai2019,Xiang2022}, due to its main overlap with the edge state. We monitor the time evolution of the excitation probabilities $P_j$ for all qubits during the QWs.
For even sites, QWs of an excitation at either boundary qubit present localization for $\varphi_\lambda\!\!=\!\!0$ [Fig.~\ref{fig:3}(g,h)] in the topological regime and dispersion for $\varphi_\lambda\!\!=\!\!\pi$ [Fig.~\ref{fig:3}(i,j)] in the trivial regime, respectively.
In comparison, as shown in Fig.~\ref{fig:3}(k--n), the QWs of an excitation at $Q_1$ ($Q_{41}$) shows localization (diffusion)
for $\varphi_\lambda\!\!=\!\!0$ and diffusion (localization) for $\varphi_\lambda\!=\!\pi$.
Thus, our experimental results assert that there always exists only one zero-energy mode localized at either edge in the commensurate off-diagonal AAH models for $\pi$-flux with odd sites.
Note that it is still challenging to observe these different behaviors of topological edge modes between even and odd sites in real materials or some other quantum simulating platforms without a fixed number of lattice sites. In our NISQ device, the individually addressable superconducting qubits assisted by Floquet engineering help to overcome these difficulties and show its potential for investigating various exotic topological phenomena.

As the $\pi$-flux off-diagonal AAH model can be mapped to the Su-Schrieffer-Heeger
(SSH) model \cite{Su1979}, the off-diagonal AAH models as a new class of topological models
are given by $b_\lambda=1/(2q)$ with an integer $q>1$ \cite{Ganeshan2013}.
Here, we apply 40 qubits to experimentally investigate the generic off-diagonal AAH model for
$b_\lambda={1}/{4}$ by tuning $g_{j,j+1}^{\textrm{eff}}\!=\!u[1\!+\!\lambda\cos(2\pi b_\lambda j+\varphi_{\lambda})]$,
with $\varphi_\lambda$ varying from 0 to $2\pi$.
This model has four energy bands,
and the top and bottom bands are fully gapped, where the quantum Hall edge
states are clearly exhibited from the measured band structure, see Fig.~\ref{fig:4}.
By tuning $\lambda\!=\!\sqrt{2}$, we see that the central gap closes as theoretically predicted in \cite{Ganeshan2013}, see Fig.~\ref{fig:4}(a),
which is difficult to be realized with a small-scale quantum simulator. Then, we tune $\lambda\!=\!1$ and measure the
band structure as shown in Fig.~\ref{fig:4}(b), where the central
two bands are shown to have four band crossing points near $\varphi_\lambda\!=\!{\pi}/{4}$, ${3\pi}/{4}$,
${5\pi}/{4}$, and ${7\pi}/{4}$. Although the mid-gap is very small to observe, we can imply from the measured energy spectrum [Fig.~\ref{fig:4}(b)] that the central two bands are gapped in the regime $(-{\pi}/{4},{\pi}/{4})$ and $({3\pi}/{4},{5\pi}/{4})$; the topological edge states appear in the regime $({\pi}/{4},{3\pi}/{4})$ and $({5\pi}/{4},{7\pi}/{4})$.
To further analyze these two central bands, we separately study the edge and bulk states
from the FT signals by only considering the boundary and bulk qubits, respectively.
In Fig.~\ref{fig:4}(c), we plot the summation of the squared FT magnitudes of two boundary qubits
$Q_1$ and $Q_{40}$ versus $\varphi_\lambda$, which mainly shows the information for both the quantum Hall edges in the top and bottom gaps and the zero-energy edges between two central bands.
We also illustrate in Fig.~\ref{fig:4}(d) the summed FT signals for selected bulk
qubits $Q_{13}$, $Q_{15}$, $Q_{26}$, and $Q_{32}$,
indicating the existence of four band crossing points.
Note that the NNN hopping merely causes the shift of zero-energy edge states to mid-gap edges, which verifies the robustness of the topological properties of the commensurate off-diagonal AAH model.

In summary, we experimentally measure the celebrated Hofstadter butterfly energy spectra of up to 41 superconducting qubits and verify the existence of topological zero-energy edge modes in the gapless commensurate AAH models. We introduce multi-qubit Floquet engineering in superconducting circuits, which can be used to realize a wider range of models in condensed matter physics than AAH models, e.g., lattice gauge theories \cite{Schweizer2019} and non-Hermitian systems \cite{Wu2019}. In addition, we provide a general automatic calibration scheme for the devices with Floquet engineering (see details in \cite{SM}), which is also adaptable to other quantum simulating platforms. Our universal 1D hybrid analogue-digital quantum simulator shows the potential to use programmable NISQ device to investigate exotic topological phases of quantum matter that is still arduous to do in real materials.

\begin{acknowledgments}
We thank S. K. Zhao for helpful discussions. This work was supported by the Synergetic Extreme Condition User Facility (SECUF). Devices were made at the Nanofabrication Facilities at Institute of Physics, CAS in Beijing. This work was supported by: the National Natural Science Foundation of China (Grant Nos.~T2121001, 11934018, 12005155, 11904393,  92065114, 12204528, and 12274142), Key Area Research and Development Program of Guangdong Province, China (Grant Nos. 2020B0303030001, 2018B030326001), Strategic Priority Research Program of Chinese Academy of Sciences (Grant No. XDB28000000) and Beijing Natural Science Foundation (Grant No.~Z200009), NTT Research, ARO (Grant No.~W911NF-18-1-0358), JST (via the Q-LEAP, and the Moonshot R\&D Grant No.~JPMJMS2061), AOARD (Grant No.~FA2386-20-1-4069), and FQXi (Grant No.~FQXi-IAF19-06).
\end{acknowledgments}


\begin{thebibliography}{43}%
	\makeatletter
	\providecommand \@ifxundefined [1]{%
		\@ifx{#1\undefined}
	}%
	\providecommand \@ifnum [1]{%
		\ifnum #1\expandafter \@firstoftwo
		\else \expandafter \@secondoftwo
		\fi
	}%
	\providecommand \@ifx [1]{%
		\ifx #1\expandafter \@firstoftwo
		\else \expandafter \@secondoftwo
		\fi
	}%
	\providecommand \natexlab [1]{#1}%
	\providecommand \enquote  [1]{``#1''}%
	\providecommand \bibnamefont  [1]{#1}%
	\providecommand \bibfnamefont [1]{#1}%
	\providecommand \citenamefont [1]{#1}%
	\providecommand \href@noop [0]{\@secondoftwo}%
	\providecommand \href [0]{\begingroup \@sanitize@url \@href}%
	\providecommand \@href[1]{\@@startlink{#1}\@@href}%
	\providecommand \@@href[1]{\endgroup#1\@@endlink}%
	\providecommand \@sanitize@url [0]{\catcode `\\12\catcode `\$12\catcode
		`\&12\catcode `\#12\catcode `\^12\catcode `\_12\catcode `\%12\relax}%
	\providecommand \@@startlink[1]{}%
	\providecommand \@@endlink[0]{}%
	\providecommand \url  [0]{\begingroup\@sanitize@url \@url }%
	\providecommand \@url [1]{\endgroup\@href {#1}{\urlprefix }}%
	\providecommand \urlprefix  [0]{URL }%
	\providecommand \Eprint [0]{\href }%
	\providecommand \doibase [0]{https://doi.org/}%
	\providecommand \selectlanguage [0]{\@gobble}%
	\providecommand \bibinfo  [0]{\@secondoftwo}%
	\providecommand \bibfield  [0]{\@secondoftwo}%
	\providecommand \translation [1]{[#1]}%
	\providecommand \BibitemOpen [0]{}%
	\providecommand \bibitemStop [0]{}%
	\providecommand \bibitemNoStop [0]{.\EOS\space}%
	\providecommand \EOS [0]{\spacefactor3000\relax}%
	\providecommand \BibitemShut  [1]{\csname bibitem#1\endcsname}%
	\let\auto@bib@innerbib\@empty
	\bibitem [{\citenamefont {Harper}(1955)}]{Harper1955}%
	\BibitemOpen
	\bibfield  {author} {\bibinfo {author} {\bibfnamefont {P.~G.}\ \bibnamefont
			{Harper}},\ }\bibfield  {title} {\bibinfo {title} {{Single band motion of
				conduction electrons in a uniform magnetic field}},\ }\href
	{https://doi.org/10.1088/0370-1298/68/10/304} {\bibfield  {journal} {\bibinfo
			{journal} {Proc. Phys. Soc. London, Sect. A}\ }\textbf {\bibinfo {volume}
			{68}},\ \bibinfo {pages} {874} (\bibinfo {year} {1955})}\BibitemShut
	{NoStop}%
	\bibitem [{\citenamefont {Aubry}\ and\ \citenamefont
		{Andr{\'e}}(1980)}]{aubry1980}%
	\BibitemOpen
	\bibfield  {author} {\bibinfo {author} {\bibfnamefont {S.}~\bibnamefont
			{Aubry}}\ and\ \bibinfo {author} {\bibfnamefont {G.}~\bibnamefont
			{Andr{\'e}}},\ }\bibfield  {title} {\bibinfo {title} {Analyticity breaking
			and {A}nderson localization in incommensurate lattices},\ }\href@noop {}
	{\bibfield  {journal} {\bibinfo  {journal} {Ann. Israel Phys. Soc}\ }\textbf
		{\bibinfo {volume} {3}},\ \bibinfo {pages} {18} (\bibinfo {year}
		{1980})}\BibitemShut {NoStop}%
	\bibitem [{\citenamefont {Hofstadter}(1976)}]{Hofstadter1976}%
	\BibitemOpen
	\bibfield  {author} {\bibinfo {author} {\bibfnamefont {D.~R.}\ \bibnamefont
			{Hofstadter}},\ }\bibfield  {title} {\bibinfo {title} {{Energy levels and
				wave functions of Bloch electrons in rational and irrational magnetic
				fields}},\ }\href {https://doi.org/10.1103/PhysRevB.14.2239} {\bibfield
		{journal} {\bibinfo  {journal} {Phys. Rev. B}\ }\textbf {\bibinfo {volume}
			{14}},\ \bibinfo {pages} {2239} (\bibinfo {year} {1976})}\BibitemShut
	{NoStop}%
	\bibitem [{\citenamefont {Hatsugai}\ and\ \citenamefont
		{Kohmoto}(1990)}]{Hatsugai1990}%
	\BibitemOpen
	\bibfield  {author} {\bibinfo {author} {\bibfnamefont {Y.}~\bibnamefont
			{Hatsugai}}\ and\ \bibinfo {author} {\bibfnamefont {M.}~\bibnamefont
			{Kohmoto}},\ }\bibfield  {title} {\bibinfo {title} {{Energy spectrum and the
				quantum Hall effect on the square lattice with next-nearest-neighbor
				hopping}},\ }\href {https://doi.org/10.1103/PhysRevB.42.8282} {\bibfield
		{journal} {\bibinfo  {journal} {Phys. Rev. B}\ }\textbf {\bibinfo {volume}
			{42}},\ \bibinfo {pages} {8282} (\bibinfo {year} {1990})}\BibitemShut
	{NoStop}%
	\bibitem [{\citenamefont {Anderson}(1958)}]{Anderson1958}%
	\BibitemOpen
	\bibfield  {author} {\bibinfo {author} {\bibfnamefont {P.~W.}\ \bibnamefont
			{Anderson}},\ }\bibfield  {title} {\bibinfo {title} {Absence of diffusion in
			certain random lattices},\ }\href {https://doi.org/10.1103/PhysRev.109.1492}
	{\bibfield  {journal} {\bibinfo  {journal} {Phys. Rev.}\ }\textbf {\bibinfo
			{volume} {109}},\ \bibinfo {pages} {1492} (\bibinfo {year}
		{1958})}\BibitemShut {NoStop}%
	\bibitem [{\citenamefont {Kraus}\ \emph {et~al.}(2012)\citenamefont {Kraus},
		\citenamefont {Lahini}, \citenamefont {Ringel}, \citenamefont {Verbin},\ and\
		\citenamefont {Zilberberg}}]{Kraus2012}%
	\BibitemOpen
	\bibfield  {author} {\bibinfo {author} {\bibfnamefont {Y.~E.}\ \bibnamefont
			{Kraus}}, \bibinfo {author} {\bibfnamefont {Y.}~\bibnamefont {Lahini}},
		\bibinfo {author} {\bibfnamefont {Z.}~\bibnamefont {Ringel}}, \bibinfo
		{author} {\bibfnamefont {M.}~\bibnamefont {Verbin}},\ and\ \bibinfo {author}
		{\bibfnamefont {O.}~\bibnamefont {Zilberberg}},\ }\bibfield  {title}
	{\bibinfo {title} {{Topological states and adiabatic pumping in
				quasicrystals}},\ }\href {https://doi.org/10.1103/PhysRevLett.109.106402}
	{\bibfield  {journal} {\bibinfo  {journal} {Phys. Rev. Lett.}\ }\textbf
		{\bibinfo {volume} {109}},\ \bibinfo {pages} {106402} (\bibinfo {year}
		{2012})}\BibitemShut {NoStop}%
	\bibitem [{\citenamefont {Hasan}\ and\ \citenamefont {Kane}(2010)}]{Hasan2010}%
	\BibitemOpen
	\bibfield  {author} {\bibinfo {author} {\bibfnamefont {M.~Z.}\ \bibnamefont
			{Hasan}}\ and\ \bibinfo {author} {\bibfnamefont {C.~L.}\ \bibnamefont
			{Kane}},\ }\bibfield  {title} {\bibinfo {title} {Colloquium: {T}opological
			insulators},\ }\href {https://doi.org/10.1103/RevModPhys.82.3045} {\bibfield
		{journal} {\bibinfo  {journal} {Rev. Mod. Phys.}\ }\textbf {\bibinfo {volume}
			{82}},\ \bibinfo {pages} {3045} (\bibinfo {year} {2010})}\BibitemShut
	{NoStop}%
	\bibitem [{\citenamefont {Qi}\ and\ \citenamefont {Zhang}(2011)}]{Qi2011}%
	\BibitemOpen
	\bibfield  {author} {\bibinfo {author} {\bibfnamefont {X.-L.}\ \bibnamefont
			{Qi}}\ and\ \bibinfo {author} {\bibfnamefont {S.-C.}\ \bibnamefont {Zhang}},\
	}\bibfield  {title} {\bibinfo {title} {Topological insulators and
			superconductors},\ }\href {https://doi.org/10.1103/RevModPhys.83.1057}
	{\bibfield  {journal} {\bibinfo  {journal} {Rev. Mod. Phys.}\ }\textbf
		{\bibinfo {volume} {83}},\ \bibinfo {pages} {1057} (\bibinfo {year}
		{2011})}\BibitemShut {NoStop}%
	\bibitem [{\citenamefont {Lahini}\ \emph {et~al.}(2009)\citenamefont {Lahini},
		\citenamefont {Pugatch}, \citenamefont {Pozzi}, \citenamefont {Sorel},
		\citenamefont {Morandotti}, \citenamefont {Davidson},\ and\ \citenamefont
		{Silberberg}}]{Lahini2009}%
	\BibitemOpen
	\bibfield  {author} {\bibinfo {author} {\bibfnamefont {Y.}~\bibnamefont
			{Lahini}}, \bibinfo {author} {\bibfnamefont {R.}~\bibnamefont {Pugatch}},
		\bibinfo {author} {\bibfnamefont {F.}~\bibnamefont {Pozzi}}, \bibinfo
		{author} {\bibfnamefont {M.}~\bibnamefont {Sorel}}, \bibinfo {author}
		{\bibfnamefont {R.}~\bibnamefont {Morandotti}}, \bibinfo {author}
		{\bibfnamefont {N.}~\bibnamefont {Davidson}},\ and\ \bibinfo {author}
		{\bibfnamefont {Y.}~\bibnamefont {Silberberg}},\ }\bibfield  {title}
	{\bibinfo {title} {Observation of a localization transition in quasiperiodic
			photonic lattices},\ }\href {https://doi.org/10.1103/PhysRevLett.103.013901}
	{\bibfield  {journal} {\bibinfo  {journal} {Phys. Rev. Lett.}\ }\textbf
		{\bibinfo {volume} {103}},\ \bibinfo {pages} {013901} (\bibinfo {year}
		{2009})}\BibitemShut {NoStop}%
	\bibitem [{\citenamefont {Li}\ \emph {et~al.}(2023)\citenamefont {Li},
		\citenamefont {Wang}, \citenamefont {Shi}, \citenamefont {Huang},
		\citenamefont {Song}, \citenamefont {Liang}, \citenamefont {Mei},
		\citenamefont {Zhou}, \citenamefont {Zhang}, \citenamefont {Zhang},
		\citenamefont {Chen}, \citenamefont {Zhao}, \citenamefont {Tian},
		\citenamefont {Yang}, \citenamefont {Xiang}, \citenamefont {Xu},
		\citenamefont {Zheng},\ and\ \citenamefont {Fan}}]{Li2022}%
	\BibitemOpen
	\bibfield  {author} {\bibinfo {author} {\bibfnamefont {H.}~\bibnamefont
			{Li}}, \bibinfo {author} {\bibfnamefont {Y.-Y.}\ \bibnamefont {Wang}},
		\bibinfo {author} {\bibfnamefont {Y.-H.}\ \bibnamefont {Shi}}, \bibinfo
		{author} {\bibfnamefont {K.}~\bibnamefont {Huang}}, \bibinfo {author}
		{\bibfnamefont {X.}~\bibnamefont {Song}}, \bibinfo {author} {\bibfnamefont
			{G.-H.}\ \bibnamefont {Liang}}, \bibinfo {author} {\bibfnamefont {Z.-Y.}\
			\bibnamefont {Mei}}, \bibinfo {author} {\bibfnamefont {B.}~\bibnamefont
			{Zhou}}, \bibinfo {author} {\bibfnamefont {H.}~\bibnamefont {Zhang}},
		\bibinfo {author} {\bibfnamefont {J.-C.}\ \bibnamefont {Zhang}}, \bibinfo
		{author} {\bibfnamefont {S.}~\bibnamefont {Chen}}, \bibinfo {author}
		{\bibfnamefont {S.~P.}\ \bibnamefont {Zhao}}, \bibinfo {author}
		{\bibfnamefont {Y.}~\bibnamefont {Tian}}, \bibinfo {author} {\bibfnamefont
			{Z.-Y.}\ \bibnamefont {Yang}}, \bibinfo {author} {\bibfnamefont
			{Z.}~\bibnamefont {Xiang}}, \bibinfo {author} {\bibfnamefont
			{K.}~\bibnamefont {Xu}}, \bibinfo {author} {\bibfnamefont {D.}~\bibnamefont
			{Zheng}},\ and\ \bibinfo {author} {\bibfnamefont {H.}~\bibnamefont {Fan}},\
	}\bibfield  {title} {\bibinfo {title} {{Observation of critical phase
				transition in a generalized Aubry-Andr{\'{e}}-Harper model with
				superconducting circuits}},\ }\href
	{https://doi.org/10.1038/s41534-023-00712-w} {\bibfield  {journal} {\bibinfo
			{journal} {npj Quantum Inform.}\ }\textbf {\bibinfo {volume} {9}},\ \bibinfo
		{pages} {40} (\bibinfo {year} {2023})}\BibitemShut {NoStop}%
	\bibitem [{\citenamefont {Xiang}\ \emph {et~al.}(2022)\citenamefont {Xiang},
		\citenamefont {Huang}, \citenamefont {Zhang}, \citenamefont {Liu},
		\citenamefont {Shi}, \citenamefont {Deng}, \citenamefont {Liu}, \citenamefont
		{Li}, \citenamefont {Liang}, \citenamefont {Mei}, \citenamefont {Yu},
		\citenamefont {Xue}, \citenamefont {Tian}, \citenamefont {Song},
		\citenamefont {Liu}, \citenamefont {Xu}, \citenamefont {Zheng}, \citenamefont
		{Nori},\ and\ \citenamefont {Fan}}]{Xiang2022}%
	\BibitemOpen
	\bibfield  {author} {\bibinfo {author} {\bibfnamefont {Z.-C.}\ \bibnamefont
			{Xiang}}, \bibinfo {author} {\bibfnamefont {K.}~\bibnamefont {Huang}},
		\bibinfo {author} {\bibfnamefont {Y.-R.}\ \bibnamefont {Zhang}}, \bibinfo
		{author} {\bibfnamefont {T.}~\bibnamefont {Liu}}, \bibinfo {author}
		{\bibfnamefont {Y.-H.}\ \bibnamefont {Shi}}, \bibinfo {author} {\bibfnamefont
			{C.-L.}\ \bibnamefont {Deng}}, \bibinfo {author} {\bibfnamefont
			{T.}~\bibnamefont {Liu}}, \bibinfo {author} {\bibfnamefont {H.}~\bibnamefont
			{Li}}, \bibinfo {author} {\bibfnamefont {G.-H.}\ \bibnamefont {Liang}},
		\bibinfo {author} {\bibfnamefont {Z.-Y.}\ \bibnamefont {Mei}}, \bibinfo
		{author} {\bibfnamefont {H.}~\bibnamefont {Yu}}, \bibinfo {author}
		{\bibfnamefont {G.}~\bibnamefont {Xue}}, \bibinfo {author} {\bibfnamefont
			{Y.}~\bibnamefont {Tian}}, \bibinfo {author} {\bibfnamefont {X.}~\bibnamefont
			{Song}}, \bibinfo {author} {\bibfnamefont {Z.-B.}\ \bibnamefont {Liu}},
		\bibinfo {author} {\bibfnamefont {K.}~\bibnamefont {Xu}}, \bibinfo {author}
		{\bibfnamefont {D.}~\bibnamefont {Zheng}}, \bibinfo {author} {\bibfnamefont
			{F.}~\bibnamefont {Nori}},\ and\ \bibinfo {author} {\bibfnamefont
			{H.}~\bibnamefont {Fan}},\ }\bibfield  {title} {\bibinfo {title} {Simulating
			{C}hern insulators on a superconducting quantum processor},\ }\href
	{https://arxiv.org/abs/2207.11797} {\bibfield  {journal} {\bibinfo  {journal}
			{arXiv:2207.11797}\ } (\bibinfo {year} {2022})}\BibitemShut {NoStop}%
	\bibitem [{\citenamefont {Roushan}\ \emph {et~al.}(2017)\citenamefont {Roushan}
		\emph {et~al.}}]{Roushan2017}%
	\BibitemOpen
	\bibfield  {author} {\bibinfo {author} {\bibfnamefont {P.}~\bibnamefont
			{Roushan}} \emph {et~al.},\ }\bibfield  {title} {\bibinfo {title}
		{{Spectroscopic signatures of localization with interacting photons in
				superconducting qubits}},\ }\href {https://doi.org/10.1126/science.aao1401}
	{\bibfield  {journal} {\bibinfo  {journal} {Science}\ }\textbf {\bibinfo
			{volume} {358}},\ \bibinfo {pages} {1175} (\bibinfo {year}
		{2017})}\BibitemShut {NoStop}%
	\bibitem [{\citenamefont {Ni}\ \emph {et~al.}(2019)\citenamefont {Ni},
		\citenamefont {Chen}, \citenamefont {Weiner}, \citenamefont {Apigo},
		\citenamefont {Prodan}, \citenamefont {Al{\`u}}, \citenamefont {Prodan},\
		and\ \citenamefont {Khanikaev}}]{Ni2019}%
	\BibitemOpen
	\bibfield  {author} {\bibinfo {author} {\bibfnamefont {X.}~\bibnamefont
			{Ni}}, \bibinfo {author} {\bibfnamefont {K.}~\bibnamefont {Chen}}, \bibinfo
		{author} {\bibfnamefont {M.}~\bibnamefont {Weiner}}, \bibinfo {author}
		{\bibfnamefont {D.~J.}\ \bibnamefont {Apigo}}, \bibinfo {author}
		{\bibfnamefont {C.}~\bibnamefont {Prodan}}, \bibinfo {author} {\bibfnamefont
			{A.}~\bibnamefont {Al{\`u}}}, \bibinfo {author} {\bibfnamefont
			{E.}~\bibnamefont {Prodan}},\ and\ \bibinfo {author} {\bibfnamefont {A.~B.}\
			\bibnamefont {Khanikaev}},\ }\bibfield  {title} {\bibinfo {title}
		{Observation of {H}ofstadter butterfly and topological edge states in
			reconfigurable quasi-periodic acoustic crystals},\ }\href
	{https://doi.org/10.1038/s42005-019-0151-7} {\bibfield  {journal} {\bibinfo
			{journal} {Commun. Phys.}\ }\textbf {\bibinfo {volume} {2}},\ \bibinfo
		{pages} {55} (\bibinfo {year} {2019})}\BibitemShut {NoStop}%
	\bibitem [{\citenamefont {Hangleiter}\ \emph {et~al.}(2021)\citenamefont
		{Hangleiter}, \citenamefont {Roth}, \citenamefont {Eisert},\ and\
		\citenamefont {Roushan}}]{Hangleiter2021}%
	\BibitemOpen
	\bibfield  {author} {\bibinfo {author} {\bibfnamefont {D.}~\bibnamefont
			{Hangleiter}}, \bibinfo {author} {\bibfnamefont {I.}~\bibnamefont {Roth}},
		\bibinfo {author} {\bibfnamefont {J.}~\bibnamefont {Eisert}},\ and\ \bibinfo
		{author} {\bibfnamefont {P.}~\bibnamefont {Roushan}},\ }\bibfield  {title}
	{\bibinfo {title} {{Precise Hamiltonian identification of a superconducting
				quantum processor}},\ }\href {http://arxiv.org/abs/2108.08319} {\bibfield
		{journal} {\bibinfo  {journal} {arXiv:2108.08319}\ } (\bibinfo {year}
		{2021})}\BibitemShut {NoStop}%
	\bibitem [{\citenamefont {Dean}\ \emph {et~al.}(2013)\citenamefont {Dean},
		\citenamefont {Wang}, \citenamefont {Maher}, \citenamefont {Forsythe},
		\citenamefont {Ghahari}, \citenamefont {Gao}, \citenamefont {Katoch},
		\citenamefont {Ishigami}, \citenamefont {Moon}, \citenamefont {Koshino},
		\citenamefont {Taniguchi}, \citenamefont {Watanabe}, \citenamefont {Shepard},
		\citenamefont {Hone},\ and\ \citenamefont {Kim}}]{Dean2013}%
	\BibitemOpen
	\bibfield  {author} {\bibinfo {author} {\bibfnamefont {C.~R.}\ \bibnamefont
			{Dean}}, \bibinfo {author} {\bibfnamefont {L.}~\bibnamefont {Wang}}, \bibinfo
		{author} {\bibfnamefont {P.}~\bibnamefont {Maher}}, \bibinfo {author}
		{\bibfnamefont {C.}~\bibnamefont {Forsythe}}, \bibinfo {author}
		{\bibfnamefont {F.}~\bibnamefont {Ghahari}}, \bibinfo {author} {\bibfnamefont
			{Y.}~\bibnamefont {Gao}}, \bibinfo {author} {\bibfnamefont {J.}~\bibnamefont
			{Katoch}}, \bibinfo {author} {\bibfnamefont {M.}~\bibnamefont {Ishigami}},
		\bibinfo {author} {\bibfnamefont {P.}~\bibnamefont {Moon}}, \bibinfo {author}
		{\bibfnamefont {M.}~\bibnamefont {Koshino}}, \bibinfo {author} {\bibfnamefont
			{T.}~\bibnamefont {Taniguchi}}, \bibinfo {author} {\bibfnamefont
			{K.}~\bibnamefont {Watanabe}}, \bibinfo {author} {\bibfnamefont {K.~L.}\
			\bibnamefont {Shepard}}, \bibinfo {author} {\bibfnamefont {J.}~\bibnamefont
			{Hone}},\ and\ \bibinfo {author} {\bibfnamefont {P.}~\bibnamefont {Kim}},\
	}\bibfield  {title} {\bibinfo {title} {Hofstadter's butterfly and the fractal
			quantum {H}all effect in moir{\'e} superlattices},\ }\href
	{https://doi.org/10.1038/nature12186} {\bibfield  {journal} {\bibinfo
			{journal} {Nature}\ }\textbf {\bibinfo {volume} {497}},\ \bibinfo {pages}
		{598} (\bibinfo {year} {2013})}\BibitemShut {NoStop}%
	\bibitem [{\citenamefont {Lu}\ \emph {et~al.}(2021)\citenamefont {Lu},
		\citenamefont {Lian}, \citenamefont {Chaudhary}, \citenamefont {Piot},
		\citenamefont {Romagnoli}, \citenamefont {Watanabe}, \citenamefont
		{Taniguchi}, \citenamefont {Poggio}, \citenamefont {MacDonald}, \citenamefont
		{Bernevig},\ and\ \citenamefont {Efetov}}]{Lu2021}%
	\BibitemOpen
	\bibfield  {author} {\bibinfo {author} {\bibfnamefont {X.}~\bibnamefont
			{Lu}}, \bibinfo {author} {\bibfnamefont {B.}~\bibnamefont {Lian}}, \bibinfo
		{author} {\bibfnamefont {G.}~\bibnamefont {Chaudhary}}, \bibinfo {author}
		{\bibfnamefont {B.~A.}\ \bibnamefont {Piot}}, \bibinfo {author}
		{\bibfnamefont {G.}~\bibnamefont {Romagnoli}}, \bibinfo {author}
		{\bibfnamefont {K.}~\bibnamefont {Watanabe}}, \bibinfo {author}
		{\bibfnamefont {T.}~\bibnamefont {Taniguchi}}, \bibinfo {author}
		{\bibfnamefont {M.}~\bibnamefont {Poggio}}, \bibinfo {author} {\bibfnamefont
			{A.~H.}\ \bibnamefont {MacDonald}}, \bibinfo {author} {\bibfnamefont {B.~A.}\
			\bibnamefont {Bernevig}},\ and\ \bibinfo {author} {\bibfnamefont {D.~K.}\
			\bibnamefont {Efetov}},\ }\bibfield  {title} {\bibinfo {title} {Multiple flat
			bands and topological {H}ofstadter butterfly in twisted bilayer graphene
			close to the second magic angle},\ }\href
	{https://doi.org/10.1073/pnas.2100006118} {\bibfield  {journal} {\bibinfo
			{journal} {PNAS}\ }\textbf {\bibinfo {volume} {118}},\ \bibinfo {pages}
		{e2100006118} (\bibinfo {year} {2021})}\BibitemShut {NoStop}%
	\bibitem [{\citenamefont {Rozhkov}\ \emph {et~al.}(2016)\citenamefont
		{Rozhkov}, \citenamefont {Sboychakov}, \citenamefont {Rakhmanov},\ and\
		\citenamefont {Nori}}]{Rozhkov2016}%
	\BibitemOpen
	\bibfield  {author} {\bibinfo {author} {\bibfnamefont {A.~V.}\ \bibnamefont
			{Rozhkov}}, \bibinfo {author} {\bibfnamefont {A.~O.}\ \bibnamefont
			{Sboychakov}}, \bibinfo {author} {\bibfnamefont {A.~L.}\ \bibnamefont
			{Rakhmanov}},\ and\ \bibinfo {author} {\bibfnamefont {F.}~\bibnamefont
			{Nori}},\ }\bibfield  {title} {\bibinfo {title} {{Electronic properties of
				graphene-based bilayer systems}},\ }\href
	{https://doi.org/10.1016/j.physrep.2016.07.003} {\bibfield  {journal}
		{\bibinfo  {journal} {Phys. Rep.}\ }\textbf {\bibinfo {volume} {648}},\
		\bibinfo {pages} {1} (\bibinfo {year} {2016})}\BibitemShut {NoStop}%
	\bibitem [{\citenamefont {Weidemann}\ \emph {et~al.}(2022)\citenamefont
		{Weidemann}, \citenamefont {Kremer}, \citenamefont {Longhi},\ and\
		\citenamefont {Szameit}}]{Weidemann2022}%
	\BibitemOpen
	\bibfield  {author} {\bibinfo {author} {\bibfnamefont {S.}~\bibnamefont
			{Weidemann}}, \bibinfo {author} {\bibfnamefont {M.}~\bibnamefont {Kremer}},
		\bibinfo {author} {\bibfnamefont {S.}~\bibnamefont {Longhi}},\ and\ \bibinfo
		{author} {\bibfnamefont {A.}~\bibnamefont {Szameit}},\ }\bibfield  {title}
	{\bibinfo {title} {Topological triple phase transition in non-{H}ermitian
			{F}loquet quasicrystals},\ }\href
	{https://doi.org/10.1038/s41586-021-04253-0} {\bibfield  {journal} {\bibinfo
			{journal} {Nature}\ }\textbf {\bibinfo {volume} {601}},\ \bibinfo {pages}
		{354} (\bibinfo {year} {2022})}\BibitemShut {NoStop}%
	\bibitem [{\citenamefont {Ganeshan}\ \emph {et~al.}(2013)\citenamefont
		{Ganeshan}, \citenamefont {Sun},\ and\ \citenamefont {{Das
				Sarma}}}]{Ganeshan2013}%
	\BibitemOpen
	\bibfield  {author} {\bibinfo {author} {\bibfnamefont {S.}~\bibnamefont
			{Ganeshan}}, \bibinfo {author} {\bibfnamefont {K.}~\bibnamefont {Sun}},\ and\
		\bibinfo {author} {\bibfnamefont {S.}~\bibnamefont {{Das Sarma}}},\
	}\bibfield  {title} {\bibinfo {title} {{Topological zero-energy modes in
				gapless commensurate Aubry-Andr{\'{e}}-Harper models}},\ }\href
	{https://doi.org/10.1103/PhysRevLett.110.180403} {\bibfield  {journal}
		{\bibinfo  {journal} {Phys. Rev. Lett.}\ }\textbf {\bibinfo {volume} {110}},\
		\bibinfo {pages} {180403} (\bibinfo {year} {2013})}\BibitemShut {NoStop}%
	\bibitem [{\citenamefont {Semeghini}\ \emph {et~al.}(2021)\citenamefont
		{Semeghini}, \citenamefont {Levine}, \citenamefont {Keesling}, \citenamefont
		{Ebadi}, \citenamefont {Wang~T.}, \citenamefont {Bluvstein}, \citenamefont
		{Verresen}, \citenamefont {Pichler}, \citenamefont {Kalinowski},
		\citenamefont {Samajdar}, \citenamefont {Omran}, \citenamefont {Sachdev},
		\citenamefont {Vishwanath}, \citenamefont {Greiner}, \citenamefont
		{Vuleti{\'c}},\ and\ \citenamefont {Lukin}}]{Semeghini2021}%
	\BibitemOpen
	\bibfield  {author} {\bibinfo {author} {\bibfnamefont {G.}~\bibnamefont
			{Semeghini}}, \bibinfo {author} {\bibfnamefont {H.}~\bibnamefont {Levine}},
		\bibinfo {author} {\bibfnamefont {A.}~\bibnamefont {Keesling}}, \bibinfo
		{author} {\bibfnamefont {S.}~\bibnamefont {Ebadi}}, \bibinfo {author}
		{\bibfnamefont {T.}~\bibnamefont {Wang~T.}}, \bibinfo {author} {\bibfnamefont
			{D.}~\bibnamefont {Bluvstein}}, \bibinfo {author} {\bibfnamefont
			{R.}~\bibnamefont {Verresen}}, \bibinfo {author} {\bibfnamefont
			{H.}~\bibnamefont {Pichler}}, \bibinfo {author} {\bibfnamefont
			{M.}~\bibnamefont {Kalinowski}}, \bibinfo {author} {\bibfnamefont
			{R.}~\bibnamefont {Samajdar}}, \bibinfo {author} {\bibfnamefont
			{A.}~\bibnamefont {Omran}}, \bibinfo {author} {\bibfnamefont
			{S.}~\bibnamefont {Sachdev}}, \bibinfo {author} {\bibfnamefont
			{A.}~\bibnamefont {Vishwanath}}, \bibinfo {author} {\bibfnamefont
			{M.}~\bibnamefont {Greiner}}, \bibinfo {author} {\bibfnamefont
			{V.}~\bibnamefont {Vuleti{\'c}}},\ and\ \bibinfo {author} {\bibfnamefont
			{M.~D.}\ \bibnamefont {Lukin}},\ }\bibfield  {title} {\bibinfo {title}
		{Probing topological spin liquids on a programmable quantum simulator},\
	}\href {https://doi.org/10.1126/science.abi8794} {\bibfield  {journal}
		{\bibinfo  {journal} {Science}\ }\textbf {\bibinfo {volume} {374}},\ \bibinfo
		{pages} {1242} (\bibinfo {year} {2021})}\BibitemShut {NoStop}%
	\bibitem [{\citenamefont {Satzinger}\ \emph {et~al.}(2021)\citenamefont
		{Satzinger} \emph {et~al.}}]{Satzinger2021}%
	\BibitemOpen
	\bibfield  {author} {\bibinfo {author} {\bibfnamefont {K.~J.}\ \bibnamefont
			{Satzinger}} \emph {et~al.},\ }\bibfield  {title} {\bibinfo {title}
		{Realizing topologically ordered states on a quantum processor},\ }\href
	{https://doi.org/10.1126/science.abi8378} {\bibfield  {journal} {\bibinfo
			{journal} {Science}\ }\textbf {\bibinfo {volume} {374}},\ \bibinfo {pages}
		{1237} (\bibinfo {year} {2021})}\BibitemShut {NoStop}%
	\bibitem [{\citenamefont {Mi}\ \emph {et~al.}(2022)\citenamefont {Mi} \emph
		{et~al.}}]{Mi2022}%
	\BibitemOpen
	\bibfield  {author} {\bibinfo {author} {\bibfnamefont {X.}~\bibnamefont {Mi}}
		\emph {et~al.},\ }\bibfield  {title} {\bibinfo {title} {{Noise-resilient edge
				modes on a chain of superconducting qubits}},\ }\href
	{https://doi.org/10.1126/science.abq5769} {\bibfield  {journal} {\bibinfo
			{journal} {Science}\ }\textbf {\bibinfo {volume} {378}},\ \bibinfo {pages}
		{785} (\bibinfo {year} {2022})}\BibitemShut {NoStop}%
	\bibitem [{\citenamefont {Preskill}(2018)}]{Preskill2018}%
	\BibitemOpen
	\bibfield  {author} {\bibinfo {author} {\bibfnamefont {J.}~\bibnamefont
			{Preskill}},\ }\bibfield  {title} {\bibinfo {title} {Quantum {C}omputing in
			the {NISQ} era and beyond},\ }\href
	{https://doi.org/10.22331/q-2018-08-06-79} {\bibfield  {journal} {\bibinfo
			{journal} {{Quantum}}\ }\textbf {\bibinfo {volume} {2}},\ \bibinfo {pages}
		{79} (\bibinfo {year} {2018})}\BibitemShut {NoStop}%
	\bibitem [{\citenamefont {Georgescu}\ \emph {et~al.}(2014)\citenamefont
		{Georgescu}, \citenamefont {Ashhab},\ and\ \citenamefont {Nori}}]{Nori2014}%
	\BibitemOpen
	\bibfield  {author} {\bibinfo {author} {\bibfnamefont {I.~M.}\ \bibnamefont
			{Georgescu}}, \bibinfo {author} {\bibfnamefont {S.}~\bibnamefont {Ashhab}},\
		and\ \bibinfo {author} {\bibfnamefont {F.}~\bibnamefont {Nori}},\ }\bibfield
	{title} {\bibinfo {title} {{Quantum simulation}},\ }\href
	{https://doi.org/10.1103/RevModPhys.86.153} {\bibfield  {journal} {\bibinfo
			{journal} {Rev. Mod. Phys.}\ }\textbf {\bibinfo {volume} {86}},\ \bibinfo
		{pages} {153} (\bibinfo {year} {2014})}\BibitemShut {NoStop}%
	\bibitem [{\citenamefont {Daley}\ \emph {et~al.}(2022)\citenamefont {Daley},
		\citenamefont {Bloch}, \citenamefont {Kokail}, \citenamefont {Flannigan},
		\citenamefont {Pearson}, \citenamefont {Troyer},\ and\ \citenamefont
		{Zoller}}]{Daley2022}%
	\BibitemOpen
	\bibfield  {author} {\bibinfo {author} {\bibfnamefont {A.~J.}\ \bibnamefont
			{Daley}}, \bibinfo {author} {\bibfnamefont {I.}~\bibnamefont {Bloch}},
		\bibinfo {author} {\bibfnamefont {C.}~\bibnamefont {Kokail}}, \bibinfo
		{author} {\bibfnamefont {S.}~\bibnamefont {Flannigan}}, \bibinfo {author}
		{\bibfnamefont {N.}~\bibnamefont {Pearson}}, \bibinfo {author} {\bibfnamefont
			{M.}~\bibnamefont {Troyer}},\ and\ \bibinfo {author} {\bibfnamefont
			{P.}~\bibnamefont {Zoller}},\ }\bibfield  {title} {\bibinfo {title}
		{Practical quantum advantage in quantum simulation},\ }\href
	{https://doi.org/10.1038/s41586-022-04940-6} {\bibfield  {journal} {\bibinfo
			{journal} {Nature}\ }\textbf {\bibinfo {volume} {607}},\ \bibinfo {pages}
		{667} (\bibinfo {year} {2022})}\BibitemShut {NoStop}%
	\bibitem [{\citenamefont {Leefmans}\ \emph {et~al.}(2022)\citenamefont
		{Leefmans}, \citenamefont {Dutt}, \citenamefont {Williams}, \citenamefont
		{Yuan}, \citenamefont {Parto}, \citenamefont {Nori}, \citenamefont {Fan},\
		and\ \citenamefont {Marandi}}]{NP_Leefmans2022}%
	\BibitemOpen
	\bibfield  {author} {\bibinfo {author} {\bibfnamefont {C.}~\bibnamefont
			{Leefmans}}, \bibinfo {author} {\bibfnamefont {A.}~\bibnamefont {Dutt}},
		\bibinfo {author} {\bibfnamefont {J.}~\bibnamefont {Williams}}, \bibinfo
		{author} {\bibfnamefont {L.}~\bibnamefont {Yuan}}, \bibinfo {author}
		{\bibfnamefont {M.}~\bibnamefont {Parto}}, \bibinfo {author} {\bibfnamefont
			{F.}~\bibnamefont {Nori}}, \bibinfo {author} {\bibfnamefont {S.}~\bibnamefont
			{Fan}},\ and\ \bibinfo {author} {\bibfnamefont {A.}~\bibnamefont {Marandi}},\
	}\bibfield  {title} {\bibinfo {title} {Topological dissipation in a
			time-multiplexed photonic resonator network},\ }\href
	{https://doi.org/10.1038/s41567-021-01492-w} {\bibfield  {journal} {\bibinfo
			{journal} {Nat. Phys.}\ }\textbf {\bibinfo {volume} {18}},\ \bibinfo {pages}
		{442} (\bibinfo {year} {2022})}\BibitemShut {NoStop}%
	\bibitem [{\citenamefont {Cheng}\ \emph {et~al.}(2023)\citenamefont {Cheng},
		\citenamefont {Deng}, \citenamefont {Gu}, \citenamefont {He}, \citenamefont
		{Hu}, \citenamefont {Huang}, \citenamefont {Li}, \citenamefont {Lin},
		\citenamefont {Lu}, \citenamefont {Lu}, \citenamefont {Qiu}, \citenamefont
		{Wang}, \citenamefont {Xin}, \citenamefont {Yu}, \citenamefont {Yung},
		\citenamefont {Zeng}, \citenamefont {Zhang}, \citenamefont {Zhong},
		\citenamefont {Peng}, \citenamefont {Nori},\ and\ \citenamefont
		{Yu}}]{Cheng2023}%
	\BibitemOpen
	\bibfield  {author} {\bibinfo {author} {\bibfnamefont {B.}~\bibnamefont
			{Cheng}}, \bibinfo {author} {\bibfnamefont {X.-H.}\ \bibnamefont {Deng}},
		\bibinfo {author} {\bibfnamefont {X.}~\bibnamefont {Gu}}, \bibinfo {author}
		{\bibfnamefont {Y.}~\bibnamefont {He}}, \bibinfo {author} {\bibfnamefont
			{G.}~\bibnamefont {Hu}}, \bibinfo {author} {\bibfnamefont {P.}~\bibnamefont
			{Huang}}, \bibinfo {author} {\bibfnamefont {J.}~\bibnamefont {Li}}, \bibinfo
		{author} {\bibfnamefont {B.-C.}\ \bibnamefont {Lin}}, \bibinfo {author}
		{\bibfnamefont {D.}~\bibnamefont {Lu}}, \bibinfo {author} {\bibfnamefont
			{Y.}~\bibnamefont {Lu}}, \bibinfo {author} {\bibfnamefont {C.}~\bibnamefont
			{Qiu}}, \bibinfo {author} {\bibfnamefont {H.}~\bibnamefont {Wang}}, \bibinfo
		{author} {\bibfnamefont {T.}~\bibnamefont {Xin}}, \bibinfo {author}
		{\bibfnamefont {S.}~\bibnamefont {Yu}}, \bibinfo {author} {\bibfnamefont
			{M.-H.}\ \bibnamefont {Yung}}, \bibinfo {author} {\bibfnamefont
			{J.}~\bibnamefont {Zeng}}, \bibinfo {author} {\bibfnamefont {S.}~\bibnamefont
			{Zhang}}, \bibinfo {author} {\bibfnamefont {Y.}~\bibnamefont {Zhong}},
		\bibinfo {author} {\bibfnamefont {X.}~\bibnamefont {Peng}}, \bibinfo {author}
		{\bibfnamefont {F.}~\bibnamefont {Nori}},\ and\ \bibinfo {author}
		{\bibfnamefont {D.}~\bibnamefont {Yu}},\ }\bibfield  {title} {\bibinfo
		{title} {Noisy intermediate-scale quantum computers},\ }\href
	{https://doi.org/10.1007/s11467-022-1249-z} {\bibfield  {journal} {\bibinfo
			{journal} {Front. Phys.}\ }\textbf {\bibinfo {volume} {18}},\ \bibinfo
		{pages} {21308} (\bibinfo {year} {2023})}\BibitemShut {NoStop}%
	\bibitem [{\citenamefont {Yan}\ \emph {et~al.}(2019)\citenamefont {Yan},
		\citenamefont {Zhang}, \citenamefont {Gong}, \citenamefont {Wu},
		\citenamefont {Zheng}, \citenamefont {Li}, \citenamefont {Wang},
		\citenamefont {Liang}, \citenamefont {Lin}, \citenamefont {Xu}, \citenamefont
		{Guo}, \citenamefont {Sun}, \citenamefont {Peng}, \citenamefont {Xia},
		\citenamefont {Deng}, \citenamefont {Rong}, \citenamefont {You},
		\citenamefont {Nori}, \citenamefont {Fan}, \citenamefont {Zhu},\ and\
		\citenamefont {Pan}}]{Yan2019}%
	\BibitemOpen
	\bibfield  {author} {\bibinfo {author} {\bibfnamefont {Z.}~\bibnamefont
			{Yan}}, \bibinfo {author} {\bibfnamefont {Y.~R.}\ \bibnamefont {Zhang}},
		\bibinfo {author} {\bibfnamefont {M.}~\bibnamefont {Gong}}, \bibinfo {author}
		{\bibfnamefont {Y.}~\bibnamefont {Wu}}, \bibinfo {author} {\bibfnamefont
			{Y.}~\bibnamefont {Zheng}}, \bibinfo {author} {\bibfnamefont
			{S.}~\bibnamefont {Li}}, \bibinfo {author} {\bibfnamefont {C.}~\bibnamefont
			{Wang}}, \bibinfo {author} {\bibfnamefont {F.}~\bibnamefont {Liang}},
		\bibinfo {author} {\bibfnamefont {J.}~\bibnamefont {Lin}}, \bibinfo {author}
		{\bibfnamefont {Y.}~\bibnamefont {Xu}}, \bibinfo {author} {\bibfnamefont
			{C.}~\bibnamefont {Guo}}, \bibinfo {author} {\bibfnamefont {L.}~\bibnamefont
			{Sun}}, \bibinfo {author} {\bibfnamefont {C.~Z.}\ \bibnamefont {Peng}},
		\bibinfo {author} {\bibfnamefont {K.}~\bibnamefont {Xia}}, \bibinfo {author}
		{\bibfnamefont {H.}~\bibnamefont {Deng}}, \bibinfo {author} {\bibfnamefont
			{H.}~\bibnamefont {Rong}}, \bibinfo {author} {\bibfnamefont {J.~Q.}\
			\bibnamefont {You}}, \bibinfo {author} {\bibfnamefont {F.}~\bibnamefont
			{Nori}}, \bibinfo {author} {\bibfnamefont {H.}~\bibnamefont {Fan}}, \bibinfo
		{author} {\bibfnamefont {X.}~\bibnamefont {Zhu}},\ and\ \bibinfo {author}
		{\bibfnamefont {J.~W.}\ \bibnamefont {Pan}},\ }\bibfield  {title} {\bibinfo
		{title} {{Strongly correlated quantum walks with a 12-qubit superconducting
				processor}},\ }\href {https://doi.org/10.1126/science.aaw1611} {\bibfield
		{journal} {\bibinfo  {journal} {Science}\ }\textbf {\bibinfo {volume}
			{364}},\ \bibinfo {pages} {753} (\bibinfo {year} {2019})}\BibitemShut
	{NoStop}%
	\bibitem [{\citenamefont {Cai}\ \emph {et~al.}(2019)\citenamefont {Cai},
		\citenamefont {Han}, \citenamefont {Mei}, \citenamefont {Xu}, \citenamefont
		{Ma}, \citenamefont {Li}, \citenamefont {Wang}, \citenamefont {Song},
		\citenamefont {Xue}, \citenamefont {Yin}, \citenamefont {Jia},\ and\
		\citenamefont {Sun}}]{Cai2019}%
	\BibitemOpen
	\bibfield  {author} {\bibinfo {author} {\bibfnamefont {W.}~\bibnamefont
			{Cai}}, \bibinfo {author} {\bibfnamefont {J.}~\bibnamefont {Han}}, \bibinfo
		{author} {\bibfnamefont {F.}~\bibnamefont {Mei}}, \bibinfo {author}
		{\bibfnamefont {Y.}~\bibnamefont {Xu}}, \bibinfo {author} {\bibfnamefont
			{Y.}~\bibnamefont {Ma}}, \bibinfo {author} {\bibfnamefont {X.}~\bibnamefont
			{Li}}, \bibinfo {author} {\bibfnamefont {H.}~\bibnamefont {Wang}}, \bibinfo
		{author} {\bibfnamefont {Y.~P.}\ \bibnamefont {Song}}, \bibinfo {author}
		{\bibfnamefont {Z.-Y.}\ \bibnamefont {Xue}}, \bibinfo {author} {\bibfnamefont
			{Z.-q.}\ \bibnamefont {Yin}}, \bibinfo {author} {\bibfnamefont
			{S.}~\bibnamefont {Jia}},\ and\ \bibinfo {author} {\bibfnamefont
			{L.}~\bibnamefont {Sun}},\ }\bibfield  {title} {\bibinfo {title}
		{{Observation of topological magnon insulator states in a superconducting
				circuit}},\ }\href {https://doi.org/10.1103/PhysRevLett.123.080501}
	{\bibfield  {journal} {\bibinfo  {journal} {Phys. Rev. Lett.}\ }\textbf
		{\bibinfo {volume} {123}},\ \bibinfo {pages} {080501} (\bibinfo {year}
		{2019})}\BibitemShut {NoStop}%
	\bibitem [{\citenamefont {Zhao}\ \emph {et~al.}(2022)\citenamefont {Zhao},
		\citenamefont {Ge}, \citenamefont {Xiang}, \citenamefont {Xue}, \citenamefont
		{Yan}, \citenamefont {Wang}, \citenamefont {Wang}, \citenamefont {Xu},
		\citenamefont {Su}, \citenamefont {Yang}, \citenamefont {Zhang},
		\citenamefont {Zhang}, \citenamefont {Guo}, \citenamefont {Xu}, \citenamefont
		{Tian}, \citenamefont {Yu}, \citenamefont {Zheng}, \citenamefont {Fan},\ and\
		\citenamefont {Zhao}}]{Zhao2022}%
	\BibitemOpen
	\bibfield  {author} {\bibinfo {author} {\bibfnamefont {S.~K.}\ \bibnamefont
			{Zhao}}, \bibinfo {author} {\bibfnamefont {Z.-Y.}\ \bibnamefont {Ge}},
		\bibinfo {author} {\bibfnamefont {Z.}~\bibnamefont {Xiang}}, \bibinfo
		{author} {\bibfnamefont {G.~M.}\ \bibnamefont {Xue}}, \bibinfo {author}
		{\bibfnamefont {H.~S.}\ \bibnamefont {Yan}}, \bibinfo {author} {\bibfnamefont
			{Z.~T.}\ \bibnamefont {Wang}}, \bibinfo {author} {\bibfnamefont
			{Z.}~\bibnamefont {Wang}}, \bibinfo {author} {\bibfnamefont {H.~K.}\
			\bibnamefont {Xu}}, \bibinfo {author} {\bibfnamefont {F.~F.}\ \bibnamefont
			{Su}}, \bibinfo {author} {\bibfnamefont {Z.~H.}\ \bibnamefont {Yang}},
		\bibinfo {author} {\bibfnamefont {H.}~\bibnamefont {Zhang}}, \bibinfo
		{author} {\bibfnamefont {Y.-R.}\ \bibnamefont {Zhang}}, \bibinfo {author}
		{\bibfnamefont {X.-Y.}\ \bibnamefont {Guo}}, \bibinfo {author} {\bibfnamefont
			{K.}~\bibnamefont {Xu}}, \bibinfo {author} {\bibfnamefont {Y.}~\bibnamefont
			{Tian}}, \bibinfo {author} {\bibfnamefont {H.~F.}\ \bibnamefont {Yu}},
		\bibinfo {author} {\bibfnamefont {D.~N.}\ \bibnamefont {Zheng}}, \bibinfo
		{author} {\bibfnamefont {H.}~\bibnamefont {Fan}},\ and\ \bibinfo {author}
		{\bibfnamefont {S.~P.}\ \bibnamefont {Zhao}},\ }\bibfield  {title} {\bibinfo
		{title} {Probing operator spreading via {F}loquet engineering in a
			superconducting circuit},\ }\href
	{https://doi.org/10.1103/PhysRevLett.129.160602} {\bibfield  {journal}
		{\bibinfo  {journal} {Phys. Rev. Lett.}\ }\textbf {\bibinfo {volume} {129}},\
		\bibinfo {pages} {160602} (\bibinfo {year} {2022})}\BibitemShut {NoStop}%
	\bibitem [{\citenamefont {Denisov}\ \emph {et~al.}(2014)\citenamefont
		{Denisov}, \citenamefont {Flach},\ and\ \citenamefont
		{H{\"{a}}nggi}}]{Denisov2014}%
	\BibitemOpen
	\bibfield  {author} {\bibinfo {author} {\bibfnamefont {S.}~\bibnamefont
			{Denisov}}, \bibinfo {author} {\bibfnamefont {S.}~\bibnamefont {Flach}},\
		and\ \bibinfo {author} {\bibfnamefont {P.}~\bibnamefont {H{\"{a}}nggi}},\
	}\bibfield  {title} {\bibinfo {title} {{Tunable transport with broken
				space-time symmetries}},\ }\href
	{https://doi.org/10.1016/j.physrep.2014.01.003} {\bibfield  {journal}
		{\bibinfo  {journal} {Phys. Rep.}\ }\textbf {\bibinfo {volume} {538}},\
		\bibinfo {pages} {77} (\bibinfo {year} {2014})}\BibitemShut {NoStop}%
	\bibitem [{\citenamefont {Wu}\ \emph {et~al.}(2018)\citenamefont {Wu},
		\citenamefont {Yang}, \citenamefont {Gong}, \citenamefont {Zheng},
		\citenamefont {Deng}, \citenamefont {Yan}, \citenamefont {Zhao},
		\citenamefont {Huang}, \citenamefont {Castellano}, \citenamefont {Munro},
		\citenamefont {Nemoto}, \citenamefont {Zheng}, \citenamefont {Sun},
		\citenamefont {Liu}, \citenamefont {Zhu},\ and\ \citenamefont {Lu}}]{Wu2018}%
	\BibitemOpen
	\bibfield  {author} {\bibinfo {author} {\bibfnamefont {Y.}~\bibnamefont
			{Wu}}, \bibinfo {author} {\bibfnamefont {L.-P.}\ \bibnamefont {Yang}},
		\bibinfo {author} {\bibfnamefont {M.}~\bibnamefont {Gong}}, \bibinfo {author}
		{\bibfnamefont {Y.}~\bibnamefont {Zheng}}, \bibinfo {author} {\bibfnamefont
			{H.}~\bibnamefont {Deng}}, \bibinfo {author} {\bibfnamefont {Z.}~\bibnamefont
			{Yan}}, \bibinfo {author} {\bibfnamefont {Y.}~\bibnamefont {Zhao}}, \bibinfo
		{author} {\bibfnamefont {K.}~\bibnamefont {Huang}}, \bibinfo {author}
		{\bibfnamefont {A.~D.}\ \bibnamefont {Castellano}}, \bibinfo {author}
		{\bibfnamefont {W.~J.}\ \bibnamefont {Munro}}, \bibinfo {author}
		{\bibfnamefont {K.}~\bibnamefont {Nemoto}}, \bibinfo {author} {\bibfnamefont
			{D.-N.}\ \bibnamefont {Zheng}}, \bibinfo {author} {\bibfnamefont {C.~P.}\
			\bibnamefont {Sun}}, \bibinfo {author} {\bibfnamefont {Y.-x.}\ \bibnamefont
			{Liu}}, \bibinfo {author} {\bibfnamefont {X.}~\bibnamefont {Zhu}},\ and\
		\bibinfo {author} {\bibfnamefont {L.}~\bibnamefont {Lu}},\ }\bibfield
	{title} {\bibinfo {title} {{An efficient and compact switch for quantum
				circuits}},\ }\href {https://doi.org/10.1038/s41534-018-0099-6} {\bibfield
		{journal} {\bibinfo  {journal} {npj Quantum Inform.}\ }\textbf {\bibinfo
			{volume} {4}},\ \bibinfo {pages} {50} (\bibinfo {year} {2018})}\BibitemShut
	{NoStop}%
	\bibitem [{\citenamefont {Reagor}\ \emph {et~al.}(2018)\citenamefont {Reagor}
		\emph {et~al.}}]{Reagor2018}%
	\BibitemOpen
	\bibfield  {author} {\bibinfo {author} {\bibfnamefont {M.}~\bibnamefont
			{Reagor}} \emph {et~al.},\ }\bibfield  {title} {\bibinfo {title}
		{{Demonstration of universal parametric entangling gates on a multi-qubit
				lattice}},\ }\href {https://doi.org/10.1126/sciadv.aao3603} {\bibfield
		{journal} {\bibinfo  {journal} {Sci. Adv.}\ }\textbf {\bibinfo {volume}
			{4}},\ \bibinfo {pages} {eaao360} (\bibinfo {year} {2018})}\BibitemShut
	{NoStop}%
	\bibitem [{\citenamefont {Lignier}\ \emph {et~al.}(2007)\citenamefont
		{Lignier}, \citenamefont {Sias}, \citenamefont {Ciampini}, \citenamefont
		{Singh}, \citenamefont {Zenesini}, \citenamefont {Morsch},\ and\
		\citenamefont {Arimondo}}]{Lignier2007}%
	\BibitemOpen
	\bibfield  {author} {\bibinfo {author} {\bibfnamefont {H.}~\bibnamefont
			{Lignier}}, \bibinfo {author} {\bibfnamefont {C.}~\bibnamefont {Sias}},
		\bibinfo {author} {\bibfnamefont {D.}~\bibnamefont {Ciampini}}, \bibinfo
		{author} {\bibfnamefont {Y.}~\bibnamefont {Singh}}, \bibinfo {author}
		{\bibfnamefont {A.}~\bibnamefont {Zenesini}}, \bibinfo {author}
		{\bibfnamefont {O.}~\bibnamefont {Morsch}},\ and\ \bibinfo {author}
		{\bibfnamefont {E.}~\bibnamefont {Arimondo}},\ }\bibfield  {title} {\bibinfo
		{title} {{Dynamical control of matter-wave tunneling in periodic
				potentials}},\ }\href {https://doi.org/10.1103/PhysRevLett.99.220403}
	{\bibfield  {journal} {\bibinfo  {journal} {Phys. Rev. Lett.}\ }\textbf
		{\bibinfo {volume} {99}},\ \bibinfo {pages} {220403} (\bibinfo {year}
		{2007})}\BibitemShut {NoStop}%
	\bibitem [{\citenamefont {Eckardt}(2017)}]{Eckardt2017}%
	\BibitemOpen
	\bibfield  {author} {\bibinfo {author} {\bibfnamefont {A.}~\bibnamefont
			{Eckardt}},\ }\bibfield  {title} {\bibinfo {title} {{\emph{Colloquium}:
				Atomic quantum gases in periodically driven optical lattices}},\ }\href
	{https://doi.org/10.1103/RevModPhys.89.011004} {\bibfield  {journal}
		{\bibinfo  {journal} {Rev. Mod. Phys.}\ }\textbf {\bibinfo {volume} {89}},\
		\bibinfo {pages} {011004} (\bibinfo {year} {2017})}\BibitemShut {NoStop}%
	\bibitem [{SM()}]{SM}%
	\BibitemOpen
	\href@noop {} {\emph {\bibinfo {title} {\emph{Supplementary Material is
					available at http://}}}}\BibitemShut {NoStop}%
	\bibitem [{\citenamefont {Satija}\ and\ \citenamefont
		{Naumis}(2013)}]{Satija2013}%
	\BibitemOpen
	\bibfield  {author} {\bibinfo {author} {\bibfnamefont {I.~I.}\ \bibnamefont
			{Satija}}\ and\ \bibinfo {author} {\bibfnamefont {G.~G.}\ \bibnamefont
			{Naumis}},\ }\bibfield  {title} {\bibinfo {title} {Chern and {M}ajorana modes
			of quasiperiodic systems},\ }\href
	{https://doi.org/10.1103/PhysRevB.88.054204} {\bibfield  {journal} {\bibinfo
			{journal} {Phys. Rev. B}\ }\textbf {\bibinfo {volume} {88}},\ \bibinfo
		{pages} {054204} (\bibinfo {year} {2013})}\BibitemShut {NoStop}%
	\bibitem [{\citenamefont {DeGottardi}\ \emph {et~al.}(2013)\citenamefont
		{DeGottardi}, \citenamefont {Sen},\ and\ \citenamefont
		{Vishveshwara}}]{Degottardi2013}%
	\BibitemOpen
	\bibfield  {author} {\bibinfo {author} {\bibfnamefont {W.}~\bibnamefont
			{DeGottardi}}, \bibinfo {author} {\bibfnamefont {D.}~\bibnamefont {Sen}},\
		and\ \bibinfo {author} {\bibfnamefont {S.}~\bibnamefont {Vishveshwara}},\
	}\bibfield  {title} {\bibinfo {title} {{Majorana fermions in superconducting
				1D systems having periodic, quasiperiodic, and disordered potentials}},\
	}\href {https://doi.org/10.1103/PhysRevLett.110.146404} {\bibfield  {journal}
		{\bibinfo  {journal} {Phys. Rev. Lett.}\ }\textbf {\bibinfo {volume} {110}},\
		\bibinfo {pages} {146404} (\bibinfo {year} {2013})}\BibitemShut {NoStop}%
	\bibitem [{\citenamefont {Thouless}\ \emph {et~al.}(1982)\citenamefont
		{Thouless}, \citenamefont {Kohmoto}, \citenamefont {Nightingale},\ and\
		\citenamefont {den Nijs}}]{Thouless1982}%
	\BibitemOpen
	\bibfield  {author} {\bibinfo {author} {\bibfnamefont {D.~J.}\ \bibnamefont
			{Thouless}}, \bibinfo {author} {\bibfnamefont {M.}~\bibnamefont {Kohmoto}},
		\bibinfo {author} {\bibfnamefont {M.~P.}\ \bibnamefont {Nightingale}},\ and\
		\bibinfo {author} {\bibfnamefont {M.}~\bibnamefont {den Nijs}},\ }\bibfield
	{title} {\bibinfo {title} {Quantized hall conductance in a two-dimensional
			periodic potential},\ }\href {https://doi.org/10.1103/PhysRevLett.49.405}
	{\bibfield  {journal} {\bibinfo  {journal} {Phys. Rev. Lett.}\ }\textbf
		{\bibinfo {volume} {49}},\ \bibinfo {pages} {405} (\bibinfo {year}
		{1982})}\BibitemShut {NoStop}%
	\bibitem [{\citenamefont {Bansil}\ \emph {et~al.}(2016)\citenamefont {Bansil},
		\citenamefont {Lin},\ and\ \citenamefont {Das}}]{Bansil2016}%
	\BibitemOpen
	\bibfield  {author} {\bibinfo {author} {\bibfnamefont {A.}~\bibnamefont
			{Bansil}}, \bibinfo {author} {\bibfnamefont {H.}~\bibnamefont {Lin}},\ and\
		\bibinfo {author} {\bibfnamefont {T.}~\bibnamefont {Das}},\ }\bibfield
	{title} {\bibinfo {title} {Colloquium: Topological band theory},\ }\href
	{https://doi.org/10.1103/RevModPhys.88.021004} {\bibfield  {journal}
		{\bibinfo  {journal} {Rev. Mod. Phys.}\ }\textbf {\bibinfo {volume} {88}},\
		\bibinfo {pages} {021004} (\bibinfo {year} {2016})}\BibitemShut {NoStop}%
	\bibitem [{\citenamefont {Su}\ \emph {et~al.}(1979)\citenamefont {Su},
		\citenamefont {Schrieffer},\ and\ \citenamefont {Heeger}}]{Su1979}%
	\BibitemOpen
	\bibfield  {author} {\bibinfo {author} {\bibfnamefont {W.~P.}\ \bibnamefont
			{Su}}, \bibinfo {author} {\bibfnamefont {J.~R.}\ \bibnamefont {Schrieffer}},\
		and\ \bibinfo {author} {\bibfnamefont {A.~J.}\ \bibnamefont {Heeger}},\
	}\bibfield  {title} {\bibinfo {title} {Solitons in polyacetylene},\ }\href
	{https://doi.org/10.1103/PhysRevLett.42.1698} {\bibfield  {journal} {\bibinfo
			{journal} {Phys. Rev. Lett.}\ }\textbf {\bibinfo {volume} {42}},\ \bibinfo
		{pages} {1698} (\bibinfo {year} {1979})}\BibitemShut {NoStop}%
	\bibitem [{\citenamefont {Schweizer}\ \emph {et~al.}(2019)\citenamefont
		{Schweizer}, \citenamefont {Grusdt}, \citenamefont {Berngruber},
		\citenamefont {Barbiero}, \citenamefont {Demler}, \citenamefont {Goldman},
		\citenamefont {Bloch},\ and\ \citenamefont {Aidelsburger}}]{Schweizer2019}%
	\BibitemOpen
	\bibfield  {author} {\bibinfo {author} {\bibfnamefont {C.}~\bibnamefont
			{Schweizer}}, \bibinfo {author} {\bibfnamefont {F.}~\bibnamefont {Grusdt}},
		\bibinfo {author} {\bibfnamefont {M.}~\bibnamefont {Berngruber}}, \bibinfo
		{author} {\bibfnamefont {L.}~\bibnamefont {Barbiero}}, \bibinfo {author}
		{\bibfnamefont {E.}~\bibnamefont {Demler}}, \bibinfo {author} {\bibfnamefont
			{N.}~\bibnamefont {Goldman}}, \bibinfo {author} {\bibfnamefont
			{I.}~\bibnamefont {Bloch}},\ and\ \bibinfo {author} {\bibfnamefont
			{M.}~\bibnamefont {Aidelsburger}},\ }\bibfield  {title} {\bibinfo {title}
		{Floquet approach to $\mathbb{Z}_2$ lattice gauge theories with ultracold
			atoms in optical lattices},\ }\href
	{https://doi.org/10.1038/s41567-019-0649-7} {\bibfield  {journal} {\bibinfo
			{journal} {Nat. Phys.}\ }\textbf {\bibinfo {volume} {15}},\ \bibinfo {pages}
		{1168} (\bibinfo {year} {2019})}\BibitemShut {NoStop}%
	\bibitem [{\citenamefont {Wu}\ \emph {et~al.}(2019)\citenamefont {Wu},
		\citenamefont {Liu}, \citenamefont {Geng}, \citenamefont {Song},
		\citenamefont {Ye}, \citenamefont {Duan}, \citenamefont {Rong},\ and\
		\citenamefont {Du}}]{Wu2019}%
	\BibitemOpen
	\bibfield  {author} {\bibinfo {author} {\bibfnamefont {Y.}~\bibnamefont
			{Wu}}, \bibinfo {author} {\bibfnamefont {W.}~\bibnamefont {Liu}}, \bibinfo
		{author} {\bibfnamefont {J.}~\bibnamefont {Geng}}, \bibinfo {author}
		{\bibfnamefont {X.}~\bibnamefont {Song}}, \bibinfo {author} {\bibfnamefont
			{X.}~\bibnamefont {Ye}}, \bibinfo {author} {\bibfnamefont {C.-K.}\
			\bibnamefont {Duan}}, \bibinfo {author} {\bibfnamefont {X.}~\bibnamefont
			{Rong}},\ and\ \bibinfo {author} {\bibfnamefont {J.}~\bibnamefont {Du}},\
	}\bibfield  {title} {\bibinfo {title} {Observation of parity-time symmetry
			breaking in a single-spin system},\ }\href
	{https://doi.org/10.1126/science.aaw8205} {\bibfield  {journal} {\bibinfo
			{journal} {Science}\ }\textbf {\bibinfo {volume} {364}},\ \bibinfo {pages}
		{878} (\bibinfo {year} {2019})}\BibitemShut {NoStop}%
\end{thebibliography}
%

\end{document}


\title{Supplementary Materials for\\``Quantum simulation of topological zero modes on a 41-qubit superconducting processor''}

\author{Yun-Hao Shi}
\thanks{These authors contributed equally to this work.}
\affiliation{Institute of Physics, Chinese Academy of Sciences, Beijing 100190, China}
\affiliation{School of Physical Sciences, University of Chinese Academy of Sciences, Beijing 100049, China}
\affiliation{Beijing Academy of Quantum Information Sciences, Beijing 100193, China}

\author{Yu Liu}
\thanks{These authors contributed equally to this work.}
\affiliation{Institute of Physics, Chinese Academy of Sciences, Beijing 100190, China}
\affiliation{School of Physical Sciences, University of Chinese Academy of Sciences, Beijing 100049, China}

\author{Yu-Ran Zhang}
\thanks{These authors contributed equally to this work.}
\affiliation{School of Physics and Optoelectronics, South China University of Technology, Guangzhou 510640, China}
\affiliation{Theoretical Quantum Physics Laboratory, Cluster for Pioneering Research, RIKEN, Wako-shi, Saitama 351-0198, Japan}
\affiliation{Center for Quantum Computing, RIKEN, Wako-shi, Saitama 351-0198, Japan}

\author{Zhongcheng Xiang}
\thanks{These authors contributed equally to this work.}
\affiliation{Institute of Physics, Chinese Academy of Sciences, Beijing 100190, China}
\affiliation{School of Physical Sciences, University of Chinese Academy of Sciences, Beijing 100049, China}

\author{Kaixuan Huang}
\affiliation{Beijing Academy of Quantum Information Sciences, Beijing 100193, China}
\affiliation{Institute of Physics, Chinese Academy of Sciences, Beijing 100190, China}

\author{Tao Liu}
\affiliation{School of Physics and Optoelectronics, South China University of Technology, Guangzhou 510640, China}

\author{Yong-Yi Wang}
\affiliation{Institute of Physics, Chinese Academy of Sciences, Beijing 100190, China}
\affiliation{School of Physical Sciences, University of Chinese Academy of Sciences, Beijing 100049, China}

\author{Jia-Chi Zhang}
\affiliation{Institute of Physics, Chinese Academy of Sciences, Beijing 100190, China}
\affiliation{School of Physical Sciences, University of Chinese Academy of Sciences, Beijing 100049, China}

\author{Cheng-Lin Deng}
\affiliation{Institute of Physics, Chinese Academy of Sciences, Beijing 100190, China}
\affiliation{School of Physical Sciences, University of Chinese Academy of Sciences, Beijing 100049, China}

\author{Gui-Han Liang}
\affiliation{Institute of Physics, Chinese Academy of Sciences, Beijing 100190, China}
\affiliation{School of Physical Sciences, University of Chinese Academy of Sciences, Beijing 100049, China}

\author{Zheng-Yang Mei}
\affiliation{Institute of Physics, Chinese Academy of Sciences, Beijing 100190, China}
\affiliation{School of Physical Sciences, University of Chinese Academy of Sciences, Beijing 100049, China}

\author{Hao Li}
\affiliation{Institute of Physics, Chinese Academy of Sciences, Beijing 100190, China}

\author{Tian-Ming Li}
\affiliation{Institute of Physics, Chinese Academy of Sciences, Beijing 100190, China}
\affiliation{School of Physical Sciences, University of Chinese Academy of Sciences, Beijing 100049, China}

\author{Wei-Guo Ma}
\affiliation{Institute of Physics, Chinese Academy of Sciences, Beijing 100190, China}
\affiliation{School of Physical Sciences, University of Chinese Academy of Sciences, Beijing 100049, China}

\author{Hao-Tian Liu}
\affiliation{Institute of Physics, Chinese Academy of Sciences, Beijing 100190, China}
\affiliation{School of Physical Sciences, University of Chinese Academy of Sciences, Beijing 100049, China}

\author{Chi-Tong Chen}
\affiliation{Institute of Physics, Chinese Academy of Sciences, Beijing 100190, China}
\affiliation{School of Physical Sciences, University of Chinese Academy of Sciences, Beijing 100049, China}

\author{Tong Liu}
\affiliation{Institute of Physics, Chinese Academy of Sciences, Beijing 100190, China}
\affiliation{School of Physical Sciences, University of Chinese Academy of Sciences, Beijing 100049, China}

\author{Ye Tian}
\affiliation{Institute of Physics, Chinese Academy of Sciences, Beijing 100190, China}

\author{Xiaohui Song}
\affiliation{Institute of Physics, Chinese Academy of Sciences, Beijing 100190, China}

\author{S. P. Zhao}
\affiliation{Institute of Physics, Chinese Academy of Sciences, Beijing 100190, China}
\affiliation{School of Physical Sciences, University of Chinese Academy of Sciences, Beijing 100049, China}
\affiliation{Songshan Lake  Materials Laboratory, Dongguan, Guangdong 523808, China}

\author{Kai Xu}
\email{kaixu@iphy.ac.cn}
\affiliation{Institute of Physics, Chinese Academy of Sciences, Beijing 100190, China}
\affiliation{School of Physical Sciences, University of Chinese Academy of Sciences, Beijing 100049, China}
\affiliation{Beijing Academy of Quantum Information Sciences, Beijing 100193, China}
\affiliation{Songshan Lake Materials Laboratory, Dongguan, Guangdong 523808, China}
\affiliation{CAS Center for Excellence in Topological Quantum Computation, UCAS, Beijing 100049, China}

\author{Dongning Zheng}
\email{dzheng@iphy.ac.cn}
\affiliation{Institute of Physics, Chinese Academy of Sciences, Beijing 100190, China}
\affiliation{School of Physical Sciences, University of Chinese Academy of Sciences, Beijing 100049, China}
\affiliation{Songshan Lake Materials Laboratory, Dongguan, Guangdong 523808, China}
\affiliation{CAS Center for Excellence in Topological Quantum Computation, UCAS, Beijing 100049, China}

\author{Franco Nori}
\email{fnori@riken.jp}
\affiliation{Theoretical Quantum Physics Laboratory, Cluster for Pioneering Research, RIKEN, Wako-shi, Saitama 351-0198, Japan}
\affiliation{Center for Quantum Computing, RIKEN, Wako-shi, Saitama 351-0198, Japan}
\affiliation{Physics Department, University of Michigan, Ann Arbor, Michigan 48109-1040, USA}

\author{Heng Fan}
\email{hfan@iphy.ac.cn}
\affiliation{Institute of Physics, Chinese Academy of Sciences, Beijing 100190, China}
\affiliation{School of Physical Sciences, University of Chinese Academy of Sciences, Beijing 100049, China}
\affiliation{Beijing Academy of Quantum Information Sciences, Beijing 100193, China}
\affiliation{Songshan Lake Materials Laboratory, Dongguan, Guangdong 523808, China}
\affiliation{CAS Center for Excellence in Topological Quantum Computation, UCAS, Beijing 100049, China}
\affiliation{Hefei National Laboratory, Hefei 230088, China}
\maketitle
\beginsupplement
\tableofcontents
\newpage
\section{\label{sec:model} Model and Hamiltonian}
\quad~We focus on the generalized one-dimensional (1D) Aubry-Andr$\acute{\mathrm{e}}$-Harper (AAH) model~\cite{Ganeshan2013,Kraus2012,Satija2013,Degottardi2013}. The total Hamiltonian of the system is
\begin{equation}\label{eq:AAH}
	\hat{H}_{\mathrm{AAH}}=\sum_{j=1}^{N}v\cos(2\pi b_vj+\varphi_v)\hat{a}^{\dagger}_{j}\hat{a}_{j}+\sum_{j=1}^{N-1}u\left[1+\lambda\cos(2\pi b_{\lambda}j+\varphi_{\lambda})\right](\hat{a}^{\dagger}_{j}\hat{a}_{j+1}+\hat{a}_{j}\hat{a}^{\dagger}_{j+1}),
\end{equation}
where $\hat{a}$ ($\hat{a}^{\dagger}$) denotes the annihilation (creation) operator, and $v$ ($\lambda$) is the strength of the cosine modulations on the frequencies (hopping couplings) with a periodicity $1/b_v$ ($1/b_{\lambda}$) and phase factor $\varphi_v$ ($\varphi_{\lambda}$). The special
case with $\lambda=0$ ($v=0$) corresponds to the diagonal (off-diagonal) AAH model. For convenience, we set $\hbar=1$ throughout the paper.

Here, we use a 1D array of 43 programmable superconducting qubits to simulate the generalized AAH model. The fabrication and experimental setup are introduced in Sec.~\ref{sec:setup}. Our system can be well described by the Hamiltonian of the Bose-Hubbard model:
\begin{equation}\label{eq:BH}
	\hat{H}_{\mathrm{BH}}=\sum_{j=1}^{N}\omega_{j}\hat{a}^{\dagger}_{j}\hat{a}_{j} + \sum_{j=1}^{N} \frac{\alpha_{j}}{2}\hat{a}^{\dagger}_{j}\hat{a}^{\dagger}_{j}\hat{a}_{j}\hat{a}_{j}+\sum_{j=1}^{N-1} g_{j,j+1}(\hat{a}^{\dagger}_{j}\hat{a}_{j+1}+\hat{a}_{j}\hat{a}^{\dagger}_{j+1}),
\end{equation}
where $\hat{a}_{j}$ ($\hat{a}^{\dagger}_{j}$) denotes the bosonic annihilation (creation) operator of the $j$-th qubit, $g_{j,j+1}$ is the nearest-neighbor (NN) coupling strength, $\omega_{j}$ denotes the qubit frequency, and $\alpha_{j}$ represents anharmonicity. With $\left|\alpha_j\right|\gg\left|g_{j,j+1}\right|$, the Hamiltonian is reduced to the XX model
\begin{equation}\label{eq:BH_reduced}
	\hat{H}=\sum_{j=1}^{N}\omega_{j}\hat{\sigma}^+_{j}\hat{\sigma}^{-}_{j} + \sum_{j=1}^{N-1} g_{j,j+1}(\hat{\sigma}^+_{j}\hat{\sigma}^{-}_{j+1}+\hat{\sigma}^-_{j}\hat{\sigma}^{+}_{j+1}),
\end{equation}
where $\hat{\sigma}^{+}_j$ ($\hat{\sigma}^{-}_j$) is the raising (lowering) operator. In this low-filling case, to realize the AAH model:
\begin{eqnarray}
	\omega_{j}=\omega_{\mathrm{ref}}+v\cos(2\pi b_vj+\varphi_v),\\
	g_{j,j+1}^{\mathrm{eff}}=u\left[1+\lambda\cos(2\pi b_{\lambda}j+\varphi_{\lambda})\right],
\end{eqnarray}
where $\omega_{\mathrm{ref}}$ denotes the reference frequency. In this work, we use Floquet engineering to adjust the effective coupling $g_{j,j+1}^{\mathrm{eff}}$. The experimental details of Floquet engineering are shown in Sec.~\ref{sec:zfield} and the automatic calibration for the frequencies and couplings are presented in Sec.~\ref{sec:qusim}.

\begin{figure}[ht!]
	\includegraphics[width=1\textwidth]{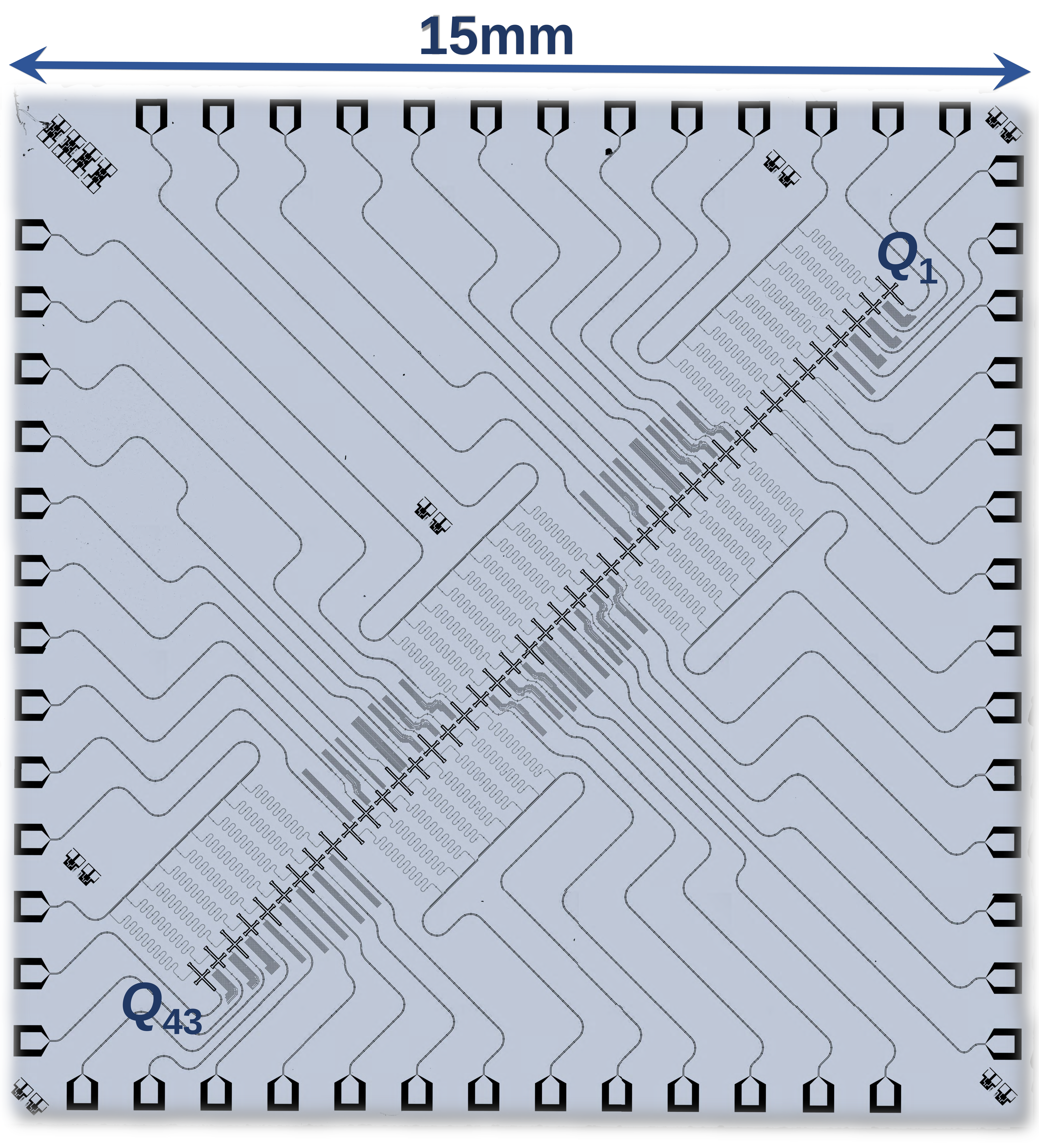}
	\caption{The optical micrograph of the 43-qubit sample. Each qubit has only one control line for the combined XY and Z controls.}\label{fig:chip_sm}
\end{figure}

\section{\label{sec:setup} Fabrication and Experimental setup}
\quad The 43-qubit \textit{Chuang-tzu} sample used in this work is fabricated on a 430 $\mu$m thick sapphire chip with standard wafer cleaning. Specifically, a layer of 100~nm Al was firstly deposited on a 15$\times$15~$\mathrm{mm}^2$ sapphire substrate and patterned with optical lithography using 0.70 $\mu$m of positive SPR955 resist. Then, we used wet etching to produce large structures, such as microwave coplanar waveguide resonators, transmission lines, control lines, and capacitors of the transmon qubits. The Josephson junctions definition process consists of patterning a bilayer of MMA and PMMA resists with electron beam lithography, which were made using double-angle evaporation with a $65$~nm thick Al layer at $+60^{\circ}$, followed by several minutes oxidation in pure oxygen, and a 100~nm thick second layer of Al at $0^{\circ}$. Finally, in order to suppress the parasitic modes, a number of airbridges~\cite{Chen2014} are constructed on the chip. The optical micrograph of the whole chip is displayed in Fig.~\ref{fig:chip_sm}.

\begin{figure}[t!]
	\includegraphics[width=6.7in]{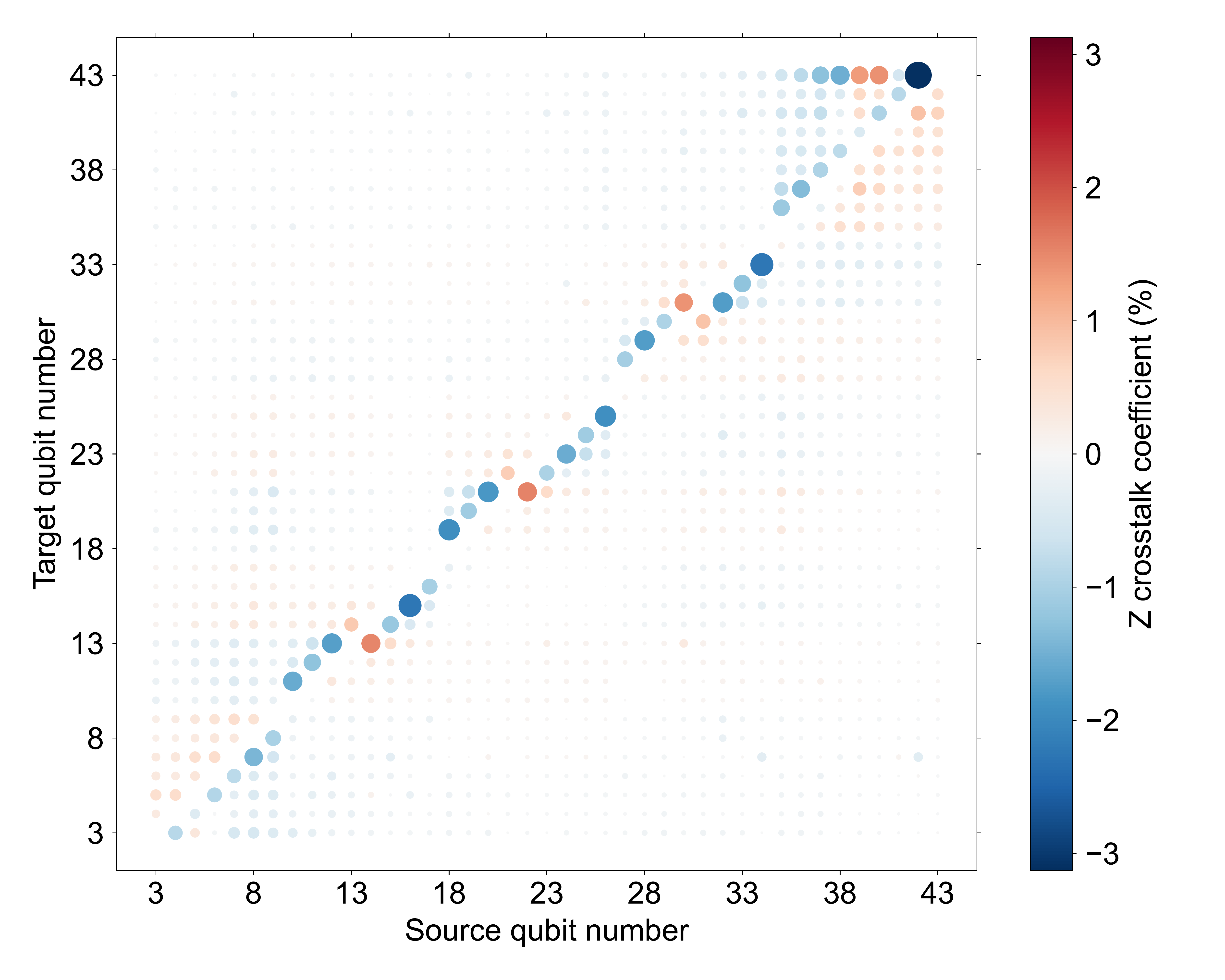}
	\caption{Z crosstalk matrix. The size of each bubble indicates the absolute value of the Z crosstalk matrix element of its corresponding source qubit to the target qubit. Except the crosstalk of $Q_{42}$ to $Q_{43}$, which is 3.13$\%$, all the remaining crosstalk coefficients are less than 2.26$\%$}\label{fig:zxtalk_mat}
\end{figure}

We optimize the end (near the qubit) of the XY/Z control lines by extending the end away from the qubits and then ground to reduce the crosstalk to other qubits due to the microwave signal and flux bias. Normally, the control line is grounded directly at the end, but the microwave signal does not disappear instantly and continues to propagate along the metal. If the spatial distance between the qubits is not far enough, it is easy to cause the microwave signal propagating to the neighboring qubits and generate the crosstalk. By extending the end of the control lines to ensure that their grounded ports are away from the qubits, the leaked microwave signal can be kept away from the qubits, reducing the microwave crosstalk. At the same time, for DC bias, the extended control lines can generate a current in the opposite direction to the incoming current, generating mutually offset magnetic flux and effectively reducing the flux crosstalk. The Z crosstalk matrix is shown in Fig.~\ref{fig:zxtalk_mat}.

\begin{figure}[t!]
	\includegraphics[width=1\textwidth]{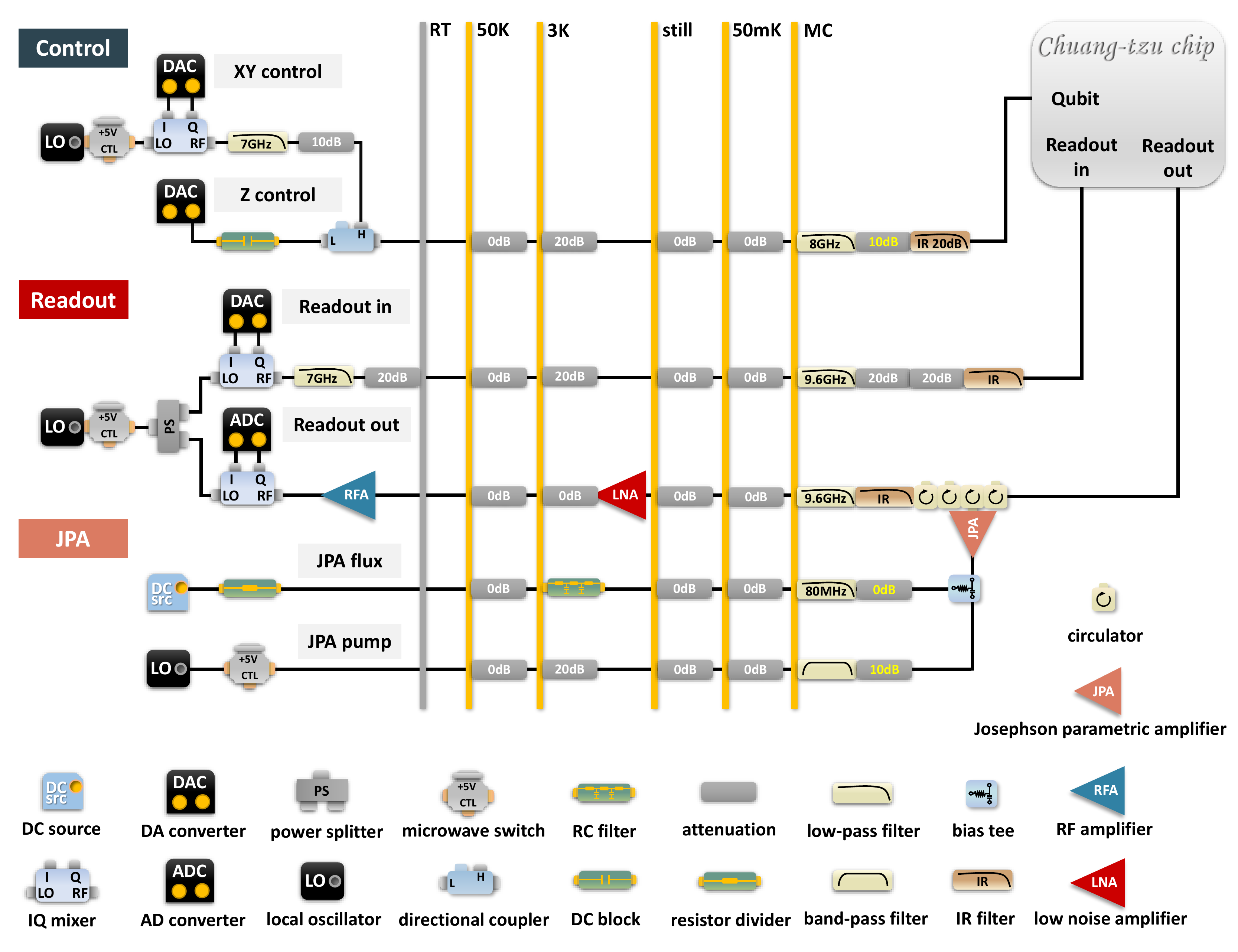}
	\caption{A schematic diagram of the experimental system and wiring information. Here, MC refers to mixing chamber, RFA is the room-temperature RF amplifier, LNA denotes the low-noise amplifier, and JPA is the Josephson parametric amplifier.}\label{fig:wiring}
\end{figure}

To implement experimental measurements, our superconducting quantum chip is placed in a BlueFors dilution refrigerator with a mixing chamber (MC) at a temperature of about 20 mK. The typical wiring of the control electronics and cryogenic equipment is shown in Fig.~\ref{fig:wiring}. There are 5 readout transmission lines, each of which is equipped with a superconducting Josephson parametric amplifier (JPA), a cryo low-noise amplifier (LNA), and a room-temperature RF amplifier (RFA). The readout pulse on the transmission line is first generated by a microwave arbitrary waveform generator (AWG) consisting of two digital-analog converter (DAC) channels and a local oscillation (LO), then amplified by the amplifiers at different temperatures after interacting with the chip, and finally modulated by the analog-digital converter (ADC).

In order to reduce the number of cryogenic control lines in the refrigerator, we combine the high-frequency microwave excitation signal with the low-frequency bias signal by using directional couplers at room temperature, so that the XY and Z control lines of the qubits are merged into one at cryogenic temperature. Moreover, we set a microwave switch at each LO port, which is controlled by the experimental trigger signal, to suppress the thermal excitation originating from the continuous microwave signal. We also perform zero calibration of every AWG to reduce intrinsic and mirror leakages.

Our superconducting quantum processor consists of 43 transmon qubits ($Q_1$, $\cdots$, $Q_{43}$) arrayed in a row, where each qubit is capacitively coupled to its nearest neighbors with a mean hopping strength of about 7.6 MHz. The typical device parameters are briefly shown in Table~\ref{tab:paras}. Here, we used $Q_3$, $\cdots$, $Q_{43}$ (relabeled as $Q_1$, $\cdots$, $Q_{41}$) for the 41-qubit experiment, and $Q_4$, $\cdots$, $Q_{43}$ (relabeled as $Q_1$, $\cdots$, $Q_{40}$) for the 40-qubit experiment, but note that their idle points are the same, and the operating points during the experiments are different. The mean energy relaxation time $\overline{T}_1$ of the 41 qubits is $21.0~\mu\text{s}$. To utilize Floquet engineering to tune the effective hopping strengths, we calibrate the qubits to idle frequencies at an average of about $660$~MHz away from their respective maximum frequencies, so that the amplitudes of the time-periodic driving can be adjusted over a wide range.
To suppress the dephasing of low-frequency qubits, all the Z control lines are equipped with DC blocks, and the mean pure dephasing time at the idle frequency is $1.2~\mu\text{s}$.

\section{Floquet Engineering of Superconducting Circuits}\label{sec:zfield}
\subsection{Time-periodic driving and effective hopping strength}
Floquet engineering has been shown as an effective method to adjust hopping strengths in superconducting quantum circuits~\cite{Cai2019,Zhao2022}. It requires the time-periodic drivings on qubits frequencies as
\begin{equation}\label{eq:floquet_freq}
	\omega(t)=\overline{\omega}+A\sin(\mu t+\varphi),
\end{equation}
where $A$, $\mu$, and $\varphi$ denote the modulation amplitude, frequency, and phase, respectively. Here, $\overline{\omega}$ is the average frequency of qubit, which holds a constant $\omega_{\mathrm{ref}}$ for all the resonant qubits in our experiments concerning the off-diagonal AAH model. For convenience, we set $\mu/2\pi=80$~MHz for all the qubits.

Without loss of generality, we first show a two-qubit example. The two-qubit Hamiltonian with nearest-neighbor hopping interaction can be written as
\begin{equation}\label{eq:floquet_h_origin}
	\hat{H}(t)=\omega_1(t)\hat{n}_1+\omega_2(t)\hat{n}_2+g(\hat{a}_{1}^{\dagger}\hat{a}_2+\hat{a}_{1}\hat{a}_2^{\dagger}),
\end{equation}
where $\hat{n}_{j}\equiv\hat{a}^{\dagger}_{j}\hat{a}_{j}$ is the number operator, $\hat{a}_{j}$ ($\hat{a}^{\dagger}_{j}$) denotes the annihilation (creation) operator of the $j$-th qubit, $\omega_j(t)$ denotes the modulated qubit frequency, and $g$ is the direct coupling strength. Here, we use the interaction picture. The corresponding unitary transformation is
\begin{equation}\label{eq:floquet_u}
	\hat{U}_{\mathrm{I}}(t)=\exp\!\left\{\mathrm{i}\left[\overline{\omega}t-\frac{A_1}{\mu}\cos{(\mu t+\varphi_1)}\right]\hat{n}_1+\mathrm{i}\left[\overline{\omega}t-\frac{A_2}{\mu}\cos{(\mu t+\varphi_2)}\right]\hat{n}_2\right\},
\end{equation} 
where the qubits frequencies take the forms of Eq.~\ref{eq:floquet_freq}. Hence, the effective Hamiltonian can be calculated as
\begin{eqnarray}
	\label{eq:floquet_h_int}
	\hat{H}_{\mathrm{I}}&=&\hat{U}_{\mathrm{I}}(t)\hat{H}(t)\hat{U}^{\dagger}_{\mathrm{I}}(t)+\mathrm{i}\left(\frac{\mathrm{d}}{\mathrm{d}t}\hat{U}_{\mathrm{I}}(t)\right)\hat{U}^{\dagger}_{\mathrm{I}}(t),\nonumber\\
	&=&g\exp\!\left[{\mathrm{i}\frac{A_1}{\mu}\cos(\mu t+\varphi_1)}\right]\exp\!\left[{-\mathrm{i}\frac{A_2}{\mu}\cos(\mu t+\varphi_2)}\right]\hat{a}_{1}^{\dagger}\hat{a}_2+\mathrm{H.c.},\nonumber\\
	&=&g\sum_{m,n=-\infty}^{+\infty}\mathrm{i}^{m+n}J_{m}\!\left(\frac{A_1}{\mu}\right)J_{n}\!\left(-\frac{A_2}{\mu}\right)\exp\!\big[{\mathrm{i}(m+n)\mu t}\big]\exp\!\big[{\mathrm{i}(m\varphi_1+n\varphi_2)}\big]\hat{a}_{1}^{\dagger}\hat{a}_2+\mathrm{H.c.},\nonumber\\
	&\approx& g\sum_{m=-\infty}^{+\infty}J_{m}\!\left(\frac{A_1}{\mu}\right)J_{-m}\!\left(-\frac{A_2}{\mu}\right)\exp\!\big[{\mathrm{i}m(\varphi_1-\varphi_2)}\big]\hat{a}_{1}^{\dagger}\hat{a}_2+\mathrm{H.c.},
\end{eqnarray}
where the second line is obtained by using the Baker-Hausdorff formula
$e^{\mathrm{i}x\hat{a}^{\dagger}\hat{a}}f(\hat{a}^{\dagger},\hat{a})e^{-\mathrm{i}x\hat{a}^{\dagger}\hat{a}}=f(\hat{a}^{\dagger}e^{\mathrm{i}x},\hat{a}e^{-\mathrm{i}x})$, the third line uses the Jacobi-Anger series $e^{\mathrm{i}x\cos{\theta}}=\sum_{m=-\infty}^{+\infty}\mathrm{i}^mJ_m(x)e^{\mathrm{i}m\theta}$, with $J_m$ being the $m$-order Bessel function of the first kind, and the last line follows from the rotating-wave approximation with the bound constraint $m+n=0$. Thus, the effective hopping strength can be expressed as
\begin{equation}\label{eq:floquet_g_full}
	g^{\mathrm{eff}}=g\sum_{m=-\infty}^{+\infty}J_{m}\!\left(\frac{A_1}{\mu}\right)J_{-m}\!\left(-\frac{A_2}{\mu}\right)\exp\!\big[{\mathrm{i}m(\varphi_1-\varphi_2)}\big].
\end{equation}
In the following, we discuss three special cases.

\begin{table}[t]
		\centering
		\begin{tabular}{lcccc}
			Parameter & Median & Mean & Stdev. & Units\\[.05cm] \hline
			\\[-.2cm]
			Qubit maximum frequency & $5.767$ & $5.914$ & $0.280$ & GHz \\[.09cm]
			Qubit idle frequency & $5.091$ & $5.254$ & $0.413$ & GHz \\[.09cm]
			Qubit anharmonicity $\alpha/2\pi$ & $-0.202$ & $-0.216$ & $0.021$ & GHz\\[.09cm]
			Readout frequency & $6.684$ & $6.680$ & $0.05$ & GHz\\[.09cm]
			Mean energy relaxation time $\overline{T}_1$ & $20.9$ & $21.0$ & $6.0$ & $\mu$s \\[.09cm]
			Pure dephasing time at idle frequency $T_2^{\ast}$ & $1.1$ & $1.2$ & $0.4$ & $\mu$s \\[.09cm]
			Qubit-resonator coupling & $35.3$ & $34.9$ & $3.8$ & MHz \\[.09cm]
			Mean fidelity of single-qubit gates & $99.2$ & $99.0$ & $1.3$ & \% \\[.09cm]
		\end{tabular}
	\caption{List of device parameters.}\label{tab:paras}
\end{table}

($i$) {$\varphi_1-\varphi_2=2n\pi$, with $n$ being an integer}. By using $J_m(x+y)=\sum_{l=-\infty}^{+\infty}J_l\left(x\right)J_{m-l}\left(y\right)$, Eq.~\ref{eq:floquet_g_full} is reduced as
\begin{equation}\label{eq:floquet_g_case1}
	g^{\mathrm{eff}}=gJ_{0}\!\left(\frac{A_1-A_2}{\mu}\right).
\end{equation} 

($ii$) {$\varphi_1-\varphi_2=(2n+1)\pi$, with $n$ being an integer}. Similarly, we can obtain
\begin{equation}\label{eq:floquet_g_case2}
	g^{\mathrm{eff}}=gJ_{0}\!\left(\frac{A_1+A_2}{\mu}\right),
\end{equation} 
based on the parity of the Bessel function $J_m(x)=J_m(-x)$ and $J_m(x)=(-1)^{m}J_{-m}(x)$.

($iii$) {$A_1=A\neq0,~A_2=0$}. In this case, there is only one time-periodic driving and the effective hopping strength is given by
\begin{equation}\label{eq:floquet_g_case3}
	g^{\mathrm{eff}}=gJ_{0}\!\left(\frac{A}{\mu}\right).
\end{equation} 
Based on the properties of the Bessel function, the decoupling point is reached when $A\approx 2.405\mu$, and the maximum coupling is $g^{\mathrm{eff}}\approx-0.403g$ with $A\approx3.832\mu$. In our devices, $g$ is around $7.6$~MHz, so the adjustable range of coupling strength is from $-3.06$~MHz to $7.6$~MHz.

Although the above results are derived for the two-qubit case, the formulas are applicable for the multi-qubit case, and the couplings of the driven qubit to readout resonator and XY microwave can be described by this simplified model.

\subsection{Amplitude modulation}

According to Eqs.~(\ref{eq:floquet_freq}, \ref{eq:floquet_g_case1}--\ref{eq:floquet_g_case3}), the effective hopping strength depends on the driving amplitude with respect to the qubit frequency. In order to accurately manipulate the hopping strength, we need to calibrate the mapping between the qubit frequency and the experimental Z pulse amplitude (Zpa).

\begin{figure}[t!]
	\includegraphics[width=1\textwidth]{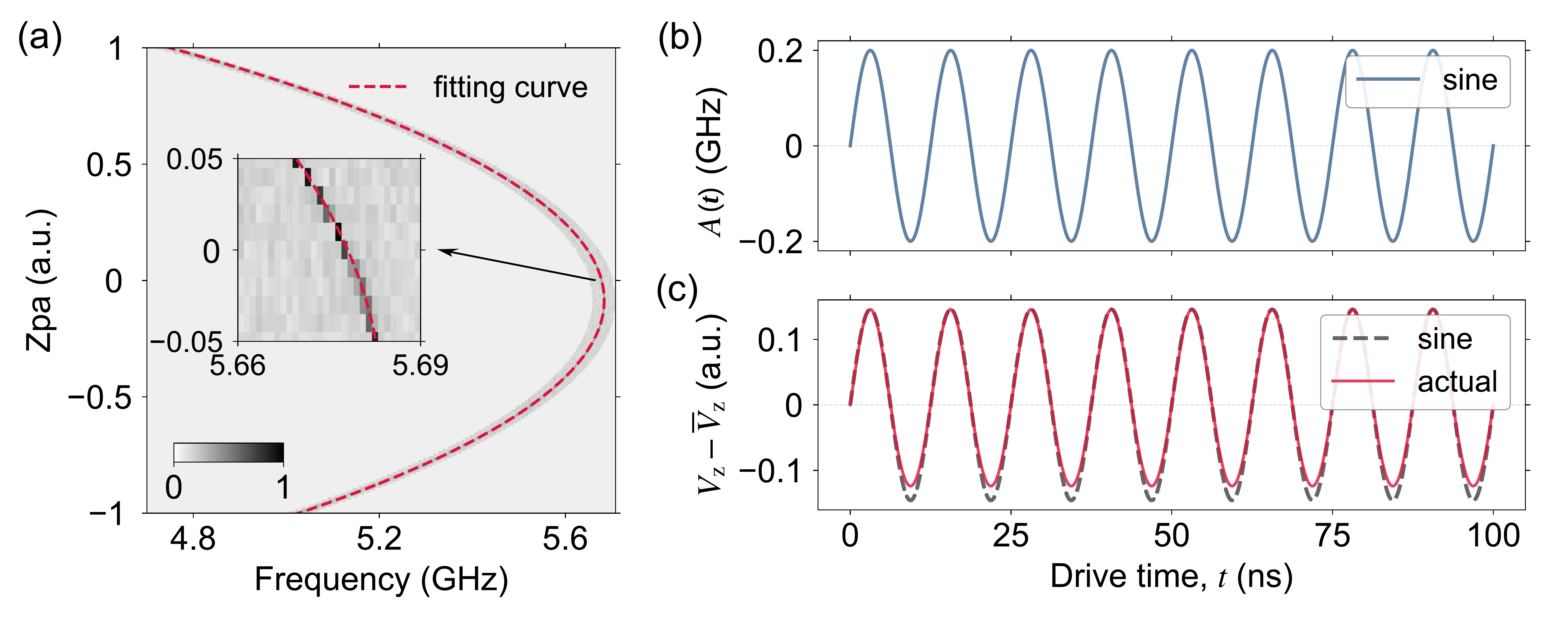}
	\caption{Single-qubit spectroscopy and Floquet Z pulse. (a) Typical experimental data of single-qubit spectroscopy. The red dashed line is for the fitting result in the form of Eq.~\ref{eq:zpa2freq}. (b) Sine modulation in the frequency with $A/(2\pi)=0.2$~GHz and $\mu/(2\pi)=80$~MHz. (c) The corresponding modulated Zpa. Here the red solid line is the actual Zpa used in the Floquet engineering, while the gray dashed line represents a sine-like Zpa.}\label{fig:spec_and_drive}
\end{figure}

For the frequency-tunable transmon qubit with symmetric Josephson junctions, the relationship between qubit frequency $\omega$ and external flux $\Phi_{\mathrm{e}}$ is given by~\cite{Koch2007}
\begin{equation}\label{eq:zpa2freq}
	\omega = \sqrt{8E_{\mathrm{JJ}}E_\mathrm{C}\left|\cos(\pi\frac{\Phi_{\mathrm{e}}}{\Phi_0})\right|}-E_\mathrm{C},
\end{equation}
where $E_{v{JJ}}$ denotes the Josephson energy when no flux enters the loop of the superconducting quantum interference device (SQUID), $E_{\mathrm{C}}$ is the charging energy, and $\Phi_0$ denotes the flux quantum. The external flux $\Phi_e$ is linearly related to the Zpa $V_z$ on the qubit produced by the DC source or AWG under the weak flux conditions, i.e., $\pi{\Phi_{\mathrm{e}}}/{\Phi_0}=kV_z+b$. Here, $E_{\mathrm{C}}$ can be determined by performing a double-photon excitation experiment, and other parameters $E_{\mathrm{JJ}}$, $k$, and $b$ can be obtained by fitting the spectroscopy measurement results of the qubit, see Fig.~\ref{fig:spec_and_drive}(a). Hence, we can obtain the relationship between the Zpa $V_z$ and the driving amplitude $A$ by substituting Eq.~\ref{eq:zpa2freq} into Eq.~\ref{eq:floquet_freq}:
\begin{eqnarray}
	\label{eq:V_z}
	V_z&=&\frac{1}{k}\arccos\left[\pm\frac{\left(\overline{\omega}+A\sin(\mu t+\varphi)+E_C\right)^2}{8E_{\mathrm{JJ}}E_\mathrm{C}}\right]-\frac{b}{k},\nonumber\\
	&=&\frac{1}{k}\arccos\left[\pm\frac{\left(\sqrt{8E_{\mathrm{JJ}}E_\mathrm{C}\left|\cos(k\overline{V}_z+b)\right|}+A\sin(\mu t+\varphi)\right)^2}{8E_{\mathrm{JJ}}E_\mathrm{C}}\right]-\frac{b}{k},
\end{eqnarray}
where $\overline{V}_z$ is the constant amplitude of a rectangular Zpa without any modulations ($A=0$~GHz), which makes the qubit frequency equal to $\overline{\omega}$. Thus, given the driving amplitude, we can then calculate the corresponding Zpa and construct the actual Z pulse waveform, see Fig.~\ref{fig:spec_and_drive}, (b) and (c).

\begin{figure*}[t]
	\includegraphics[width=1\textwidth]{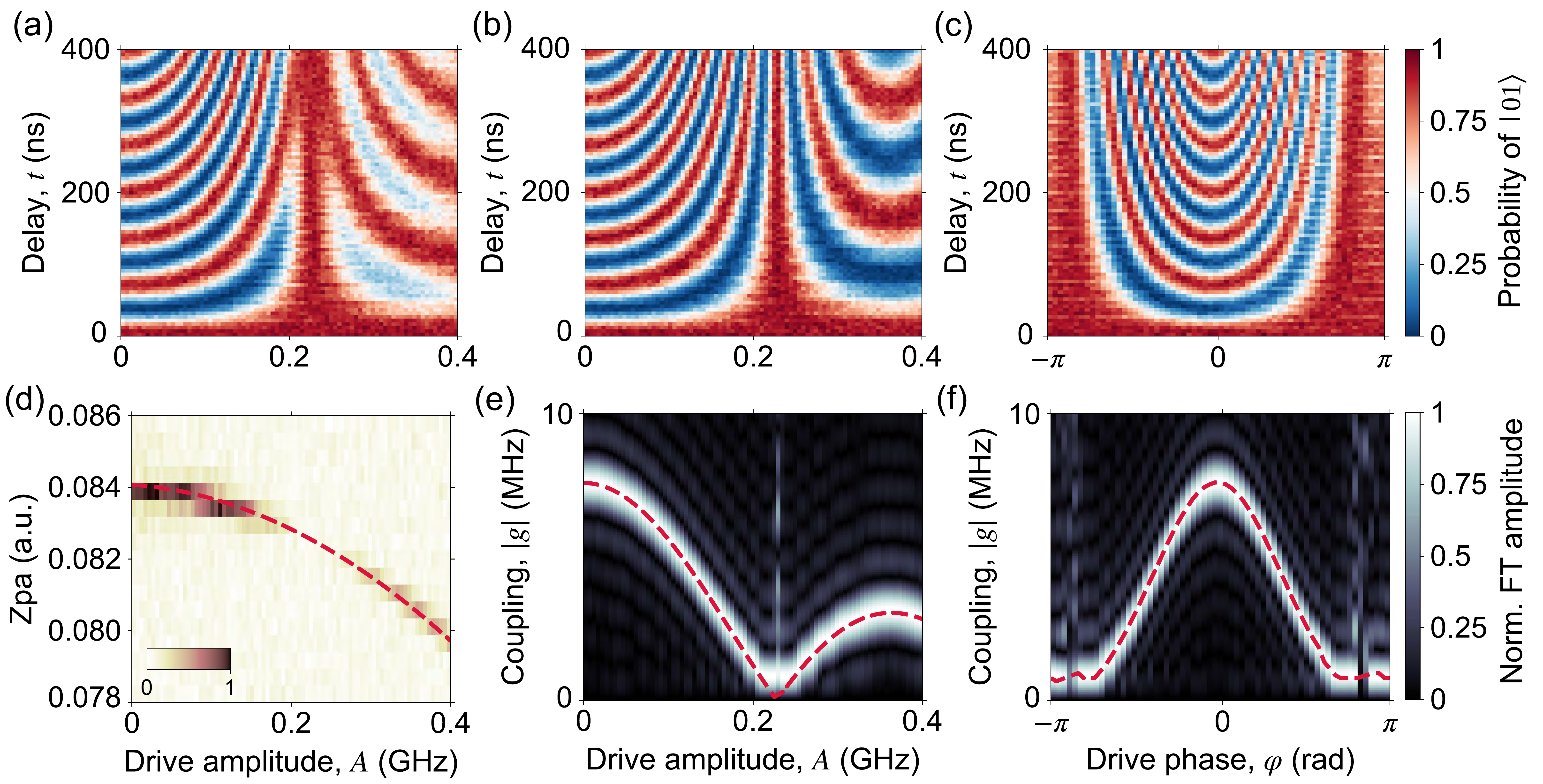}
	\caption{Amplitude modulation and phase alignment of Floquet engineering. Here we fix $\mu/(2\pi)=80$~MHz. (a) Swap probability of $\ket{10}$ with the time-periodic driving before correcting the qubit frequency detuning. (b) Swap probability of $\ket{10}$ after correcting the qubit frequency detuning. (c) Phase alignment between two qubits. Both driving amplitudes are set to $0.1$~GHz. (d) Spectroscopic measurement of the qubit frequency detuning induced by the modulation of the Z pulse. The red dashed curve is the polynomial fitting result of the relationship between the driving amplitude and the compensation Zpa. (e) Coupling strength versus the driving amplitude, which is extracted from the FT of (b). The red dashed curve denotes the fitting result in the form of the Bessel function. (f) Coupling strength versus drive phase, which is extracted from the FT of (c). The phase corresponding to the maximum coupling strength is the phase difference to be compensated. Here we determine this phase by a polynomial fitting shown by the red dashed curve.}
	\label{fig:zFieldAll}
\end{figure*}

To evaluate the actual coupling strength between two qubits, we perform a swap-like experiment. First, we identify the Zpa of each qubit to make them resonant and then add the time-periodic driving on one of the qubits. As the driving amplitude $A$ increases, the period of the swap changes, but at the same time, the swap probability becomes smaller, as shown in Fig.~\ref{fig:zFieldAll}(a). This is due to the frequency of the driven qubit gradually deviating from the resonance point. We correct this frequency detuning by fitting the results of scanning the compensation Zpa and the driving amplitude with a polynomial, as shown in Fig.~\ref{fig:zFieldAll}(d). Note that the probability of readout becomes smaller as the driving amplitude approaches the decoupling point. Because the couplings of the driven qubit to readout resonator and XY microwave are also modulated, except that the resonance conditions are not satisfied here and the rotation wave approximation may be excluded. 

After this correction, we obtain a better swap probability result, as shown in Fig.~\ref{fig:zFieldAll}(b). In the experiments, we use Eq.~\ref{eq:V_z} to estimate the relationship between the driving amplitude and the Zpa, and apply the Bessel function in Eq.~\ref{eq:floquet_g_case3} to calculate the hopping strength. However, considering the control deviation, we need to add a scale factor $\eta$ ($\approx 1$) into the Bessel function, and the effective hopping strength is expressed as
\begin{equation}\label{eq:floquet_g_case3_modified}
	g^{\mathrm{eff}}=gJ_{0}\!\left(\eta\frac{A}{\mu}\right).
\end{equation} 
The above formula is used to fit the mapping between the driving amplitude and the coupling strength, as displayed in Fig.~\ref{fig:zFieldAll}(e). The experimental couplings are calculated as half the Fourier frequency of the swap probability of $\ket{01}$ (or $\ket{10}$).

\subsection{Phase alignment}

In our experiments, we modulate the coupling strengths between each pair of neighboring qubits to realize the off-diagonal AAH model. All qubits are simultaneously driven by the time-periodic Z pulse. The modulated coupling strength satisfies
\begin{equation}\label{eq:floquet_g_full_modified}
	g_{j,j+1}^{\mathrm{eff}}=g_{j,j+1}\sum_{m=-\infty}^{+\infty}J_{m}\!\left(\frac{\eta_jA_j}{\mu}\right)J_{-m}\!\left(-\frac{\eta_{j+1}A_{j+1}}{\mu}\right)\exp\!\big[{\mathrm{i}m(\varphi_j-\varphi_{j+1})}\big]\,
\end{equation}
where the phase difference is $\Delta\varphi_j=\varphi_j-\varphi_{j+1}$. Compared with Eq.~\ref{eq:floquet_g_full}, Eq.~\ref{eq:floquet_g_full_modified} is modified by two scale factors $\eta_j$ and $\eta_{j+1}$, which are determined by fitting the amplitude modulation curves. Similar to Eq.~\ref{eq:floquet_g_case1}, the effective coupling is given by the following formula with $\Delta\varphi_j=0$
\begin{equation}\label{eq:floquet_g_case1_modified}
	g_{j,j+1}^{\mathrm{eff}}=g_{j,j+1}J_{0}\!\left(\frac{\eta_jA_j-\eta_{j+1}A_{j+1}}{\mu}\right).
\end{equation}
Note that when the driving amplitudes of the two qubits meet $\eta_jA_j=\eta_{j+1}A_{j+1}$, the coupling strength reaches its maximum value $g_{j,0}$. We perform the phase alignment calibration in this condition and scan the drive phase of one qubit to measure the period of the two-qubit swap. If the phase is aligned, the coupling strength will be the maximum $g_{j,0}$. We figure out this point and compensate this drive phase into the time-periodic driving of the corresponding qubit. The typical experimental data is shown in Fig.~\ref{fig:zFieldAll}, (c) and (f). It can be seen that for the devices that have been calibrated for the timing of the Z pulses, the compensation of the drive phase is almost zero.

\begin{figure}[t!]
	\begin{center}
		\includegraphics[width=1\textwidth]{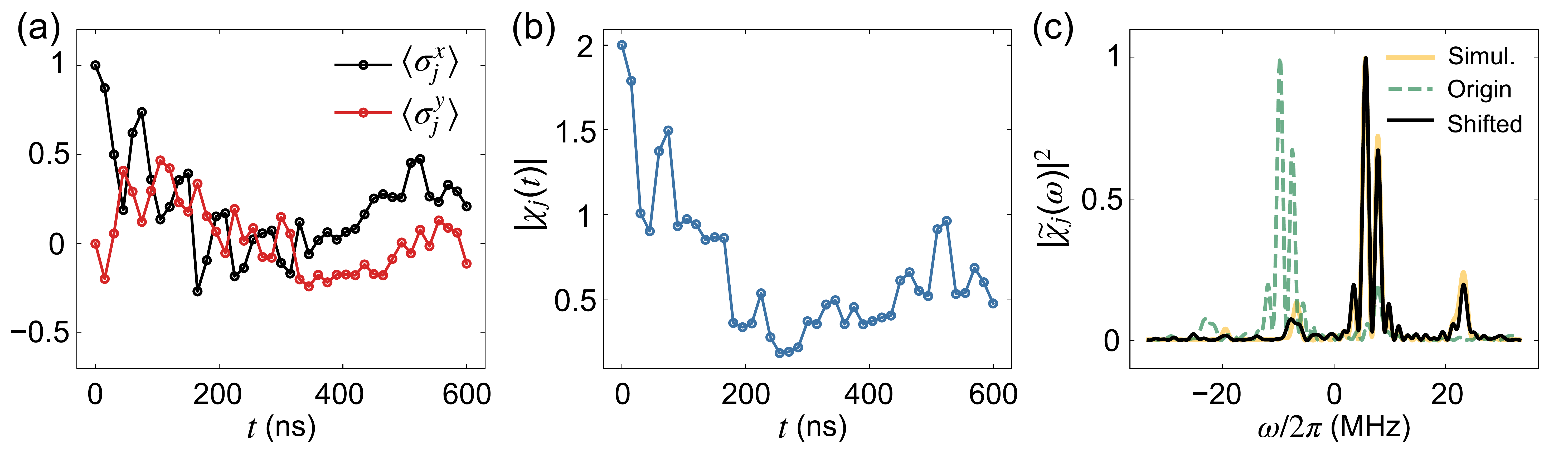}
		\caption{Typical experimental data of the single-qubit Loschmidt echo. (a) Pauli expectations $\langle\hat{\sigma}^{x}(t)\rangle$ and $\langle\hat{\sigma}^{y}(t)\rangle$ versus the evolution time $t$. (b) The magnitude of the Loschmidt echo amplitude versus the evolution time $t$. (c) Fourier spectrum of the Loschmidt echo amplitude.
		}
		\label{fig:LE_fft_integrate}
	\end{center}
\end{figure}

\section{\label{sec:loschmidt_echo} Energy Spectrum Measurement}
\subsection{Loschmidt echo and Fourier spectrum}

The Loschmidt echo (LE) of qubit $j$ is defined as $\mathcal{L}_j(t)=|\chi_j(t)|^2$
, where the amplitude is given by
\begin{equation}\label{eq:chi_j_t}	\chi_j(t)=\langle2\hat{\sigma}_j^+(t)\rangle=\langle\hat{\sigma}_j^x(t)\rangle+\mathrm{i}\langle\hat{\sigma}_j^y(t)\rangle.
\end{equation}
Here the average is taken with respect to the time-evolved state
\begin{equation}\label{eq:Psi_t}
	\ket{\Psi(t)}_{j}=e^{-\mathrm{i}\hat{H}t}\ket{\Psi(0)}_{j}=\sum_{n}C_{n,j}e^{-\mathrm{i}E_{n}t}\ket{\phi_{n}},
\end{equation}
by virtue of the time-dependent Schr$\ddot{\mathrm{o}}$dinger equation, where $C_{n,j}=\langle\phi_{n}|\Psi(0)\rangle_{j}$ and the spectral decomposition of the total Hamiltonian is $H=\sum_{n}E_{n}\ket{\phi_{n}}\bra{\phi_{n}}$. 

Considering the single-excitation case, we choose the initial state that is prepared by a $Y_{{\pi}/{2}}$ gate on qubit $j$:
\begin{equation}\label{eq:Psi_0}
	\ket{\Psi(0)}_{j}=\ket{0}_{1}\otimes\ket{0}_{2}\otimes\cdots\frac{1}{\sqrt{2}}\left(\ket{0}_j+\ket{1}_j\right)\otimes\cdots\ket{0}_{N-1}\otimes\ket{0}_N,
\end{equation}
and measure the expectation $\langle\hat{\sigma}_j^{x}(t)\rangle$ and $\langle\hat{\sigma}_j^y(t)\rangle$ to obtain the amplitude of LE by Eq.~\ref{eq:chi_j_t} (see typical experimental data in Fig.~\ref{fig:LE_fft_integrate}, (a) and (b)). In the absence of any decoherence or noise, $\chi_j(t)$ can be calculated as
\begin{equation}\label{eq:chi_j_t_ideal}
	\chi_j(t)=\sum_{n}|C_{n,j}|^2e^{-\mathrm{i}E_{n}t},
\end{equation}
and the information about the energy spectrum is encoded in its Fourier transform (FT)
\begin{equation}\label{eq:chiFFT_ideal}
	\widetilde{\chi}_j(\omega)=\mathscr{F}\{\chi_j(t)\}=\sum_{n}|C_{n,j}|^2\delta(\omega-E_{n}),
\end{equation}
where $\mathscr{F}\{\cdot\}$ denotes the Fourier transform and $\delta(\cdot)$ is the Dirac delta function. Although the corresponding peak of each eigenenergy $E_{n}$ is included in Eq.~\ref{eq:chiFFT_ideal}, we are still unable to directly obtain all the spectral information just from one single-qubit result because the coefficient $C_{n,j}$ could be too small, and for some particular eigenenergy $E_n$ this coefficient may be too large that it overwhelms other peaks. Therefore, we select a few qubits and sum the squared magnitudes of their $\widetilde{\chi}_j(\omega)$ as
\begin{equation}\label{eq:I_w}
	I(\omega)=\sum_{j}\left|\widetilde{\chi}_j(\omega)\right|^2,
\end{equation}
thus the positions of the peaks in $I(\omega)$ indicate the energy spectrum of the total Hamiltonian. In fact, these peaks are widened due to the measurement time window and decoherence, as explained in the following.

\begin{figure}[t!]
	\centering
	\includegraphics[width=4in]{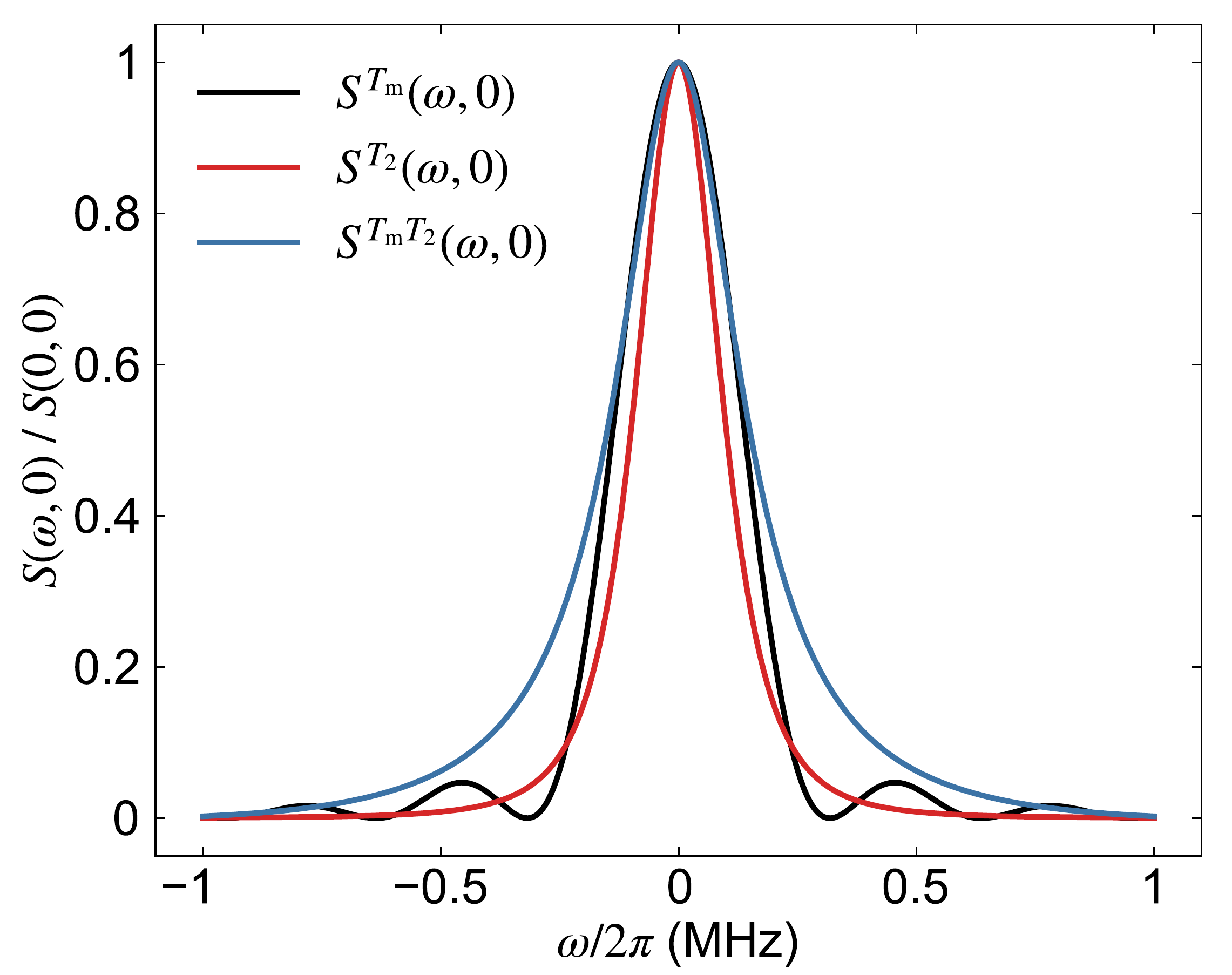}
	\caption{Effects of both the measurement time window and decoherence on the Fourier spectrum. Here we set $T_{\mathrm{m}}=T_{2}=1~\mu\mathrm{s}$. }\label{fig:LEamp_pfunc}
\end{figure}

\subsection{Measurement time window and decoherence}
\textbf{(\textit{i}) Only considering the effects of the measurement time window}. Considering the finite length of the measurement time window $[0, T_{\mathrm{m}}]$ for acquiring experimental data, we rewrite the LE amplitude as
\begin{equation}\label{eq:chi_j_t_m}
	\chi^{T_{\mathrm{m}}}_j(t)=\sum_{n}|C_{n,j}|^2e^{-\mathrm{i}E_{n}t}\mathrm{rect}\!\left(\frac{t-\frac{T_{\mathrm{m}}}{2}}{T_{\mathrm{m}}}\right),
\end{equation}
where $\mathrm{rect}(\cdot)$ is the rectangular time window function. Hence, its FT can be expressed as

\begin{eqnarray}
	\label{eq:chiFFT_j_t_m}
	\widetilde{\chi}^{T_{\mathrm{m}}}_j(\omega)&=&\mathscr{F}\left\{\sum_{n}|C_{n,j}|^2e^{-\mathrm{i}E_{n}t}\right\}\ast\mathscr{F}\left\{\mathrm{rect}\!\left(\frac{t-\frac{T_{\mathrm{m}}}{2}}{T_{\mathrm{m}}}\right)\right\}\nonumber\\
	&=&\sum_{n}|C_{n,j}|^2\delta(\omega-E_{n})\ast \exp\!\left({-\mathrm{i}\omega \frac{T_{\mathrm{m}}}{2}}\right)T_{\mathrm{m}}\mathrm{sinc}\!\left(\omega \frac{T_{\mathrm{m}}}{2}\right)\nonumber\\
	&=&\sum_{n}|C_{n,j}|^2\exp\!\left[{-\mathrm{i}(\omega-E_{n})\frac{T_{\mathrm{m}}}{2}}\right]T_{\mathrm{m}}\mathrm{sinc}\!\left[(\omega-E_{n})\frac{T_{\mathrm{m}}}{2}\right],
\end{eqnarray}
where $\ast$ denotes the convolution, the first line follows from the convolution theorem of the FT and the third line follows from $\delta(x-x_0)\ast f(x)=f(x-x_0)$. To evaluate the peak broadening of $I(\omega)$ around $\omega=E_{n}$ due to the measurement time window, we define the profile function
\begin{equation}\label{eq:S_m}
	S^{T_{\mathrm{m}}}(\omega, E_{n})\!=\!\left|\exp\!\left[{-\mathrm{i}(\omega\!-\!E_{n})\frac{T_{\mathrm{m}}}{2}}\right]T_{\mathrm{m}}\mathrm{sinc}\!\left[(\omega\!-\!E_{n})\frac{T_{\mathrm{m}}}{2}\right]\right|^2=T_{\mathrm{m}}^2\mathrm{sinc}^2\!\left[(\omega\!-\!E_{n})\frac{T_{\mathrm{m}}}{2}\right],
\end{equation}
and calculate its full width at half maximum (FWHM):
\begin{equation}\label{eq:FWHM_m}
	\Delta\omega_{T_{\mathrm{m}}}\approx\frac{5.57}{T_{\mathrm{m}}},\quad\Delta f_{T_{\mathrm{m}}}=\frac{\Delta\omega_{T_{\mathrm{m}}}}{2\pi}\approx\frac{0.89}{T_{\mathrm{m}}}.
\end{equation}

\textbf{(\textit{ii}) Only considering the effects of decoherence}. As the evolution time increases, one can observe an exponential decay of the LE amplitude due to the decoherence process in the experiments with a decaying rate $\gamma$
\begin{equation}\label{eq:chi_j_t_2}
	\chi^{T_{2}}_j(t)=\sum_{n}|C_{n,j}|^2e^{-\mathrm{i}E_{n}t}\exp\!\left({-\gamma|t|}\right).
\end{equation}
Here $\gamma=1/T_2=1/(2T_1)+1/T^{\ast}_2$, with $T_1$ and $T_2^{\ast}$ being the energy relaxation time and the pure dephasing time of the measured qubit $j$, respectively. The FT of $\chi^{T_{2}}_j(t)$ is given by
\begin{eqnarray}
	\label{eq:chiFFT_j_t_2}
	\widetilde{\chi}^{T_{2}}_j(\omega)&=&\mathscr{F}\left\{\sum_{n}|C_{n,j}|^2e^{-\mathrm{i}E_{n}t}\right\}\ast\mathscr{F}\Bigg\{\exp\!\left({-\gamma|t|}\right)\Bigg\}\nonumber\\
	&=&\sum_{n}|C_{n,j}|^2\delta(\omega-E_{n})\ast \frac{2\gamma}{\gamma^2+\omega^2}\nonumber\\
	&=&\sum_{n}|C_{n,j}|^2\frac{2\gamma}{\gamma^2+(\omega-E_{n})^2}\nonumber\\
	&=&\sum_{n}|C_{n,j}|^2\frac{\frac{2}{T_2}}{(\frac{1}{T_2})^2+(\omega-E_{n})^2},
\end{eqnarray}
and the corresponding profile function becomes
\begin{equation}\label{eq:S_2}
	S^{T_{2}}(\omega, E_{n})=\frac{(\frac{2}{T_2})^2}{\big[(\frac{1}{T_2})^2+(\omega-E_{n})^2\big]^2},
\end{equation}
with the FWHM
\begin{equation}\label{eq:FWHM_2}
	\Delta\omega_{T_{\mathrm{2}}}=\frac{2\sqrt{\sqrt{2}-1}}{T_2}\approx\frac{1.29}{T_2},\quad\Delta f_{T_{\mathrm{2}}}=\frac{\Delta\omega_{T_{\mathrm{2}}}}{2\pi}=\frac{\sqrt{\sqrt{2}-1}}{\pi T_2}\approx\frac{0.2}{T_2}.
\end{equation}

\begin{figure}[t!]
	\begin{center}
		\includegraphics[width=1\textwidth]{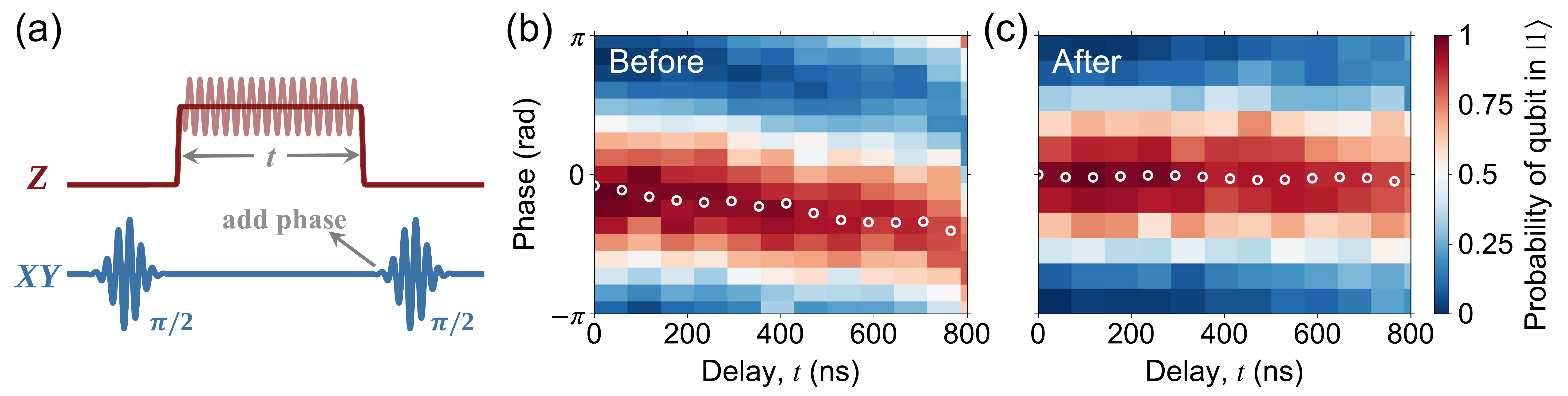}
		\caption{Phase calibration. (a) The Ramsey-like pulse sequence of measuring the phase accumulation.
			The experimental results show the  $|1\rangle$-state probability evolution before (b) and after (c) the phase compensation, respectively.
		}
		\label{fig:LEcaliphase}
	\end{center}
\end{figure}

\textbf{(\textit{iii}) Considering both the effects of measurement time window and decoherence}. Taking into account both the effects of measurement time window and decoherence, we rewrite the FT of LE amplitude as
\begin{eqnarray}
	\label{eq:chiFFT_j_t_m2}
	\widetilde{\chi}^{T_{\mathrm{m}}T_2}_j(\omega)\!\!\!\!&=&\!\!\!\!\mathscr{F}\left\{\sum_{n}|C_{n,j}|^2e^{-\mathrm{i}E_{n}t}\right\}\ast\mathscr{F}\left\{\mathrm{rect}\!\left(\frac{t-\frac{T_{\mathrm{m}}}{2}}{T_{\mathrm{m}}}\right)\exp\!\left({-\gamma|t|}\right)\right\}\nonumber\\
	\!\!\!\!&=&\!\!\!\!\sum_{n}|C_{n,j}|^2\frac{\frac{2}{T_2}} {(\frac{1}{T_2})^2\!+\!(\omega\!-\!E_{n})^2}\exp{\left(-\mathrm{i}\omega\frac{T_{\mathrm{m}}}{2}\right)}\nonumber\\
	&~&\times\left[1\!+\!e^{- \frac{T_{\mathrm{m}}}{2T_2}}\left((\omega\!-\!E_{n})T_2\sin\!\left[(\omega\!-\!E_{n})\frac{T_{\mathrm{m}}}{2}\right]-\cos\!\left[(\omega\!-\!E_{n})\frac{T_{\mathrm{m}}}{2}\right]\right)\right]
\end{eqnarray}
with the corresponding profile function
\begin{equation}\label{eq:S_2}
	S^{T_{\mathrm{m}}T_{2}}(\omega, E_{n})\!=\!\frac{\left(\frac{2}{T_2}\right)^2\left[\!1\!+\!e^{-\frac{T_{\mathrm{m}}}{2T_2}}\left((\omega\!-\!E_{n})T_2\sin\!\left[(\omega\!-\!E_{n})\frac{T_{\mathrm{m}}}{2}\right]\!-\!\cos\!\left[(\omega\!-\!E_{n})\frac{T_{\mathrm{m}}}{2}\right]\right)\right]^2}{\big[(\frac{1}{T_2})^2\!+\!(\omega-E_{n})^2\big]^2}.
\end{equation}
It can be verified that
\begin{equation} \lim_{T_2\rightarrow\infty}S^{T_{\mathrm{m}}T_{2}}=S^{T_{\mathrm{m}}},\quad \lim_{T_{\mathrm{m}}\rightarrow\infty}S^{T_{\mathrm{m}}T_{2}}=S^{T_{2}}.
\end{equation}
In Fig.~\ref{fig:LEamp_pfunc}, we show the effects of the measurement time window and decoherence on the Fourier spectrum.

According to the above discussion, the FWHM (the precision in the energy spectrum measurement) is limited by the measurement time length and the decoherence time. In our experiments, the decoherence time only limits how long we can track the LE amplitude oscillations, which will ultimately bound the maximum precision in the Fourier spectrum. If the decoherence time is determined, the precision can be improved by extending the measurement time appropriately.

\subsection{Dynamical phase calibration}

In our experiments, we tune all qubits on the operating frequencies from their idle points with rectangular Z pulses after preparing the initial states. The total dynamical phase can be split into two parts. One comes from the difference between the idle frequency $\omega_{\mathrm{idle}}$ and the target operating frequency $\omega_{\mathrm{targ}}$
\begin{equation}\label{eq:DPhase1}
	\Delta \phi_1=(\omega_{\mathrm{targ}}-\omega_{\mathrm{idle}})t,
\end{equation}
where $t$ is the evolution time. The other results from the imperfect rectangle Z pulse (e.g.~rising and falling edges, and distortions), which can be expressed as
\begin{equation}\label{eq:DPhase2}
	\Delta \phi_2=\int_0^t\big[\omega_{\mathrm{actu}}(t)-\omega_{\mathrm{targ}}\big]\mathrm{d}t,
\end{equation}
with $\omega_{\mathrm{actu}}(t)$ being the time-independent qubit frequency biased by the actual Z pulse. The total phase accumulation is thus given by $\Delta\phi=\Delta\phi_1+\Delta \phi_2$.

For an experiment like quantum walks without phase measurement, the phase accumulation does not affect the expectations of observables such as the populations or probabilities. However, in the Loschmidt echo measurement, we need to compensate for phase accumulation to obtain the correct Pauli expectations $\langle\hat{\sigma}^x(t)\rangle$ and $\langle\hat{\sigma}^y(t)\rangle$, which represents the phase (off-diagonal) information of the qubit. We perform a Ramsey-like experiment to measure the phase accumulation via scanning the phase of the second $\pi/2$ pulse. The pulse sequence is shown in Fig.~\ref{fig:LEcaliphase}(a). For experiments without tuning couplings, we just apply the rectangular Z pulse on the qubit, while a periodic driving on the Z control line is added for the experiments that require tunable couplings. Compared with the experimental data before compensation (Fig.~\ref{fig:LEcaliphase}(b)), the phase accumulation can be almost completely compensated by adding the corresponding phase $\Delta\phi$, as seen in Fig.~\ref{fig:LEcaliphase}(c).

Note that the above experimental compensation is carried out in the frequency reference frame of the qubit itself. In the experimental data processing, one may consider the frame of reference frequency $\omega_{\mathrm{ref}}$ and multiply the measured off-diagonal observable $\langle\hat{\sigma}^{+}(t)\rangle=\langle\hat{\sigma}^{x}(t)\rangle+\mathrm{i}\langle\hat{\sigma}^{y}(t)\rangle$ by an additional phase factor $\exp({-\mathrm{i}\Delta\omega_{\mathrm{ref}}t})$ with $\Delta\omega_{\mathrm{ref}}=\omega_{\mathrm{ref}}-\omega_{\mathrm{targ}}$. Hence, the Fourier results are convoluted with $\delta(\omega-\Delta\omega_{\mathrm{ref}})$, and the corresponding energy spectrum will be shifted to the reference frame of $\omega_{\mathrm{ref}}$. As shown in Fig.~\ref{fig:LE_fft_integrate}(c), the shifted experimental results agree with the simulation results that minus the reference frequency. In fact, the shift simply depends on the chosen reference frequency and the relative values of the eigenenergies remain the same. 

\section{\label{sec:qusim} Automatic Calibration Scheme}

\begin{figure}[t!]
	\begin{center}
		\includegraphics[width=1\textwidth]{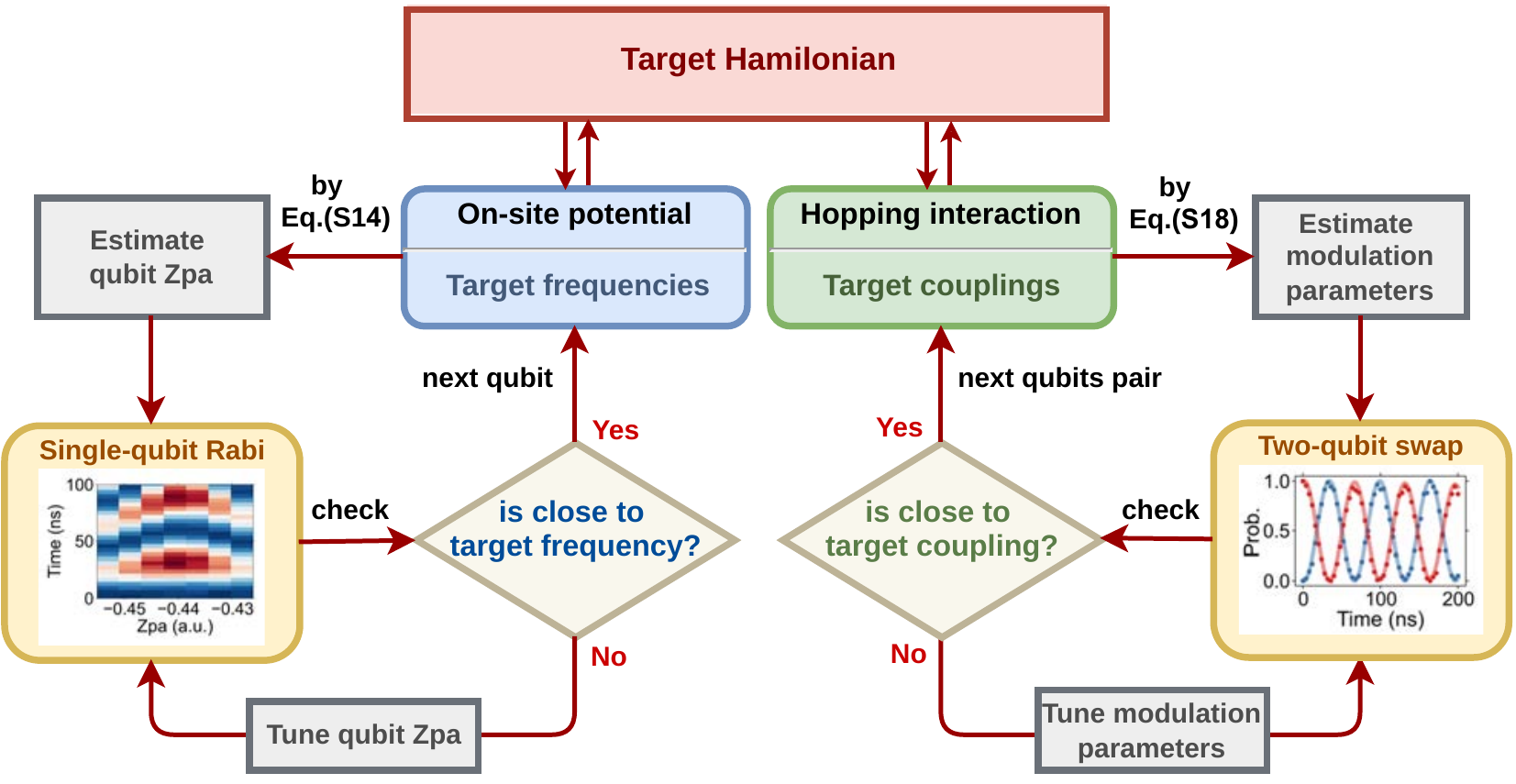}
		\caption{Automatic calibration for the general quantum simulation schemes assisted by the Floquet engineering in superconducting circuits. 
		}
		\label{fig:freq_g_cali}
	\end{center}
\end{figure}

In this section, we will give a brief introduction to the automatic calibration scheme for the general quantum simulation tasks assisted by Floquet engineering in superconducting circuits. The whole calibration diagram is briefly shown in Fig.~\ref{fig:freq_g_cali}.

For a general quantum simulation task, the core is to construct the target Hamiltonian. In most superconducting circuits, the total Hamiltonian can be split into two terms, namely the on-site potential (target frequencies) and the hopping interaction (target couplings). Our aim is to engineer these two parts. For the on-site potential, we apply the rectangular Z pulse on the qubit, and its frequency can be controlled by the amplitude (Zpa) of this fast bias. For the hopping interaction, we apply the external periodic driving on the qubits Z line to adjust the neighboring couplings within a certain range. As mentioned in Sec.~\ref{sec:zfield}, the hopping strength corresponds to the amplitude and frequency of the periodic driving.

After measuring the individual spectroscopic and Floquet parameters of each qubit, we obtain a rough mapping between the experimental parameters and the target Hamiltonian. However, to simulate the target Hamiltonian with high accuracy, we cannot directly use the parameters from the individual measurement to simultaneously manipulate the qubits. Because the crosstalk and complex interactions in multi-qubit quantum devices will lead to the deviation of experimental control parameters. Therefore, it is necessary to tune these parameters under synchronous control. In our experiments, we set this synchronous control environment for the frequencies of all non-target qubits being staggered approximately $\pm80$~MHz around the reference frequency $\omega_{\mathrm{ref}}$. In this environment, we sequentially calibrate the control parameters for each qubit or pair of qubits. The qubit frequency is measured by the vacuum Rabi oscillation and the coupling is characterized by the two-qubit swap experiment.

To simulate the diagonal AAH model, the hopping strength is fixed to about $7.6$~MHz. For each $b_{v}$, we calibrate the frequencies to satisfy $\omega_j=\omega_{\mathrm{ref}}+v\cos(2\pi b_{v}j)$,  with $v/(2\pi)=15.2$~MHz and $\omega_{\mathrm{ref}}/(2\pi)\approx5.107$~GHz. For the realization of the off-diagonal AAH model, the qubit frequencies are set as $\omega_j/(2\pi)=\omega_{\mathrm{ref}}/(2\pi)\approx5.02$~GHz, and the couplings obey $g_{j}=u\left[1+\lambda\cos(2\pi b_{\lambda}j+\varphi_{\lambda})\right]$ with $\omega_{\mathrm{ref}}/(2\pi)\approx5.02$~GHz. We calibrate $u/(2\pi)=4.78$~MHz and $\lambda=0.4$ for $b_{\lambda}=1/2$, while for $b_{\lambda}=1/4$ we adjust $u/(2\pi)=3.35$~MHz, with $\lambda=1.0$  and $u/(2\pi)=2.77$~MHz, with $\lambda=\sqrt{2}$.

Note that the Hamiltonian parameters used in the numerical simulation have not been adjusted to fit the experimental results. Once the experimental control parameters, corresponding to the frequencies and couplings of the target Hamiltonian, were calibrated, we built experimental circuits and performed measurements. It is also important that in the energy spectrum measurement, the calibration of the dynamical phase is required; otherwise, the measured spectrum data will have an overall shift relative to the numerical simulation data. Therefore, we must eliminate the accumulation of the dynamical phase by inserting a virtual Z gate before the experimental measurement, see Sec.~\ref{sec:loschmidt_echo}.

\begin{figure}[ht!]
	\begin{center}
		\includegraphics[width=0.99\textwidth]{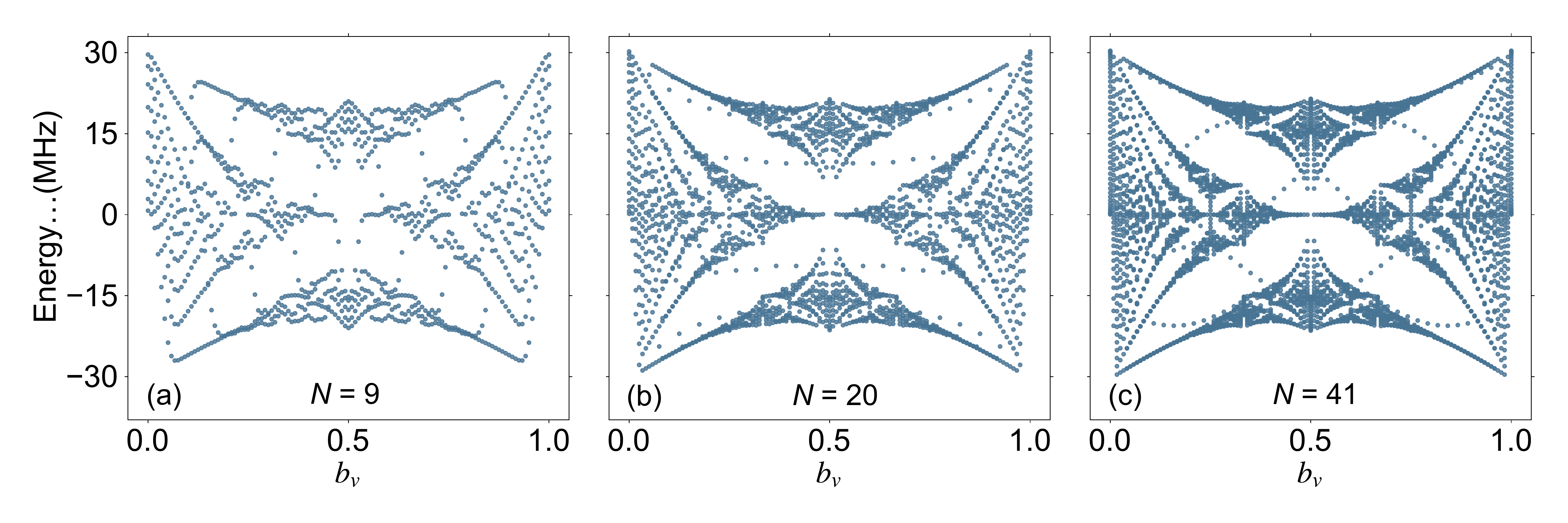}
		\caption{Finite-size effects on the Hofstadter butterfly energy spectrum of diagonal AAH models with $u/(2\pi)=7.6$~MHz and $v=2u$.
		}
		\label{fig:size_effect_butterfly}
	\end{center}
\end{figure}

\section{\label{sec:extended} Additional Discussion}

\begin{figure}[hb!]
	\begin{center}
		\includegraphics[width=0.99\textwidth]{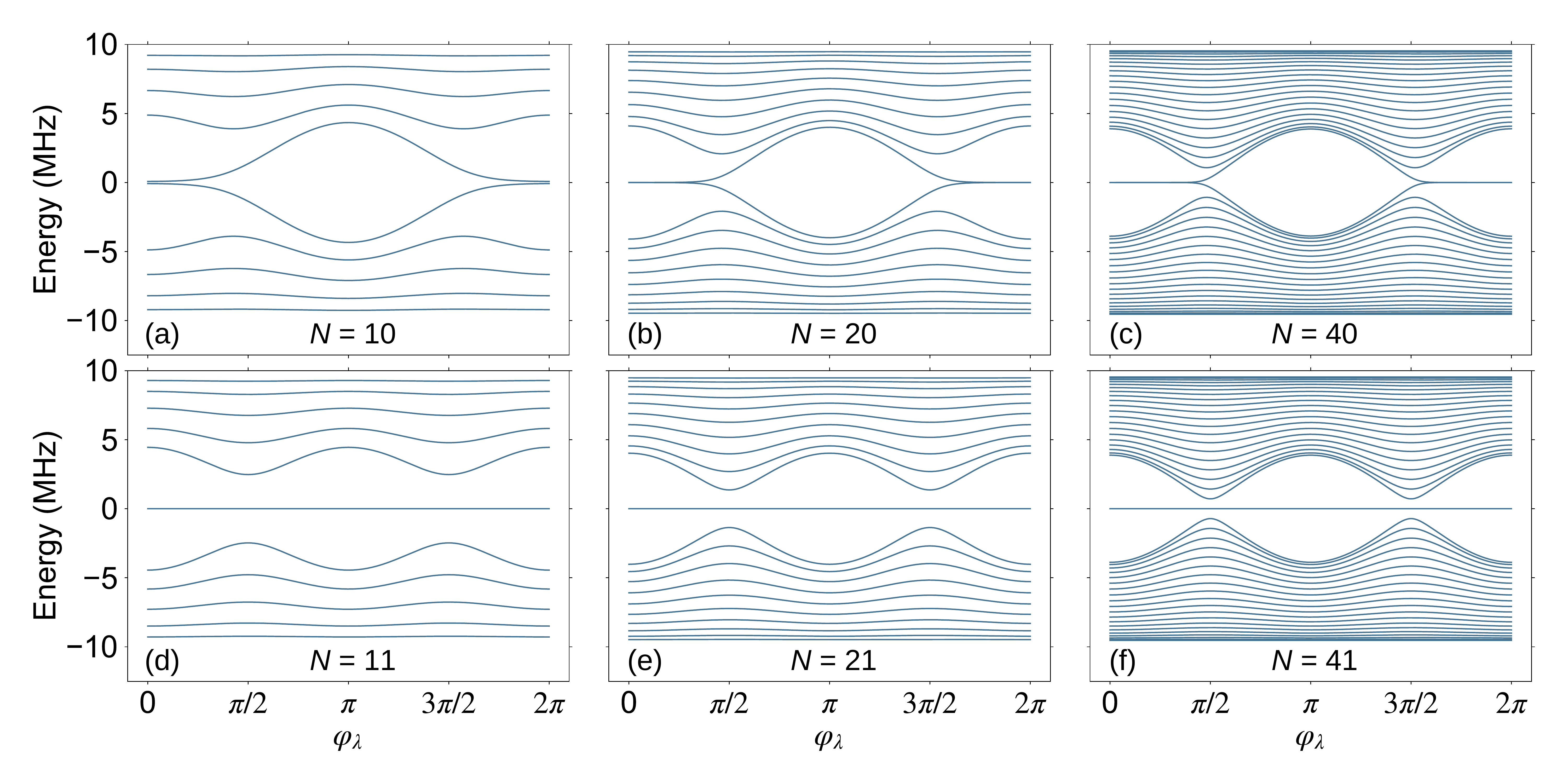}
		\caption{Finite-size effects on the energy spectrum of commensurate off-diagonal AAH models for $\pi$-flux ($b_\lambda={1}/{2}$). Here we set $u/(2\pi)=4.78$~MHz and $\lambda=0.4$.
		}
		\label{fig:size_effect_topo_40_41}
	\end{center}
\end{figure}

\begin{figure}[hb!]
	\begin{center}
		\includegraphics[width=0.99\textwidth]{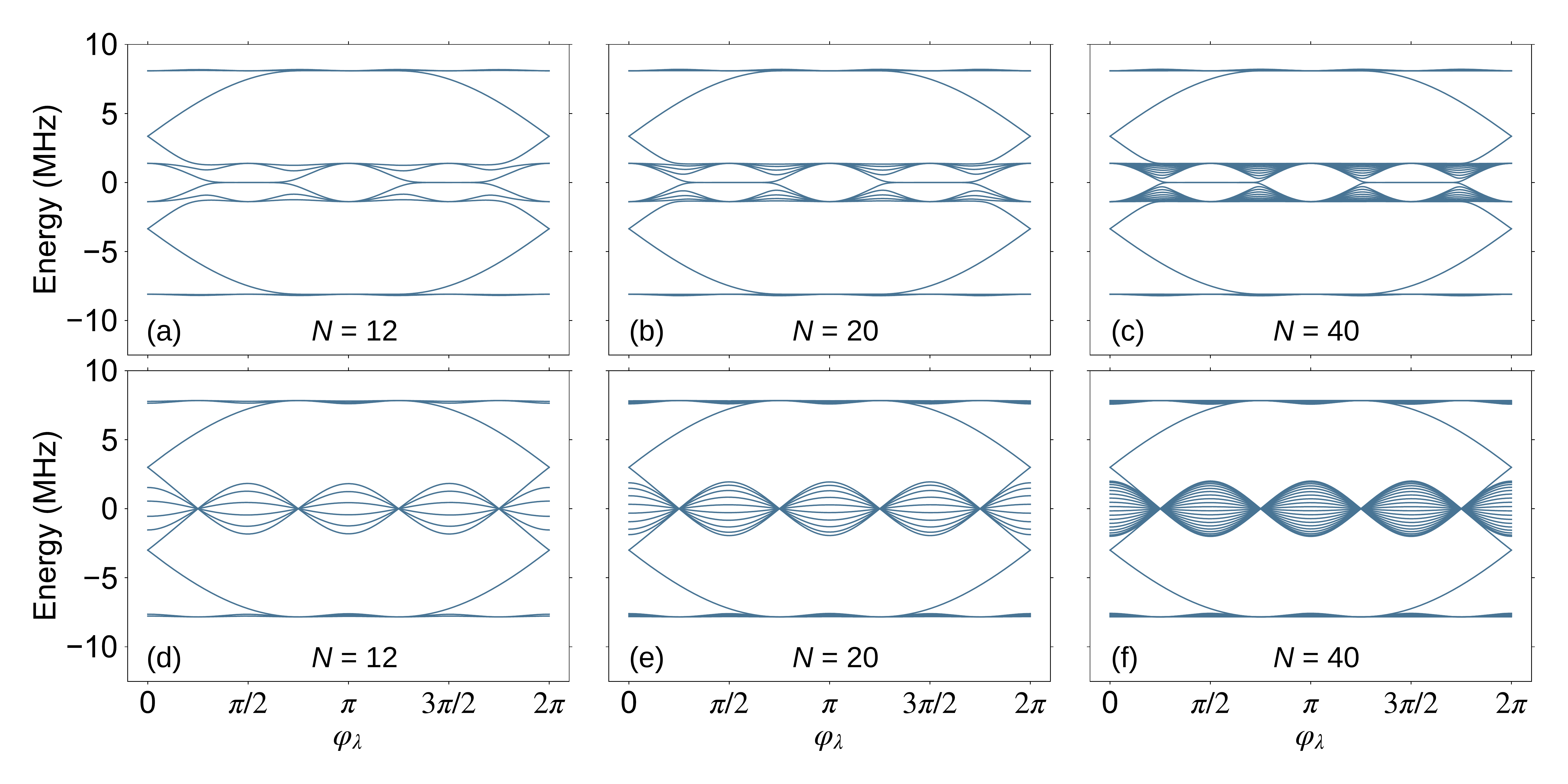}
		\caption{Finite-size effects on the energy spectrum of generic commensurate off-diagonal AAH models with $b_\lambda={1}/{4}$. (a)-(c), Band structure for $\lambda=1$ and $u/(2\pi)=3.35$~MHz. (d)-(e), Band structure for $\lambda=\sqrt{2}$ and $u/(2\pi)=2.77$~MHz.
		}
		\label{fig:size_effect_topo40_1_2}
	\end{center}
\end{figure}

In  Figs.~\ref{fig:size_effect_butterfly}--\ref{fig:size_effect_topo40_1_2}, we show numerical results for the band structures of generalized AAH models with different numbers of qubits.  The clear and rich features in the band structures are attributed to the sufficiently large qubit number of our quantum processor. For instance, the fractal structure of Hofstadter butterfly spectrum is very unclear by using 9 qubits. Even with 20 qubits, this band structure cannot be captured clearly, and moreover, these results are just ideal numerical simulations. Furthermore, our work experimentally simulates topological phases on a NISQ device when the noise and decoherence cannot be ignored.  The exact numerical simulation of the effects of decoherence is beyond the capability of a classical supercomputer.

\begin{figure}[ht!]
	\begin{center}
		\includegraphics[width=0.86\textwidth]{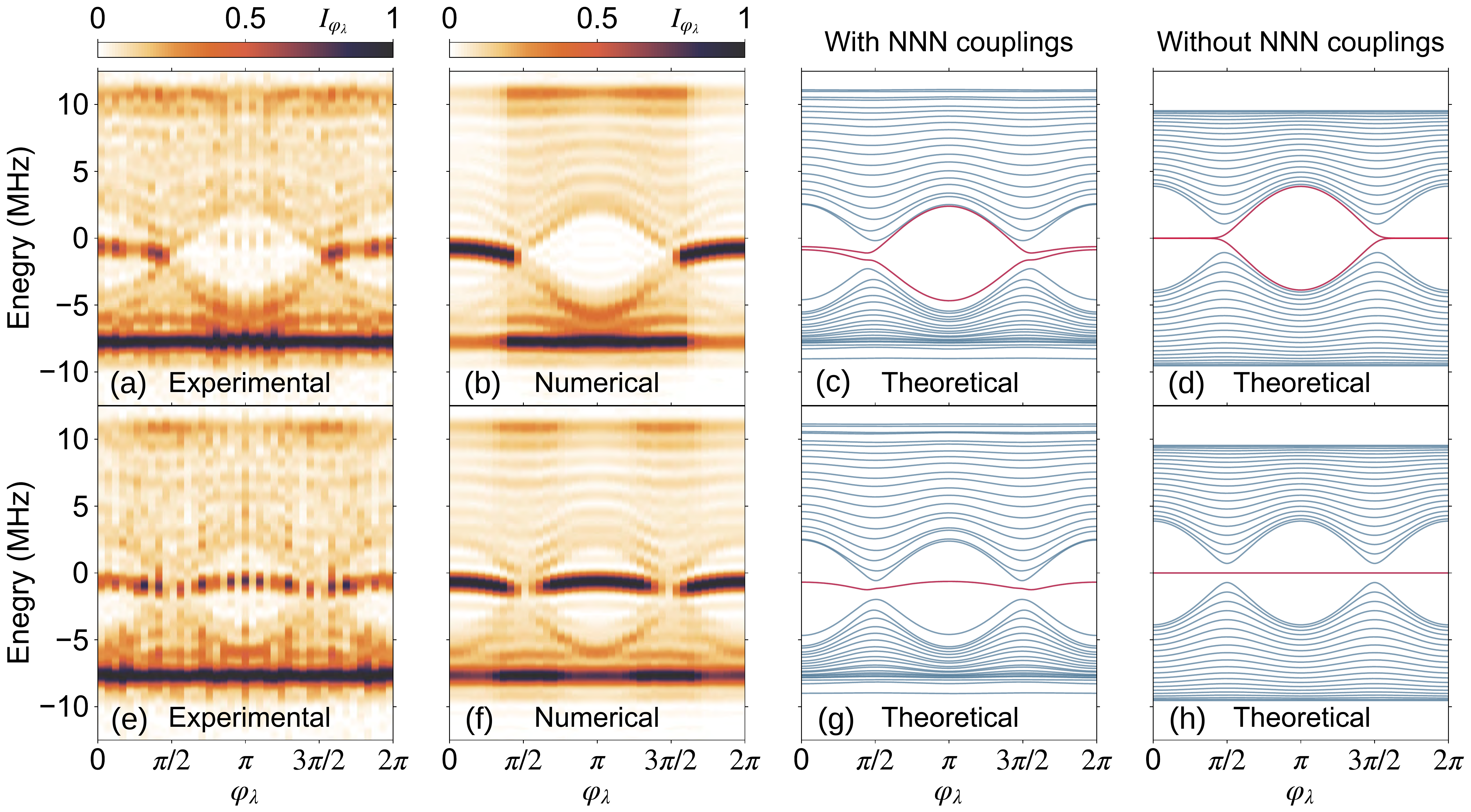}
		\caption{Topological zero-energy edge modes in commensurate off-diagonal AAH models for $\pi$-flux ($b_\lambda={1}/{2}$). (a)-(e) Band structure spectroscopy of off-diagonal AAH models with even number $N=40$ (a-d) and odd number $N=41$ (e)-(h) of sites. Here we set $u/(2\pi)=4.78$~MHz and $\lambda=0.4$. (a) and (e), Experimental data for $I_{\varphi_{\lambda}}$. (b) and (f), Numerical data for $I_{\varphi_{\lambda}}$ considering NNN couplings. (c) and (g), Theoretical eigenenergy spectrum with NNN couplings. (d) and (h), Theoretical eigenenergy spectrum without NNN couplings. 
			The topological edge modes (highlighted in red) are verified to be robust, although the energies are slightly shifted away from zero.
		}
		\label{fig:extended_data_topo_40_41}
	\end{center}
\end{figure}

\begin{figure}[ht!]
	\begin{center}
		\includegraphics[width=0.86 \textwidth]{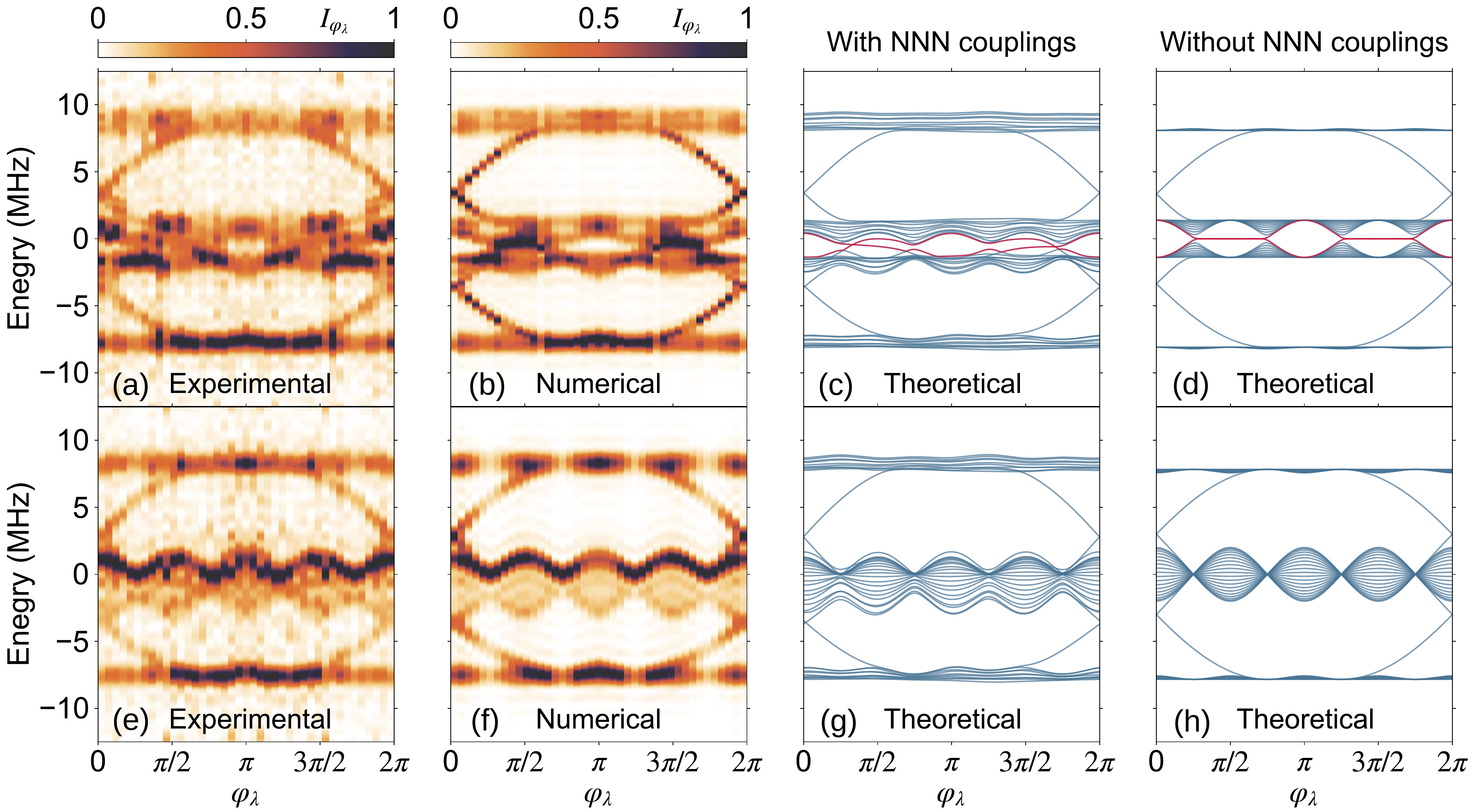}
		\caption{
			Band structure spectroscopy of generic commensurate off-diagonal AAH models
			with $N=40$ sites for $b_\lambda={1}/{4}$.
			(a)-(d), Band structure for $\lambda=1$ and $u/(2\pi)=3.35$~MHz. The gap between two central
			bands opens near $\varphi_\lambda=0$ and $\pi$. The topologically nontrivial zero-energy modes are observed between two central bands.
			(e)-(h), Band structure for $\lambda=\sqrt{2}$ and $u/(2\pi)=2.77$~MHz. The gap between the two central bands closes, and no topological edge states between these two bands are observed.
			(a) and (e), Experimental data for  $I_{\varphi_{\lambda}}$. 
			(b) and (f), Numerical data for $I_{\varphi_{\lambda}}$ considering NNN couplings.
			(c) and (g), Theoretical eigenenergy spectrum with NNN couplings. 
			(d) and (h), Theoretical eigenenergy spectrum without NNN couplings. 
		}
		\label{fig:extended_data_topo40_1_2}
	\end{center}
\end{figure}

In the generalized AAH model, the particle-hole symmetry is usually protected, contributing to the stability of the edge modes~\cite{Ryu2002,Ganeshan2013}. However, the existence of weak next-nearest-neighboring hopping of our sample slightly breaks the particle-hole symmetry. In our sample, this NNN hopping is site-dependent and disordered. The corresponding Hamiltonian is $\hat{H}_{\mathrm{NNN}}=\sum_{j=1}^{N-2}g_{j,j+2}\hat{a}^{\dagger}_{j}\hat{a}_{j+2}+\textrm{H.c.}$, where the average of $g_{j,j+2}/(2\pi)$ is about $0.7$~MHz (standard deviation $0.3$~MHz). In Fig.~\ref{fig:extended_data_topo_40_41}, we compare the experimental results with the theoretical results with and without NNN couplings in the commensurate off-diagonal AAH models with $b_{\lambda}=1/2$. The topological edge modes (highlighted in red) have a small shift of the zero energy due to the existence of weak NNN couplings. However, the experimental results still show the robustness of the topological zero-energy edge states in the commensurate off-diagonal AAH models.
For the generic commensurate off-diagonal AAH models with $b_{\lambda}=1/4$, the NNN coupling opens a gap between the two central bands, leading to the shift of the topological zero-energy edge states to mid-gap edges, as shown in Fig.~\ref{fig:extended_data_topo40_1_2}. This also verifies the robustness of the topological properties of the generic commensurate off-diagonal AAH model.

%